# Homogenization of Rough Surfaces: Effective Surface Stress and Superficial Elasticity

P.Mohammadi[1], L. Liu[1], P. Sharma[1,2,♣], R. V. Kukta[3]
[1]*Department of Mechanical Engineering,*
[2]*Department of Physics*
*University of Houston, Houston, TX, 77204, U.S.A*
[3]*Department of Mechanical Engineering*
*Stony Brook University, Stony Brook, NY, U.S.A.*

**Abstract:** Relating microstructure to properties, electromagnetic, mechanical, thermal and their couplings has been a major focus of mechanics, physics and materials science. The majority of the literature focuses on deriving homogenized constitutive responses for macroscopic composites relating effective properties to various microstructural details. Due to large surface to volume ratio, phenomena at the nanoscale require consideration of surface energy effects and the latter are frequently used to interpret size-effects in material behavior. Elucidation of the effect of surface roughness on the surface stress and elastic behavior is relatively under-studied and quite relevant to the behavior of nanostructures. In this work, we present derivations that relate both periodic and random roughness to the effective surface elastic behavior. We find that the residual surface stress is hardly affected by roughness while the superficial elasticity properties are dramatically altered and, importantly, may also result in a change in its sign----this has ramifications in interpretation of sensing based on frequency measurement changes due to surface elasticity. We show that the square of resonance frequency of a cantilever beam with rough surface decreases as much as three times of its value for flat surface.

## I. Introduction

For a cubic piece of copper with 1 nm sides, nearly 64% of atoms reside on the surface. This simple fact makes apparent the enormous role surfaces play at the nanoscale. Surface atoms have different coordination numbers, charge distribution and subsequently different physical, mechanical and chemical properties. These differences are manifested phenomenologically in that the various bulk properties such as elastic modulus, melting temperature, electromagnetic properties among others are different for surfaces. For example, experiments show that some surfaces are elastically softer (Goudeau et al., 2001; Hurley et al., 2001; Villain et al., 2002; Sun and Zhang, 2003; Workum and Pablo, 2003), while others stiffer (Renault et al., 2003). These differences play an increasing role as the material characteristic size is shrunk smaller and smaller----for example, leading to size-dependency in the elastic modulus of nanostructures.

Surface energy effects are usually accounted for via recourse to a theoretical framework proposed by Gurtin and Murdoch (1975, 1978). The surface is treated as a zero-thickness deformable elastic entity possessing non-trivial elasticity as well as a residual stress (the so-called "surface stress"). It is worthwhile to indicate that while fundamentally similar, a parallel line of works exists that are more materials oriented: Cahn (1989), Streitz (1994), Weissmuller and Cahn (1997), Johnson (2000), Voorhees and Johnson (2004) and Cammarata (1994, 2009a, 2009b) among others. The reader is referred to an extensive recent review by Cammarata (2009) on the literature. Steigmann and Ogden (1997) later generalized the Gurtin–Murdoch theory and

---

♣ Corresponding author: psharma@uh.edu

incorporated curvature dependence on surface energy as well thus resolving some important issues related to the use of Gurtin-Murdoch theory in the context of compressive stress states and for wrinkling type behavior.

A fair amount of literature has appeared that explain various interesting size-effects due to surface energy effects, e.g. nanoinclusions (Sharma et al. 2003, Sharma and Ganti, 2003, Sharma, 2004; Sharma and Ganti, 2004; Duan et al., 2005a, 2005b; He and Li, 2006; Lim et al., 2005; Mi and Kouris 2007; Sharma and Wheeler, 2007; Tian and Rajapakse, 2007, 2008; Hui and Chen, 2010), quantum dots ( Sharma et al. 2002, 2003; Peng et al., 2006), nanoscale beams and plates (Miller and Shenoy, 2000; Jing et al. 2006; Bar et al. 2010; Liu and Rajapakse, 2010), nano particles, wires and films (Streitz et al. 1994; Diao et al. 2003, 2004; Villain et al., 2004; Dingreville et al., 2005; Diao et al., 2006) on sensing and vibration (Lim and He, 2004; Wang and Feng, 2007; Park and Klein, 2008; Park, 2009), composites (Mogilevskaya et al., 2008) and studies on surface properties (Shenoy, 2005; Shodja and Tehranchi, 2010; Mi et al., 2008).

Some recent works are worth mentioning as they provide clarifications and guidance on some of the theories underlying surface energy effects, e.g. Ru (2010), Mogielvskaya (2008, 2010) and Schiavone and Ru (2009). Huang and co-workers (Wang et al., 2010; Huang and Sun, 2007) have pointed out the importance of residual surface stress on elastic properties of nanostructures and composites.

Surfaces of real materials, even the most thoroughly polished ones, will typically exhibit random roughness across different lateral length scales. How are the surface properties renormalized due to such roughness? Can the surface roughness be artificially tailored to obtain desired surface characteristics? These questions are at the heart of the present manuscript. We provide a homogenization scheme for both periodically and randomly rough surface duly incorporating both surface stress and surface elasticity. Very little work has appeared that addresses effect of roughness on both surface stress and surface elasticity. Notable exceptions are the following recent works: Wiessmuller and Duan (2008) who focus on deriving the effective residual stress for the rough surface of a cantilever beam and their follow-up work by Wang et al., (2010) who generalized it to the anisotropic case. We will present a comparison of our work to theirs in due course. One specific difference is that we also derive effective superficial elasticity constants and not just the residual surface stress. The outline of this paper is as follows. In Section II we briefly summarize the Gurtin-Murdoch surface elasticity theory and formulate the problem while in Section III we present our general homogenization strategy. In Section IV, specializing to the 2D case, we present results for both randomly and periodically rough surfaces. Discussion of our results is in Section V. In this section, we compare our work with that of Duan and co-workers (2008) as well as discuss some of the important ramifications of our results including the interpretation of resonance frequency shift of a cantilever beam used for sensing. Several of the derivation details are spread across five appendices.

## II. Surface Elasticity

Consider a semi-infinite elastic media that occupies the region $B = \{(x, y, z) : y < h(x, z)\}$, where the function $h(x, z)$ describes the surface roughness. We denote the bulk and boundary of the media by $B$ and $\partial B$, respectively. Let $\mathbb{C}$ be the fourth-rank bulk stiffness tensor and assume there is no applied body force. In linearized elasticity, the displacement $u : B \to \mathbb{R}^3$ satisfies the equilibrium equation

$$\text{div}[\mathbb{C}\nabla u] = 0 \quad \text{in } B. \tag{1}$$

These equations will be supplemented by traction boundary conditions on the rough surface, which we describe below in detail.

We employ the linearized surface elasticity theory of Gurtin and Murdoch (Gurtin and Murdoch, 1975; Gurtin et. al., 1998). In this theory the surface is modeled as a deformable elastic membrane that adheres to the bulk material without slipping. Let $e_n$ be the outward unit normal to the surface,

$$\mathbb{P} = \mathbb{I} - e_n \otimes e_n \tag{2}$$

be the projection from $\mathbb{R}^3$ to the subspace orthogonal to $e_n$, and

$$M_1 = \{M : Me_n = 0, \quad M^T e_n = 0, \quad M \in \mathbb{R}^{3\times 3}\} \tag{3}$$

be the subspace, where surface strains belong to. Then the surface strain $\varepsilon^s$ is given by

$$\nabla_s u = (\nabla u)\mathbb{P}, \quad Du = \mathbb{P}\nabla_s u, \quad \varepsilon^s = \mathbb{P}\varepsilon\mathbb{P} = \frac{1}{2}\left(Du + (Du)^T\right) \quad \text{on } \partial B, \tag{4}$$

where $\nabla_s$ denotes the surface gradient. We remark that the above equations follow from the kinematic assumption that displacements are continuous up to the surface.

Let $\tau^0 \in \mathbb{R}$ be the magnitude of the residual isotropic stress tensor (often referred to as the surface tension), $\mathbb{I}_s = \mathbb{P}\mathbb{I}\mathbb{P}$ be the identity mapping from $M_1$ to $M_1$, $\lambda^s$ and $\mu^s$ be the surface elastic constants (Láme parameters), and symmetric matrix $\varepsilon_s^0 \in M_1$ be the residual/eigen surface strain such that $\mathbb{C}_s \varepsilon_s^0 = -\tau^0 \mathbb{I}_s$ on $\partial B$. We adopt the linear isotropic surface constitutive law from Gurtin and Murduch (1975), equation (8.6)

$$S = \mathbb{C}_s \left( \varepsilon^s - \varepsilon_s^0 \right) \quad \text{on} \quad \partial B, \tag{5}$$

where $\varepsilon_s^0 = \dfrac{-\tau^0}{2(\lambda^s + \mu^s)} \mathbb{I}_s$, $S$ is the (first) Piola-Kirchhoff surface stress tensor, and $\mathbb{C}_s$ is the isotropic surface elasticity tensor such that for any symmetric $E \in M_1$,

$$\mathbb{C}_s(E) = \lambda^s Tr(E) \mathbb{I}_s + 2\mu^s E \quad \text{on} \quad \partial B. \tag{6}$$

We remark that the surface constitutive law used here (equation 5) is different than Gurtin and Murdoch by a term of $\tau^0 \nabla_s u$. This term leads to asymmetry of the surface stress tensor and quite a few works have chosen to ignore its presence completely (as justified in some cases). The reader is referred to Ru (2010), Mogilevskaya et al. (2008) and Huang (2010) for further discussions on this subject. We anticipate that if $\tau^0 \ll \lambda^s, \mu^s$, the effect of this term is negligible. In Appendix E, we will assess its impact in detail and for the remainder of the calculations, this term will be ignored.

The equilibrium of any sub-surface of $\partial B$ implies that

$$(\mathbb{C} \nabla u) e_n = \text{div}_s \left[ \mathbb{C}_s \left( \varepsilon^s - \varepsilon_s^0 \right) \right] \quad \text{on} \quad \partial B. \tag{7}$$

The above equation can also be regarded as boundary condition for (1). In summary, equations (1) and (7) form the boundary value problem for linearized elasticity with surface effects.

Further, within a non-consequential constant, the elastic energy contributed by the surface is given by (Gurtin and Murdoch, 1975, equation 9.3 and theorem 9.1)

$$\Gamma[u] = \frac{1}{2} \int_{\partial B} \left[ \left( \varepsilon^s - \varepsilon_s^0 \right) \cdot \mathbb{C}_s \left( \varepsilon^s - \varepsilon_s^0 \right) \right]. \tag{8}$$

Below we consider the effective behaviors of a rough surface.

## III. Homogenization Strategy and Problem Formulation

In this section we outline our homogenization strategy for a rough surface. We assume the amplitude of the roughness $h$ is small compared with the average distance $\lambda$ between successive 'peaks' or 'valleys' on the surface, $\delta = \dfrac{h}{\lambda} \ll 1$. This dimensionless number will be the small parameter used in our subsequent perturbation calculations. The overall half space is subject to a uniform in-plane stress $\sigma = \sigma^\infty \hat{\sigma}$, where $\sigma^\infty$ is the magnitude of $\sigma$ and $\hat{\sigma}$ with $|\hat{\sigma}|=1$ is any plane-stress ($xz$-plane) tensor. By (1) and (7) our original problem is to solve for $u: B \to \mathbb{R}^3$:

$$\begin{cases} \text{div}(\mathbb{C}\nabla u) = 0 & \text{in } B, \\ (\mathbb{C}\nabla u)e_n = \text{div}_s\left[\mathbb{C}_s\left(\varepsilon^s - \varepsilon_s^0\right)\right] & \text{on } \partial B, \\ (\mathbb{C}\nabla u)e_x = \sigma_\infty e_x & \text{as } |y| \to \infty. \end{cases} \qquad (9)$$

Due to the presence of the non-trivial boundary condition (9)$_2$, we will find the solution via perturbation theory. We assume that the solution to (9) can be expanded as

$$u = u^{(0)} + \delta u^{(1)} + \delta^2 u^{(2)} + \cdots \qquad (10)$$

Inserting (10) into (9), by (9)$_1$ and (9)$_3$ we have

$$\begin{cases} \text{div}\left[\mathbb{C}\nabla u^{(i)}\right] = 0 \quad i=0,1,2 & \text{in } B^0, \\ \left(\mathbb{C}\nabla u^{(0)}\right)e_x = \sigma^\infty e_x & \text{as } |y| \to \infty, \\ \left(\mathbb{C}\nabla u^{(i)}\right)e_x = 0 \quad i=1,2,\cdots & \text{as } |y| \to \infty, \end{cases} \qquad (11)$$

where $B^0 = \{(x,y,z): y < 0\}$. We notice that the boundary conditions at the infinity are homogenous unless $i=0$.

The boundary conditions on the rough surface, i.e., (9)$_2$, can be converted to an effective boundary condition on the nominal flat surface $\partial B^0$. To this end, we assume that the displacement on $\partial B$ can be obtained by extrapolating from the displacement and their derivatives on $\partial B^0$ through Taylor series expansion. Upon tedious calculations that are outlined in section IV, we find the boundary conditions on the nominal flat surface as

$$\left(\mathbb{C}\nabla u^{(i)}\right)e_2 = t^{(i)} \quad i=0,1,2 \quad \text{on } \partial B^0, \qquad (12)$$

where the detailed expressions for surface traction $t^{(i)}$ are presented in section IV.I. for sinusoidal rough surface and random rough surface. Upon solving (11) and (12) for $u^{(i)}$ (i=0,1,2), we can find the total elastic energy of the half-space as a function of the applied far field stress

$$\begin{aligned} E^{act}(\sigma^\infty) &= \frac{1}{2}\int_B \nabla u \cdot \mathbb{C}(\nabla u) + \frac{1}{2}\int_{\partial B}\left(\varepsilon^s - \varepsilon_s^0\right) \cdot \mathbb{C}_s\left(\varepsilon^s - \varepsilon_s^0\right) \\ &= \frac{1}{2}\int_B \left(\nabla u^{(0)} + \delta\nabla u^{(1)} + \delta^2\nabla u^{(2)}\right) \cdot \mathbb{C}\left(\nabla u^{(0)} + \delta\nabla u^{(1)} + \delta^2\nabla u^{(2)}\right) \\ &\quad + \frac{1}{2}\int_{\partial B}\left[\left(\nabla_s u^{(0)} + \delta\nabla_s u^{(1)} + \delta^2\nabla_s u^{(2)} - \varepsilon^0\right) \cdot \mathbb{C}_s\left(\nabla_s u^{(0)} + \delta\nabla_s u^{(1)} + \delta^2\nabla_s u^{(2)} - \varepsilon^0\right)\right] + O(\delta^3), \end{aligned} \qquad (13)$$

where $u^{(i)}$ (i=0,1,2), the solution of (9), depends on the far applied stress $\sigma^\infty$, and the first and second term on the right hand side of (13) is the elastic energy contributed by the bulk and surface, respectively.

We will approximate this elastic body with rough surface by a half-space solid with a flat surface where the flat surface has effective properties different from the original rough surface. To define the effective properties of the surface, we propose to equate the total elastic energy of the rough-surface half space $\left(E^{act}\right)$ to the total elastic energy of a half space with a nominal "effective" flat surface $\left(E^{eff}\right)$

$$E^{eff}\left(\sigma^\infty\right) = \frac{1}{2}\int_{B^0} \nabla u^{(0)} \cdot \mathbb{C}\left(\nabla u^{(0)}\right) + \frac{1}{2}\int_{\partial B^0} \left(\nabla u^{(0)} - \left(\varepsilon_s^0\right)^{eff}\right) \cdot \mathbb{C}_s^{eff}\left(\nabla u^{(0)} - \left(\varepsilon_s^0\right)^{eff}\right), \tag{14}$$

where $\left(\varepsilon_s^0\right)^{eff}$ is the effective surface residual strain and $\mathbb{C}_s^{eff}$ is the effective surface elasticity tensor. By

$$E^{act}\left(\sigma^\infty\right) = E^{eff}\left(\sigma^\infty\right), \tag{15}$$

we can find the effective surface stress and effective surface elastic modulus.

Next we consider the case when the surface roughness profile is random. Because of the randomness, it is useful to introduce the operator $P$ (called "smoothing operator", Eguiluz and Maradudin, 1983) and $Q$ such that

$$Pu_i = \langle u_i \rangle, \qquad P + Q = 1, \qquad u_i \equiv (P+Q)u_i = \langle u_i \rangle + Qu_i. \tag{16}$$

In order to find the displacement $u_i (i = x, y, z)$, we need to find its average $\langle u_i \rangle$ over the ensemble of realization of the surface roughness and its fluctuation component $Qu_i$. Direct calculations show that the average field satisfies an effective (non-stochastic) problem that is formally similar to the flat surface problem. So in this case we have

$$div\left(\mathbb{C}\left\langle\nabla u^{(i)}\right\rangle\right) = 0 \quad i = 0,1,2 \quad \text{in } B^0 \tag{17}$$

with ensemble average of boundary conditions

$$\left(\mathbb{C}\left\langle\nabla u^{(i)}\right\rangle\right).e_2 = \left\langle t^{(i)}\right\rangle \quad i = 0,1,2 \quad \text{on } \partial B^0. \tag{18}$$

To solve these boundary value problems as will be seen later, we need to find the $Q$-terms which similarly satisfy the equilibrium equations.

$$div\left(\mathbb{C}Q\left(\nabla u^{(i)}\right)\right) = 0 \quad i = 0,1,2 \quad \text{in } B^0 \tag{19}$$

with boundary conditions

$$\left(\mathbb{C}Q\left(\nabla u^{(i)}\right)\right)e_2 = Q\left(t^{(i)}\right) \quad i = 0,1,2 \quad \text{on } \partial B^0. \tag{20}$$

To define the effective surface properties of the surface, we propose to equate the ensemble average of the total elastic energy of the half space with rough surface, $\langle E^{act}(\sigma^\infty)\rangle$, with the total elastic energy of a half space with a nominal "effective" flat surface, $E^{eff}(\sigma^\infty)$, endowed with effective surface stress and surface elasticity constants:

$$E^{eff}(\sigma^\infty) = \langle E^{act}(\sigma^\infty)\rangle. \tag{21}$$

The above equation will enable us to find the effective surface stress and effective surface elastic constants in a similar manner as for nonrandom cases.

## IV. Solution

### IV.I. General Procedure

We now specialize to two dimensions. Our work can be readily extended to three dimensions. However, the calculations are quite tedious with relatively little prospects for (additional) novel insights. We will employ two coordinate systems. The first one is the standard Cartesian frame $(e_1, e_2)$ aligned along the nominally flat surface while the second one $(e_n, e_s)$ is the unit normal and unit tangent along the curve. (See Figure 1)

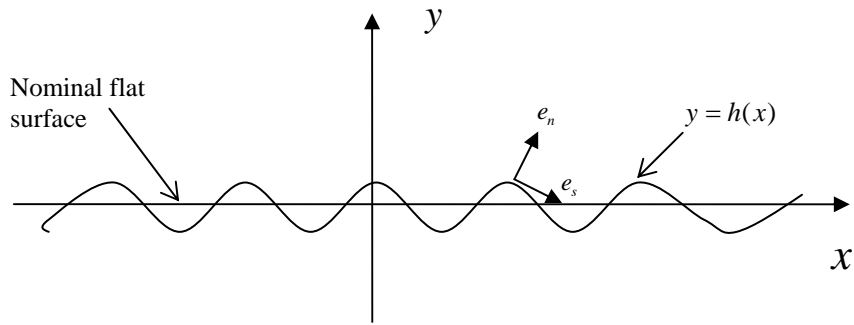

Figure 1: A rough surface profile

If the surface is given by $y = h(x)$, we have the following differential geometry results for plane curves (Frenet formulae):

$$e_s = \frac{e_1 + e_2 h_x}{\sqrt{1+h_x^2}}, \quad e_n = \frac{-h_x e_1 + e_2}{\sqrt{1+h_x^2}}, \quad \frac{de_s}{ds} = \kappa e_n, \quad \frac{de_n}{ds} = -\kappa e_s, \quad (22)$$

where $\kappa = \dfrac{h_{xx}}{(1+h_x^2)^{3/2}}$ is the curvature. Let $u : B \to \mathbb{R}^2$ be the displacement. On the surface $\partial B$, let $u_s$ and $u_n$ be the displacements in the tangential and normal directions,

$$u = u_s e_s + u_n e_n. \quad (23)$$

By (4), the surface strain is given by

$$\varepsilon^s = \varepsilon_{ss} e_s \otimes e_s, \quad \varepsilon_{ss} = \left(\frac{\partial u_s}{\partial s} - \kappa u_n\right) \quad (24)$$

By (5) and (6), the surface stress is given by

$$S = \tau^o e_s \otimes e_s + k^s \varepsilon_{ss} e_s \otimes e_s, \quad k^s \equiv \lambda^s + 2\mu^s. \quad (25)$$

We remark that the above surface strain-stress relation is reminiscent of the familiar bulk strain-stress relation for plain strain. By (25), the boundary condition (7) is now re-written as

$$(\mathbb{C}\nabla u) e_n = \left(\tau^o \kappa + k^s \kappa \varepsilon_{ss}\right) e_n + \left(k^s \frac{\partial \varepsilon_{ss}}{\partial s}\right) e_s \quad \text{on } \partial B \quad (26)$$

Below we convert the above boundary condition to Cartesian coordinates using expressions in Appendix A. In regard of the assumption of small-roughness, the displacement on the rough surface may be approximated by

$$[u(x,y)]_{y=h(x)} = u(x,0) + \delta h_0(x) \left[\frac{\partial u(x,y)}{\partial y}\right]_{y=0} + \frac{1}{2}\delta^2 h_0^2(x) \left[\frac{\partial^2 u(x,y)}{\partial y^2}\right]_{y=0} + \cdots, \quad (27)$$

where we have assumed that the Taylor expansion is valid around the surface. Inserting (10) and (27) into (26) and keeping terms up to $O(\delta^2)$, we find the boundary conditions on the nominal flat surface for $u^{(i)}$ ($i = 0,1,2$), i.e., the right hand side of (12) as follows.

$$t^{(0)} = \left(t_x^{(0)}, t_y^{(0)}\right), \qquad t_x^{(0)} = k^s \frac{\partial \varepsilon_{xx}^{(0)}}{\partial x}, \qquad t_y^{(0)} = 0, \tag{28}$$

$$\begin{aligned}
t^{(1)} &= \left(t_x^{(1)}, t_y^{(1)}\right), \\
t_x^{(1)} &= h_{0x}\sigma_{xx}^{(0)} - h_0 \frac{\partial \sigma_{xy}^{(0)}}{\partial y} + k^s \frac{\partial \varepsilon_{xx}^{(1)}}{\partial x} + k^s h_0 \frac{\partial^2 \varepsilon_{xx}^{(0)}}{\partial x \partial y} \\
&\quad + 2k^s h_{0xx}\varepsilon_{xy}^{(0)} + 2k^s h_{0x} \frac{\partial \varepsilon_{xy}^{(0)}}{\partial x} + k^s h_{0x} \frac{\partial \varepsilon_{xx}^{(0)}}{\partial y}, \\
t_y^{(1)} &= h_{0x}\sigma_{yx}^{(0)} - h_0 \frac{\partial \sigma_{yy}^{(0)}}{\partial y} + \tau^o h_{0xx} + (k^s) h_{0xx}\varepsilon_{xx}^{(0)} + (k^s) h_{0x} \frac{\partial \varepsilon_{xx}^{(0)}}{\partial x}
\end{aligned} \tag{29}$$

and

$$\begin{aligned}
t^{(2)} &= \left(t_x^{(2)}, t_y^{(2)}\right), \\
t_x^{(2)} &= h_{0x}\sigma_{xx}^{(1)} + h_0 h_{0x} \frac{\partial \sigma_{xx}^{(0)}}{\partial y} + \frac{1}{2}h_{0x}^2 \sigma_{xy}^{(0)} - h_0 \frac{\partial \sigma_{xy}^{(1)}}{\partial y} - \frac{1}{2}h_0^2 \frac{\partial^2 \sigma_{xy}^{(0)}}{\partial y^2} - \tau^o h_{0x}h_{0xx} \\
&\quad - k^s h_{0x}h_{0xx}\varepsilon_{xx}^{(0)} + k^s \frac{\partial \varepsilon_{xx}^{(2)}}{\partial x} + k^s h_0 \frac{\partial^2 \varepsilon_{xx}^{(1)}}{\partial x \partial y} + \frac{1}{2}k^s \left(h_0\right)^2 \frac{\partial^3 \varepsilon_{xx}^{(0)}}{\partial x \partial y^2} - k^s h_{0x}^2 \frac{\partial \varepsilon_{xx}^{(0)}}{\partial x} \\
&\quad + 2k^s h_{0xx}\varepsilon_{xy}^{(1)} + 2k^s h_{0xx}h_0 \frac{\partial \varepsilon_{xy}^{(0)}}{\partial y} + 2k^s h_{0x} \frac{\partial \varepsilon_{xy}^{(1)}}{\partial x} + 2k^s h_{0x}h_0 \frac{\partial^2 \varepsilon_{xy}^{(0)}}{\partial x \partial y} \\
&\quad + 2k^s h_{0x}h_{0xx}\left(\varepsilon_{yy}^{(0)} - \varepsilon_{xx}^{(0)}\right) + k^s h_{0x}^2\left(\frac{\partial \varepsilon_{yy}^{(0)}}{\partial x} - \frac{1}{2}\frac{\partial \varepsilon_{xx}^{(0)}}{\partial x}\right) + k^s h_{0x} \frac{\partial \varepsilon_{xx}^{(1)}}{\partial y} \\
&\quad + k^s h_{0x} h_0 \frac{\partial^2 \varepsilon_{xx}^{(0)}}{\partial y^2} + 2k^s \left(h_{0x}\right)^2 \frac{\partial \varepsilon_{xy}^{(0)}}{\partial y}, \\
t_y^{(2)} &= h_{0x}\sigma_{yx}^{(1)} + h_{0x}h_0 \frac{\partial \sigma_{yx}^{(0)}}{\partial y} + \frac{1}{2}h_{0x}^2 \sigma_{yy}^{(0)} - h_0 \frac{\partial \sigma_{yy}^{(1)}}{\partial y} - \frac{1}{2}h_0^2 \frac{\partial^2 \sigma_{yy}^{(0)}}{\partial y^2} + k^s h_{0xx}\varepsilon_{xx}^{(1)} \\
&\quad + k^s h_{0xx}h_0 \frac{\partial \varepsilon_{xx}^{(0)}}{\partial y} + 2k^s h_{0x}h_{0xx}\varepsilon_{xy}^{(0)} + h_{0x}k^s \frac{\partial \varepsilon_{xx}^{(1)}}{\partial x} + h_{0x}h_0 k^s \frac{\partial^2 \varepsilon_{xx}^{(0)}}{\partial x \partial y} + 2k^s h_{0x}h_{0xx}\varepsilon_{xy}^{(0)} \\
&\quad + 2k^s h_{0x}h_{0x} \frac{\partial \varepsilon_{xy}^{(0)}}{\partial x} + k^s h_{0x}h_{0x} \frac{\partial \varepsilon_{xx}^{(0)}}{\partial y} + \tau^o h_{0x}h_0 \frac{\partial^3 u_y^{(0)}}{\partial x \partial y^2} + h_{0x}k^s \frac{\partial \varepsilon_{xx}^{(1)}}{\partial x} + h_{0x}h_0 k^s \frac{\partial^2 \varepsilon_{xx}^{(0)}}{\partial x \partial y} \\
&\quad + 2k^s h_{0x}h_{0xx}\varepsilon_{xy}^{(0)} + 2k^s h_{0x}h_{0x} \frac{\partial \varepsilon_{xy}^{(0)}}{\partial x} + k^s h_{0x}h_{0x} \frac{\partial \varepsilon_{xx}^{(0)}}{\partial y} - \tau^o h_{0x}h_{0xx} \frac{\partial u_y^{(0)}}{\partial x},
\end{aligned} \tag{30}$$

where $h_{0x} = \frac{\partial h_0}{\partial x}$, $h_{0xx} = \frac{\partial^2 h_0}{\partial x^2}$, etc. We remark that the elasticity problems (11) and (12) for $u^{(i)}$ $(i = 0, 1, 2)$ are now the classical Cerruti-Boussinesq half-space problems whose solutions can be found in text books, e.g., Johnson, 1985. Upon specifying the surface roughness profile $h_0(x)$, we can solve (11) and (12) for the elastic fields, compute the total elastic energy (13) and (14) and find the effective properties of the nominal flat surface by using (15). Below we present the detailed calculations for a sinusoidal surface and a random surface.

**Sinusoidal roughness profile**

To fix the idea we first consider a sinusoidal rough surface. Let the surface be given by $h(x) = a\cos kx$ ($\delta = ak \ll 1$). This rough surface may be regarded as a perturbation of the flat surface $\{(x,y): y = 0\}$:

$$h(x) = 0 + \frac{ak}{k}\cos kx = \delta h_0, \quad h_0 = \frac{\cos kx}{k}, \quad \delta = ak \ll 1 \qquad (31)$$

Assume that the far-field stress is given by $\sigma = \sigma^\infty e_1 \otimes e_1$. By (28)-(30),

$$t_x^{(0)} = 0, \quad t_y^{(0)} = 0, \qquad (32)$$

$$t_x^{(1)} = -\sigma^\infty \sin kx, \quad t_y^{(1)} = -k\left(\tau^o + k^s \frac{(1-\nu^2)}{E}\sigma^\infty\right)\cos kx, \qquad (33)$$

$$t_x^{(2)} = \beta \sin 2kx, \quad t_y^{(2)} = \gamma \cos 2kx, \qquad (34)$$

where the constants $\beta$ and $\gamma$ are given by

$$\beta = 2\sigma^\infty + \frac{1}{2}\tau^o k + \frac{5}{2}\frac{1-\nu^2}{E}k^s k\sigma^\infty + 2\frac{(1-\nu^2)}{E}k^2 k^s\left(\tau^o + k^s\frac{(1-\nu^2)}{E}\sigma^\infty\right) + \frac{2(1+\nu)}{E}k^s k\sigma^\infty,$$

and $\qquad (35)$

$$\gamma = -\sigma^\infty + \frac{(1-2\nu)(1+\nu)}{E}k^2 k^s\left(\tau^o + k^s\frac{(1-\nu^2)}{E}\sigma^\infty\right) + 2\sigma^\infty kk^s\frac{(1-\nu^2)}{E}.$$

The detailed calculations are presented in Appendix C.

By solving (11) and (12), and the right hand side (12) given by (32), we obtain the zeroth order strain field

$$\varepsilon_{xx}^{(0)} = \frac{1-\nu^2}{E}\sigma^\infty, \quad \varepsilon_{yy}^{(0)} = \frac{-\nu(1+\nu)}{E}\sigma^\infty \quad \text{in } B^0 \qquad (36)$$

By solving (11) and (12), with the right hand side (12) given by (33), the first order strain field is given by

$$\varepsilon_{xx}^{(1)} = \frac{1-v^2}{E}\left\{-\sigma^\infty(2+ky)e^{ky} - k\left(\tau^o + k^s\frac{(1-v^2)}{E}\sigma^\infty\right)(1+ky)e^{ky}\right\}\cos kx -$$

$$-\frac{v(1+v)}{E}\left\{\sigma^\infty kye^{ky} + k\left(\tau^o + k^s\frac{(1-v^2)}{E}\sigma^\infty\right)(ky-1)e^{ky}\right\}\cos kx \quad \text{in } B^0,$$

$$\varepsilon_{yy}^{(1)} = -\frac{v(1+v)}{E}\left\{-\sigma^\infty(2+ky)e^{ky} - k\left(\tau^o + k^s\frac{(1-v^2)}{E}\sigma^\infty\right)(1+ky)e^{ky}\right\}\cos kx \quad (37)$$

$$+\frac{1-v^2}{E}\left\{\sigma^\infty(ky)e^{ky} + k\left(\tau^o + k^s\frac{(1-v^2)}{E}\sigma^\infty\right)(ky-1)e^{ky}\right\}\cos kx \quad \text{in } B^0,$$

$$\varepsilon_{xy}^{(1)} = \frac{(1+v)}{E}\left\{-\sigma^\infty(ky+1)e^{ky} - k\left(\tau^o + k^s\frac{(1-v^2)}{E}\sigma^\infty\right)kye^{ky}\right\}\sin kx \quad \text{in } B^0.$$

By solving (11) and (12), with the right hand side (12) given by (34), the second order solution is obtained

$$\varepsilon_{xx}^{(2)} = \frac{1-v^2}{E}\left\{2\beta(1+ky)e^{2ky}\cos 2kx + \gamma(1+2ky)e^{2ky}\cos 2kx\right\} -$$
$$-\frac{v(1+v)}{E}\left\{-2\beta kye^{2ky}\cos 2kx + \gamma(1-2ky)e^{2ky}\cos 2kx\right\} \quad \text{in } B^0. \quad (38)$$

We remark that the energy contributed by strain fields, $\varepsilon_{xy}^{(2)}$ and $\varepsilon_{yy}^{(2)}$, are negligible compared with $\delta^2$. We therefore do not present them here. By using the relations in Appendix A and the Taylor series extrapolation, surface strain on the rough surface is given by

$$[\varepsilon_{ss}]_{y=h(x)} = \frac{(1-v^2)}{E}\sigma^\infty + \delta\left\{-\frac{(1-2v)(1+v)}{E}k\left(\tau^o + k^s\frac{(1-v^2)}{E}\sigma^\infty\right) - 2\sigma^\infty\frac{(1-v^2)}{E}\right\}\cos kx$$

$$+\delta^2\left\{\begin{bmatrix}-3\frac{(1-v^2)}{E}\sigma^\infty - 2\frac{(1-v^2)}{E}k\left(\tau^o + k^s\frac{(1-v^2)}{E}\sigma^\infty\right) - \frac{v(1+v)}{E}\sigma^\infty\end{bmatrix}\cos^2 kx + \\ +\frac{(1+v)}{E}\sigma_\infty\sin^2 kx + \eta\cos 2kx\end{bmatrix}\right\}, \quad (39)$$

where $\eta = \frac{(1-2v)(1+v)}{E}\gamma + \frac{2(1-v^2)}{E}\beta$.

As mentioned in Sec. III, our homogenization scheme requires calculation of the total energy under the action of the applied stress $\sigma^\infty$. Inserting (36)-(39) into (13), we obtain

$$E^{act}\left(\sigma^{\infty}\right)=\frac{1}{2}\left(\frac{1}{\lambda}\right)\int_{0}^{\lambda}\int_{-\infty}^{h(x)}\left(\begin{array}{l}\varepsilon^{(0)}\cdot\mathbb{C}\varepsilon^{(0)}+2\delta\varepsilon^{(1)}\cdot\mathbb{C}\varepsilon^{(0)}+\\ +\delta^{2}\left(2\varepsilon^{(2)}\cdot\mathbb{C}\varepsilon^{(0)}+\varepsilon^{(1)}\cdot\mathbb{C}\varepsilon^{(1)}\right)\end{array}\right)\sqrt{1+\delta^{2}\sin^{2}kx}dxdy+$$
$$+\frac{1}{2}\left(\frac{1}{\lambda}\right)\int_{0}^{\lambda}\left(\left(\varepsilon_{ss}-\varepsilon_{ss}^{0}\right)\cdot\mathbb{C}_{s}\left(\varepsilon_{ss}-\varepsilon_{ss}^{0}\right)\right)\sqrt{1+\delta^{2}\sin^{2}kx}dx.$$
(40)

where $\varepsilon_{ss}$ is given by (39). By (14),

$$E^{eff}\left(\sigma^{\infty}\right)=\frac{1}{2}\left(\frac{1}{\lambda}\right)\int_{0}^{\lambda}\int_{-\infty}^{0}\varepsilon^{(0)}\cdot\mathbb{C}\varepsilon^{(0)}dydx$$
$$+\frac{1}{\lambda}\int_{0}^{\lambda}\left(\left(\varepsilon_{xx}^{(0)}e_{1}\otimes e_{1}-\left(\varepsilon_{s}^{0}\right)^{eff}\right)\cdot\mathbb{C}_{s}^{eff}\left(\varepsilon_{xx}^{(0)}e_{1}\otimes e_{1}-\left(\varepsilon_{s}^{0}\right)^{eff}\right)\right)dx,$$
(41)

where $\varepsilon_{xx}^{(0)}=\frac{\left(1-v^{2}\right)}{E}\sigma^{\infty}$ is given by (36). Let

$$\left(\tau^{0}\right)^{eff}=e_{1}\otimes e_{1}\cdot\mathbb{C}_{s}^{eff}\left(\varepsilon_{s}^{0}\right)^{eff}$$
(42)

and

$$\left(k^{s}\right)^{eff}=e_{1}\otimes e_{1}\cdot\mathbb{C}_{s}^{eff}e_{1}\otimes e_{1}.$$
(43)

Therefore by $E^{act}=E^{eff}$, we have the effective properties of the rough surface given by

$$\left(\tau^{0}\right)^{eff}=\left(\frac{E}{1-v^{2}}\right)\frac{\partial E^{act}\left(\sigma^{\infty}\right)}{\partial\sigma^{\infty}}\bigg|_{\sigma^{\infty}=0}\quad\text{and}\quad\left(k^{s}\right)^{eff}=\left(\frac{E}{1-v^{2}}\right)^{2}\frac{\partial^{2}E^{act}\left(\sigma^{\infty}\right)}{\left(\partial\sigma^{\infty}\right)^{2}}\bigg|_{\sigma^{\infty}=0}.$$
(44)

Inserting (36)-(39) into (40), by (44), we obtain

$$\left(\tau^{0}\right)^{eff}=\tau^{o}\left[1+\delta^{2}\left(-\frac{3}{4}-kk^{s}\frac{\left(1+8v\right)\left(1+v\right)}{8E}+\frac{\left(1-2v\right)^{2}\left(1+v\right)^{2}}{2E^{2}}k^{2}\left(k^{s}\right)^{2}\right)\right],$$
(45)

$$\left(k^{s}\right)^{eff}=k^{s}+\delta^{2}\left(\begin{array}{l}-\frac{E}{k\left(1-v^{2}\right)}\frac{\left(9-8v\right)}{8\left(1-v\right)}+\frac{1}{4}k^{s}+\frac{\left(1-2v\right)^{2}\left(1+v\right)^{2}}{2E^{2}}k^{2}\left(k^{s}\right)^{3}+\\ \frac{\left(1+v\right)}{E}\frac{\left(-24v+7\right)}{8}k\left(k^{s}\right)^{2}\end{array}\right).$$
(46)

If $\dfrac{kk^s}{E} \ll 1$, equations (45) and (46) can be further simplified as

$$\left(\tau^0\right)^{eff} = \tau^o\left(1 - \dfrac{3}{4}\delta^2\right), \qquad (47)$$

and

$$\left(k^s\right)^{eff} = k^s - \delta^2 \dfrac{E}{k(1-v^2)}\dfrac{(9-8v)}{8(1-v)}, \qquad (48)$$

respectively.

**Random roughness profile**

For random rough surfaces, which are typical for materials that are not artificially engineered, we need to find the average displacement $\langle u^{(i)}\rangle$ and the fluctuating displacement $Qu^{(i)}$ over an ensemble of the surface roughness (Eguiluz and Maradudin, 1983a, 1983b). Recall two useful operators P and Q, which are such that

$$Pu^{(i)} = \langle u^{(i)}\rangle, \qquad P + Q = 1, \qquad u^{(i)} \equiv (P+Q)u^{(i)} = \langle u^{(i)}\rangle + Qu^{(i)}. \qquad (49)$$

We remark that the fluctuation $Qu^{(i)}$ is at the order of $\delta Pu^{(i)}$ $\forall\ i = 0,1,2,\cdots$, which will be repeatedly used below. We further assume that the random roughness satisfies that for a constant $\eta > 0$,

$$Ph(x) = 0, \qquad Ph^2(x) = \eta^2, \qquad (50)$$

where $\eta$ is the standard deviation of the roughness.

Applying operator $P$ to equations (28)-(30) and assuming that $h = \delta h_0$, $h_0 \sim 1$ lead to the following results

$$\langle t_x^{(0)}\rangle = k^s \dfrac{\partial \langle \varepsilon_{xx}^{(0)}\rangle}{\partial x}, \qquad \langle t_y^{(0)}\rangle = 0, \qquad (51)$$

$$\langle t_x^{(1)}\rangle = -P\left[h_0 \dfrac{\partial Q\sigma_{xy}^{(0)}}{\partial y}\right] + P\left[h_{0x} Q\sigma_{xx}^{(0)}\right] + k^s \dfrac{\partial \langle \varepsilon_{xx}^{(1)}\rangle}{\partial x} + k^s P\left[h_0 \dfrac{\partial^2 Q\varepsilon_{xx}^{(0)}}{\partial x \partial y}\right] + $$
$$+ 2k^s P\left[h_{0xx} Q\varepsilon_{xy}^{(0)}\right] + 2k^s P\left[h_{0x} \dfrac{\partial Q\varepsilon_{xy}^{(0)}}{\partial x}\right] + k^s P\left[h_{0x} \dfrac{\partial Q\varepsilon_{xx}^{(0)}}{\partial y}\right], \qquad (52)$$

$$\left\langle t_y^{(1)} \right\rangle = -P\left[ h_0 \frac{\partial Q\sigma_{yy}^{(0)}}{\partial y} \right] + P\left[ h_{0x} Q\sigma_{yx}^{(0)} \right] + \tau^o \left\langle h_{0xx} \right\rangle + k^s P\left[ h_{0xx} Q\varepsilon_{xx}^{(0)} \right] + k^s P\left[ h_{0x} \frac{\partial Q\varepsilon_{xx}^{(0)}}{\partial x} \right],$$

$$\left\langle t_x^{(2)} \right\rangle = P\left[ h_{0x} Q\sigma_{xx}^{(1)} \right] + P\left[ h_0 h_{0x} \right] \frac{\partial \left\langle \sigma_{xx}^{(0)} \right\rangle}{\partial y} + \frac{1}{2} P\left[ h_0 h_{0x} \right] \left\langle \sigma_{xy}^{(0)} \right\rangle - P\left[ h_0 \frac{\partial Q\sigma_{xy}^{(1)}}{\partial y} \right] -$$

$$- \frac{1}{2} \eta^2 \frac{\partial^2 \left\langle \sigma_{xy}^{(0)} \right\rangle}{\partial y^2} - \tau^o P\left[ h_{0x} h_{0xx} \right] - k^s P\left[ h_{0x} h_{0xx} \right] \left\langle \varepsilon_{xx}^{(0)} \right\rangle + k^s \frac{\partial \left\langle \varepsilon_{xx}^{(2)} \right\rangle}{\partial x} + k^s P\left[ h_0 \frac{\partial^2 Q\varepsilon_{xx}^{(1)}}{\partial x \partial y} \right]$$

$$+ \frac{1}{2} \eta^2 k^s \frac{\partial^3 \left\langle \varepsilon_{xx}^{(0)} \right\rangle}{\partial x \partial y^2} - k^s P\left[ h_{0x} h_{0x} \right] \frac{\partial \left\langle \varepsilon_{xx}^{(0)} \right\rangle}{\partial x} + 2k^s P\left[ h_{0xx} Q\varepsilon_{xy}^{(1)} \right] + 2k^s P\left[ h_{0xx} h_0 \right] \frac{\partial \left\langle \varepsilon_{xy}^{(0)} \right\rangle}{\partial y}$$

$$+ 2k^s P\left[ h_{0x} \frac{\partial Q\varepsilon_{xy}^{(1)}}{\partial x} \right] + 2k^s P\left[ h_{0x} h_0 \right] \frac{\partial^2 \left\langle \varepsilon_{xy}^{(0)} \right\rangle}{\partial x \partial y} + 2k^s P\left[ h_{0x} h_{0xx} \right] \left( \left\langle \varepsilon_{yy}^{(0)} \right\rangle - \left\langle \varepsilon_{xx}^{(0)} \right\rangle \right) +$$

$$+ k^s P\left[ h_{0x} h_{0x} \right] \left( \frac{\partial \left\langle \varepsilon_{yy}^{(0)} \right\rangle}{\partial x} - \frac{1}{2} \frac{\partial \left\langle \varepsilon_{xx}^{(0)} \right\rangle}{\partial x} \right) + k^s P\left[ h_{0x} \frac{\partial Q\varepsilon_{xx}^{(1)}}{\partial y} \right] + k^s P\left[ h_0 h_{0x} \right] \frac{\partial^2 \left\langle \varepsilon_{xx}^{(0)} \right\rangle}{\partial y^2} +$$

$$+ 2k^s P\left[ h_{0x} h_{0x} \right] \frac{\partial \left\langle \varepsilon_{xy}^{(0)} \right\rangle}{\partial y}, \tag{53}$$

$$\left\langle t_y^{(2)} \right\rangle = -P\left[ h_0 \frac{\partial Q\sigma_{yy}^{(1)}}{\partial y} \right] - \frac{1}{2} \eta^2 \frac{\partial^2 \left\langle \sigma_{yy}^{(0)} \right\rangle}{\partial y^2} + P\left[ h_{0x} Q\sigma_{yx}^{(1)} \right] + P\left[ h_0 h_{0x} \frac{\partial Q\sigma_{yx}^{(0)}}{\partial y} \right] +$$

$$+ k^s P\left[ h_{0xx} Q\varepsilon_{xx}^{(1)} \right] + k^s P\left[ h_{0x} \frac{\partial Q\varepsilon_{xx}^{(1)}}{\partial x} \right] + 2k^s \varepsilon^2 \frac{\partial \left\langle \varepsilon_{xy}^{(0)} \right\rangle}{\partial x} + k^s \varepsilon^2 \frac{\partial \left\langle \varepsilon_{xx}^{(0)} \right\rangle}{\partial y} +$$

$$k^s P\left[ h_{0xx} h_0 \frac{\partial Q\varepsilon_{xx}^{(0)}}{\partial y} \right] + 2k^s P\left[ h_{0x} h_{0xx} Q\varepsilon_{xy}^{(0)} \right] + k^s P\left[ h_{0x} h_0 \frac{\partial^2 Q\varepsilon_{xx}^{(0)}}{\partial x \partial y} \right] +$$

$$+ 2k^s P\left[ h_{0x} h_{0xx} Q\varepsilon_{xy}^{(0)} \right].$$

Equation (18) with $\left\langle t^{(i)} \right\rangle$ given by (51)-(53) prescribe the traction boundary conditions for $\left\langle u^{(i)} \right\rangle$ on the average, nominal surface. The average displacement $\left\langle u^{(i)} \right\rangle$ can then be obtained by solving (17) and (18) which are again the Cerruti-Bossinesq problems. From (52) and (53) we observe that in order to find the average displacement $\left\langle u^{(1)} \right\rangle$ ($\left\langle u^{(2)} \right\rangle$), we have to a priori find the fluctuation $Qu^{(0)}$ ($Qu^{(1)}$). The fluctuations $Qu^{(i)}$ satisfy the boundary value problems (19) and (20). To find the right hand side of (20), we act on (28)-(30) with the operator $Q$. The results are as follows.

$$Qt_x^{(0)} = k^s \frac{\partial Q\varepsilon_{xx}^{(0)}}{\partial x}, \qquad Qt_y^{(0)} = 0, \tag{54}$$

$$Qt_x^{(1)} = -h_0 \frac{\partial \langle \sigma_{xy}^{(0)} \rangle}{\partial y} + h_{0x} \langle \sigma_{xx}^{(0)} \rangle + k^s \frac{\partial Q\varepsilon_{xx}^{(1)}}{\partial x} + k^s h_0 \frac{\partial^2 \langle \varepsilon_{xx}^{(0)} \rangle}{\partial x \partial y} + 2k^s h_{0xx} \langle \varepsilon_{xy}^{(0)} \rangle$$

$$+ 2k^s h_{0x} \frac{\partial \langle \varepsilon_{xy}^{(0)} \rangle}{\partial x} + k^s h_{0x} \frac{\partial \langle \varepsilon_{xx}^{(0)} \rangle}{\partial y}, \tag{55}$$

$$Qt_y^{(1)} = -h_0 \frac{\partial \langle \sigma_{yy}^{(0)} \rangle}{\partial y} + h_{0x} \langle \sigma_{yx}^{(0)} \rangle + \tau^o h_{0xx} + k^s h_{0xx} \langle \varepsilon_{xx}^{(0)} \rangle + k^s h_{0x} \frac{\partial \langle \varepsilon_{xx}^{(0)} \rangle}{\partial x},$$

$$Qt_x^{(2)} = h_{0x} Q\sigma_{xx}^{(1)} + h_{0x} \langle \sigma_{xx}^{(1)} \rangle + h_{0x} h_0 \frac{\partial \langle \sigma_{xx}^{(0)} \rangle}{\partial y} + \frac{1}{2}(h_{0x})^2 \langle \sigma_{xy}^{(0)} \rangle - h_0 \frac{\partial \langle \sigma_{xy}^{(1)} \rangle}{\partial y}$$

$$- h_0 \frac{\partial Q\sigma_{xy}^{(1)}}{\partial y} - \frac{1}{2}(h_0)^2 \frac{\partial^2 \langle \sigma_{xy}^{(0)} \rangle}{\partial y^2} - \tau^o h_{0x} h_{0xx} - k^s h_{0x} h_{0xx} \langle \varepsilon_{xx}^{(0)} \rangle + k^s \frac{\partial Q\varepsilon_{xx}^{(2)}}{\partial x} +$$

$$+ k^s h_0 \frac{\partial^2 \langle \varepsilon_{xx}^{(1)} \rangle}{\partial x \partial y} + k^s h_0 \frac{\partial^2 Q\varepsilon_{xx}^{(1)}}{\partial x \partial y} + \frac{1}{2} k^s (h_0)^2 \frac{\partial^3 \langle \varepsilon_{xx}^{(0)} \rangle}{\partial x \partial y^2} - k^s (h_{0x})^2 \frac{\partial \langle \varepsilon_{xx}^{(0)} \rangle}{\partial x} +$$

$$+ 2k^s h_{0xx} \langle \varepsilon_{xy}^{(1)} \rangle + 2k^s h_{0xx} Q\varepsilon_{xy}^{(1)} + 2k^s h_{0xx} h_0 \frac{\partial \langle \varepsilon_{xy}^{(0)} \rangle}{\partial y} + 2k^s h_{0x} \frac{\partial \langle \varepsilon_{xy}^{(1)} \rangle}{\partial x} + 2k^s h_{0x} \frac{\partial Q\varepsilon_{xy}^{(1)}}{\partial x}$$

$$+ 2k^s h_0 h_{0x} \frac{\partial^2 Q\varepsilon_{xy}^{(0)}}{\partial x \partial y} + 2k^s h_{0x} h_{0xx} \left( \langle \varepsilon_{yy}^{(0)} \rangle - \langle \varepsilon_{xx}^{(0)} \rangle \right) + k^s (h_{0x})^2 \left( \frac{\partial \langle \varepsilon_{yy}^{(0)} \rangle}{\partial x} - \frac{1}{2} \frac{\partial \langle \varepsilon_{xx}^{(0)} \rangle}{\partial x} \right)$$

$$+ k^s h_{0x} \frac{\partial \langle \varepsilon_{xx}^{(1)} \rangle}{\partial y} + k^s h_{0x} \frac{\partial Q\varepsilon_{xx}^{(1)}}{\partial y} + k^s h_{0x} h_0 \frac{\partial^2 \langle \varepsilon_{xx}^{(0)} \rangle}{\partial y^2} + 2k^s (h_{0x})^2 \frac{\partial \langle \varepsilon_{xy}^{(0)} \rangle}{\partial y}, \tag{56}$$

$$Qt_y^{(2)} = h_{0x} Q\sigma_{yx}^{(1)} + h_{0x} \langle \sigma_{yx}^{(1)} \rangle + h_{0x} h_0 \frac{\partial \langle \sigma_{yx}^{(0)} \rangle}{\partial y} + \frac{1}{2} h_{0x}^2 \langle \sigma_{yy}^{(0)} \rangle - h_0 \frac{\partial \langle \sigma_{yy}^{(1)} \rangle}{\partial y} - h_0 \frac{\partial Q\sigma_{yy}^{(1)}}{\partial y}$$

$$- \frac{1}{2} h_0^2 \frac{\partial^2 \langle \sigma_{yy}^{(0)} \rangle}{\partial y^2} + k^s h_{0xx} \langle \varepsilon_{xx}^{(1)} \rangle + k^s h_{0xx} Q\varepsilon_{xx}^{(1)} + k^s h_{0xx} h_0 \frac{\partial \langle \varepsilon_{xx}^{(0)} \rangle}{\partial y} + 2k^s h_{0x} h_{0xx} \langle \varepsilon_{xy}^{(0)} \rangle$$

$$+ k^s h_{0x} \frac{\partial \langle \varepsilon_{xx}^{(1)} \rangle}{\partial x} + k^s h_{0x} \frac{\partial Q\varepsilon_{xx}^{(1)}}{\partial x} + h_{0x} h_0 k^s \frac{\partial^2 \langle \varepsilon_{xx}^{(0)} \rangle}{\partial x \partial y} + 2k^s h_{0x} h_{0xx} \langle \varepsilon_{xy}^{(0)} \rangle +$$

$$+ 2k^s h_{0x} h_{0x} \frac{\partial \langle \varepsilon_{xy}^{(0)} \rangle}{\partial x} + k^s h_{0x} h_{0x} \frac{\partial \langle \varepsilon_{xx}^{(0)} \rangle}{\partial y}.$$

Similar to the sinusoidal rough surface, we assume that a far-field stress $\sigma = \sigma^\infty e_1 \otimes e_1$ is applied on the half-space. Below we show the solutions for average $\langle u^{(i)} \rangle$ and fluctuating $Qu^{(i)}$ for $i = 0, 1, 2$ with the following sequence. Solving (17) and (18) for $i = 0$ with the right hand side of (18) given by (51), we obtain $\langle u^{(0)} \rangle$, then solving (19), (20) for $i = 0$ with the right hand side of (20) given by (54), we obtain $Qu^{(0)}$. Similarly, solving (17) and (18) for $i = 1$ with the right hand side of (18) given by (52), we obtain $\langle u^{(1)} \rangle$ and solving (19), (20) for $i = 1$ with the right hand side of (20) given by (55), we obtain $Qu^{(1)}$. Again, solving (17) and (18) for $i = 2$ with the right hand side of (18) given by (53), we obtain $\langle u^{(2)} \rangle$ and solving (19), (20) for $i = 2$ with the right hand side of (20) given by (56), we obtain $Qu^{(2)}$. We remark that the detailed calculations are presented in Appendix D.

By (51),
$$\langle t_x^{(0)} \rangle = 0, \quad \langle t_y^{(0)} \rangle = 0 \qquad \text{on} \quad \partial B^0. \tag{57}$$

Then the zeroth order average strain field is given by

$$\langle \varepsilon_{xx}^{(0)} \rangle = \frac{1-\nu^2}{E}\sigma^\infty, \quad \langle \varepsilon_{yy}^{(0)} \rangle = \frac{-\nu(1+\nu)}{E}\sigma^\infty, \quad \langle \varepsilon_{xy}^{(0)} \rangle = 0 \qquad \text{in } B^0. \tag{58}$$

In order to find the first order average strain field, we need first to find the zeroth order fluctuating strain field. By (54),

$$Qt_x^{(0)} = 0, \quad Qt_x^{(0)} = 0, \qquad \text{on} \quad \partial B^0. \tag{59}$$

So the zeroth order fluctuating strain field is obtained as

$$Q\sigma_{xx}^{(0)} = Q\sigma_{xy}^{(0)} = Q\sigma_{yy}^{(0)} = 0 \qquad \text{in } B^0. \tag{60}$$

Consequently by (52),

$$\langle t_x^{(1)} \rangle = 0, \quad \langle t_y^{(1)} \rangle = 0 \qquad \text{on} \quad \partial B^0. \tag{61}$$

Then the first order average strain field is obtained as

$$\langle \varepsilon_{xx}^{(1)} \rangle = \langle \varepsilon_{xy}^{(1)} \rangle = \langle \varepsilon_{yy}^{(1)} \rangle = 0. \qquad \text{in } B^0. \tag{62}$$

Similarly, in order to find the second order average strain field, we need first to find the first order fluctuating strain field. By (55),

$$Qt_x^{(1)} = h_{0x}\sigma_\infty, \quad Qt_y^{(1)} = \left(\tau^o + \frac{(1-\nu^2)}{E}(k^s - \tau^o)\sigma^\infty\right)h_{0xx} \qquad \text{on} \quad \partial B^0. \tag{63}$$

The first order fluctuating strain field is obtained as

$$Q\varepsilon_{xx}^{(1)}(x,y) = \frac{1-\nu^2}{E} \frac{1}{2\pi} \int_{-\infty}^{\infty} \left[ f(\alpha)\left(-2\frac{|\alpha|}{i\alpha} + i\alpha y\right) + p(\alpha)(1+|\alpha|y) \right] e^{i\alpha x+|\alpha|y} d\alpha -$$

$$- \frac{\nu(1+\nu)}{E} \frac{1}{2\pi} \int_{-\infty}^{\infty} \left[ -i\alpha f(\alpha)y + p(\alpha)(1-|\alpha|y) \right] e^{i\alpha x+|\alpha|y} d\alpha,$$

$$Q\varepsilon_{xy}^{(1)}(x,y) = \frac{(1+\nu)}{2\pi E} \int_{-\infty}^{\infty} \left[ f(\alpha)(1+|\alpha|y) - i\alpha p(\alpha)y \right] e^{i\alpha x+|\alpha|y} d\alpha, \qquad \text{in } B^0 \qquad (64)$$

$$Q\varepsilon_{yy}^{(1)}(x,y) = \frac{1-\nu^2}{E} \frac{1}{2\pi} \int_{-\infty}^{\infty} \left[ -i\alpha f(\alpha)y + p(\alpha)(1-|\alpha|y) \right] e^{i\alpha x+|\alpha|y} d\alpha -$$

$$- \frac{\nu(1+\nu)}{E} \frac{1}{2\pi} \int_{-\infty}^{\infty} \left[ f(\alpha)\left(-2\frac{|\alpha|}{i\alpha} + i\alpha y\right) + p(\alpha)(1+|\alpha|y) \right] e^{i\alpha x+|\alpha|y} d\alpha,$$

where $f(\alpha) = \sigma_\infty(i\alpha)h_0(\alpha)$ and $p(\alpha) = \left( \tau^o + \frac{(1-\nu^2)}{E}(k^s - \tau^o)\sigma^\infty \right)(i\alpha)^2 h_0(\alpha)$.

The second order average boundary conditions are summarized to

$$\left\langle t_x^{(2)} \right\rangle = 0, \qquad \left\langle t_y^{(2)} \right\rangle = 0 \qquad \text{on } \partial B^0. \qquad (65)$$

and the second order average strain field is obtained as

$$\left\langle \varepsilon_{xx}^{(2)} \right\rangle = \left\langle \varepsilon_{xy}^{(2)} \right\rangle = \left\langle \varepsilon_{yy}^{(2)} \right\rangle = 0. \qquad \text{in } B^0. \qquad (66)$$

We also need to find the second order $Q$-terms in order to find the total energy later. By (56),

$$Qt_x^{(2)} = h_{0x}\left(-2\sigma_\infty \frac{1}{2\pi}\int_{-\infty}^{\infty}|\alpha|h_0(\alpha)e^{i\alpha x}d\alpha + \left(\tau^o + \frac{(1-\nu^2)}{E}k^s\sigma^\infty\right)h_{0xx}(x)\right) -$$

$$- h_0\left(2\sigma_\infty \frac{1}{2\pi}\int_{-\infty}^{\infty}i\alpha|\alpha|h_0(\alpha)e^{i\alpha x}d\alpha - \left(\tau^o + \frac{(1-\nu^2)}{E}k^s\sigma^\infty\right)h_{0xxx}(x)\right)$$

$$- \tau^o h_{0x}h_{0xx} - k^s h_{0x}h_{0xx}\frac{(1-\nu^2)}{E}\sigma^\infty + \qquad \text{on } \partial B^0$$

$$+k^s h_0 \begin{pmatrix} 3\dfrac{1-\nu^2}{E}\sigma_\infty h_{0xxx}(x) - 2\dfrac{1-\nu^2}{E}\left(\tau^o + \dfrac{(1-\nu^2)}{E}k^s\sigma^\infty\right)\dfrac{1}{2\pi}\int_{-\infty}^{\infty}i\alpha^3|\alpha|h_0(\alpha)e^{i\alpha x}d\alpha \\ +\dfrac{\nu(1+\nu)}{E}\sigma_\infty h_{0xxx}(x) \end{pmatrix} \qquad (67)$$

$$+ 2k^s h_{0xx}\left(\frac{(1+\nu)}{E}\sigma_\infty h_{0x}(x)\right) + 2k^s h_{0x}\left(\frac{(1+\nu)}{E}\sigma_\infty h_{0xx}(x)\right) +$$

$$+ 2k^s h_{0x}h_{0xx}\left(-\frac{\nu(1+\nu)}{E}\sigma_\infty - \frac{(1-\nu^2)}{E}\sigma^\infty\right) +$$

$$+k^s h_{0x} \begin{pmatrix} 3\sigma_\infty \dfrac{1-\nu^2}{E}h_{0xx}(x) - 2\dfrac{1-\nu^2}{E}\left(\tau^o + \dfrac{(1-\nu^2)}{E}k^s\sigma^\infty\right)\dfrac{1}{2\pi}\int_{-\infty}^{\infty}\alpha^2|\alpha|h_0(\alpha)e^{i\alpha x}d\alpha \\ +\sigma_\infty \dfrac{\nu(1+\nu)}{E}h_{0xx}(x) \end{pmatrix},$$

$$Qt_y^{(2)} = h_{0x}(x)Q\sigma_{yx}^{(1)} + h_0(x)\sigma_\infty h_{0xx}(x) +$$

$$+ k^s h_{0xx}(x)\begin{pmatrix} -2\sigma_\infty \dfrac{1-\nu^2}{E}\dfrac{1}{2\pi}\int_{-\infty}^{\infty}|\alpha|h_0(\alpha)e^{i\alpha x}d\alpha + \\ +\dfrac{(1+\nu)(1-2\nu)}{E}\left(\tau^o + \dfrac{(1-\nu^2)}{E}k^s\sigma^\infty\right)h_{0xx}(x) \end{pmatrix} + \qquad \text{on } \partial B^0. \qquad (68)$$

$$+ k^s h_{0x}(x)\begin{pmatrix} -2\dfrac{1-\nu^2}{E}\dfrac{\sigma_\infty}{2\pi}\int_{-\infty}^{\infty}i\alpha|\alpha|h_0(\alpha)e^{i\alpha x}d\alpha + \\ +\dfrac{(1+\nu)(1-2\nu)}{E}\left(\tau^o + \dfrac{(1-\nu^2)}{E}k^s\sigma^\infty\right)h_{0xxx}(x) \end{pmatrix}$$

The solutions for the second order $Q$-terms are

$$Q\varepsilon_{xx}^{(2)}(x,y) = \frac{1-\nu^2}{E}\frac{1}{2\pi}\int_{-\infty}^{\infty}\left[g(\alpha)\left(-2\frac{|\alpha|}{i\alpha}+i\alpha y\right)+q(\alpha)(1+|\alpha|y)\right]e^{i\alpha x+|\alpha|y}d\alpha$$

$$-\frac{\nu(1+\nu)}{E}\frac{1}{2\pi}\int_{-\infty}^{\infty}\left[-i\alpha g(\alpha)y+q(\alpha)(1-|\alpha|y)\right]e^{i\alpha x+|\alpha|y}d\alpha \qquad \text{in } B^0, \quad (69)$$

$$Q\varepsilon_{xy}^{(2)}(x,y) = \frac{(1+\nu)}{2\pi E}\int_{-\infty}^{\infty}\left[g(\alpha)(1+|\alpha|y)-i\alpha q(\alpha)y\right]e^{i\alpha x+|\alpha|y}d\alpha \qquad \text{in } B^0,$$

where $g(\alpha)$ and $q(\alpha)$ are the Fourier transform of (67) and (68), respectively.

By using the relations in Appendix A and performing the Taylor series extrapolation, the surface strain on the rough surface is obtained as

$$[\varepsilon_{ss}]_{y=h(x)} = \frac{1-v^2}{E}\sigma^\infty + \delta \left\{ \begin{bmatrix} \frac{1}{2\pi}\int_{-\infty}^{\infty} -2|\alpha|\sigma_\infty h_0(\alpha)e^{i\alpha x}d\alpha + \left(\tau^o + \frac{(1-v^2)}{E}k^s\sigma^\infty\right)h_{0xx}(x) \end{bmatrix} \frac{(1-v^2)}{E} \\ + \left(\tau^o + \frac{(1-v^2)}{E}k^s\sigma^\infty h_{0xx}(x)\right)\frac{-v(1+v)}{E} \right\}$$

$$+ \delta^2 \left\{ h_0(x)\begin{bmatrix} 3\sigma_\infty \frac{1-v^2}{E}h_{0xx}(x) - 2\frac{1-v^2}{E}\left(\tau^o + \frac{(1-v^2)}{E}k^s\sigma^\infty\right)\frac{1}{2\pi}\int_{-\infty}^{\infty}\alpha^2|\alpha|h_0(\alpha)e^{i\alpha x}d\alpha + \\ \sigma_\infty \frac{v(1+v)}{E}h_{0xx}(x) \end{bmatrix} \right.$$
$$\left. -(h_{0x})^2\frac{(1-v^2)}{E}\sigma^\infty + Q\varepsilon_{xx}^{(2)} + 2h_{0x}(x)\frac{(1+v)}{E}\sigma_\infty h_{0x}(x) + (h_{0x})^2\frac{-v(1+v)}{E}\sigma^\infty \right\}. \quad (70)$$

Similar to the sinusoidal rough surface, our homogenization scheme requires calculation of the total energy under the action of the applied stress $\sigma^\infty$. Using the total strain fields as $\varepsilon_{ij} = \langle\varepsilon_{ij}\rangle + Q\varepsilon_{ij}$ $i,j=x,y$ and inserting into (13) we can find the ensemble average of the total energy

$$\langle E^{act}(\sigma^\infty)\rangle = \frac{1}{2}P\left[\int_{-\infty}^{h(x)}(\varepsilon^{(0)} + \delta\varepsilon^{(1)} + \delta^2\varepsilon^{(2)}).\mathbb{C}(\varepsilon^{(0)} + \delta\varepsilon^{(1)} + \delta^2\varepsilon^{(2)})dy\right]$$
$$+ \frac{1}{2}P\left[(\varepsilon_{ss} - \varepsilon_{ss}^0)\cdot\mathbb{C}_s(\varepsilon_{ss} - \varepsilon_{ss}^0)\right]. \quad (71)$$

We remark that the detailed calculations are presented in Appendix D. By (14) and (41)-(43) and using $\langle E^{act}\rangle = E^{eff}$, we have the effective

$$\left(\tau^0\right)^{eff} = \left(\frac{E}{1-v^2}\right)\frac{\partial\langle E^{act}\rangle(\sigma^\infty)}{\partial\sigma^\infty}\bigg|_{\sigma^\infty=0} \quad \text{and} \quad \left(k^s\right)^{eff} = \left(\frac{E}{1-v^2}\right)^2\frac{\partial^2\langle E^{act}\rangle(\sigma^\infty)}{(\partial\sigma^\infty)^2}\bigg|_{\sigma^\infty=0}. \quad (72)$$

Consequently,

$$\left(\tau^0\right)^{eff} = \tau^o - \delta^2\tau^o\left(\begin{array}{c} \frac{4\eta^2}{a^2} + \frac{1}{\sqrt{\pi}}k^s\frac{(1+8v)(1+v)}{E}\left(\frac{2\eta^2}{a^3}\right) - \\ -(k^s)^2\sigma^\infty\frac{(1+v)^2(1-2v)^2}{E^2}\left(\frac{12\eta^2}{a^4}\right) \end{array}\right), \quad (73)$$

and

$$\left(k^s\right)^{\text{eff}} = \left(k^s\right) + \delta^2 \begin{pmatrix} -\dfrac{(9-8\nu)E}{4(1-\nu)(1-\nu^2)}\left(\dfrac{1}{\sqrt{\pi}}\right)\left(\dfrac{\eta^2}{a}\right) + k^s\left(\dfrac{8\eta^2}{a^2}\right) \\ +\dfrac{(-57\nu+11)(1+\nu)}{4E}\left(k^s\right)^2\left(\dfrac{1}{\sqrt{\pi}}\right)\left(\dfrac{\eta^2}{a^3}\right) \\ +\dfrac{(1-2\nu)^2(1+\nu)^2}{E}\left(k^s\right)^3\left(\dfrac{6\eta^2}{a^4}\right) \end{pmatrix}. \qquad (74)$$

If $\dfrac{k^s}{aE} \ll 1$, equations (73) and (74) can be further simplified as

$$\left(\tau^0\right)^{\text{eff}} = \tau^o\left(1 - \delta^2 \dfrac{4\eta^2}{a^2}\right), \qquad (75)$$

$$\left(k^s\right)^{\text{eff}} = \left(k^s\right) - \delta^2 \dfrac{(9-8\nu)E}{4(1-\nu)(1-\nu^2)}\left(\dfrac{1}{\sqrt{\pi}}\right)\left(\dfrac{\eta^2}{a}\right). \qquad (76)$$

## V. Discussion

We can use the simple expressions we have derived to make some assessments on the effect of roughness on the surface properties. Taking Copper as an example, with Young's modulus $E$ of 115 GPa, Poisson's ratio $\nu$ of 0.34, surface stress $\tau^o \approx 1.04$ N/m and surface elastic constant $k^s \approx -3.16$ N/m for the (001) crystal face (Shenoy, 2005). If we consider a sinusoidal roughness with $ak = 0.2$ and wave length $\lambda$ equal to at least 10 nm, $k$ will be of the order of $\dfrac{2\pi}{\lambda} = \dfrac{2\pi}{10^{-8}} = 6.28 \times 10^8$ m$^{-1}$ and by (47) the effective surface stress can be calculated to be

$$\left(\tau^0\right)^{\text{eff}} = 0.97\tau^o \qquad (77)$$

So $\left(\tau^0\right)^{\text{eff}}$ for this rough surface is barely 3 percent less than the pristine value, $\tau^o$.

Likewise by (48), $\left(k^s\right)^{\text{eff}}$ is obtained as

$$\left(k^s\right)^{\text{eff}} = k^s - \delta^2 \dfrac{E}{k(1-\nu^2)}\left(\dfrac{9-8\nu}{8(1-\nu)}\right) \cong -13.01 \qquad (78)$$

A dramatic change from the flat surface value of -3.16 N/m! Therefore, we can conclude that while residual surface stress is hardly affected by the roughness, the surface elastic parameters undergo a dramatic shift. It should be noted here that surface roughness can cause even change of sign in surface elastic depending on the extent of the roughness. Finally, as evident from the expressions for the both the periodic and random roughness case, even if the bare surface possesses zero surface elasticity i.e. $k^s \approx 0$ roughness will "create" surface elasticity i.e. effective value of $k^s$ will be non-zero.

**Comparison with Weissmuller and Duan's (2008) Results**

Weissmuller and Duan (2008) showed that the response of the curvature of cantilevers to changes in their surface stress in the presence of the surface roughness is different from nominally planar surfaces. Considering surface residual stress for cantilevers, they concluded that deliberate structuring of the surface allows the magnitude and even its sign to be tuned. They have concluded that bending of the substrate is controlled by changes in in-plane component of the surface-induced stress, $T$ only. Their calculation shows that $T$ for the isotropic solid with a nearly planar surface $\theta^2 \ll 1$ (assuming isotropy) is equal to

$$T = \frac{\langle f \rangle_s}{h^l}\left(1 - \frac{v^l}{1-v^l}\langle \theta^2 \rangle\right) \qquad (79)$$

where $f$ is surface residual stress and $v^l$ is the Poisson's ratio. Through their calculations, they assumed that $f$ depends on the surface orientation but this assumption does not have any contribution in creating $\left(1 - \frac{v^l}{1-v^l}\langle \theta^2 \rangle\right)$ term that shows the apparent action of $f$ will be reduced by a geometric effect that scales with the root-mean-square of $\theta$.

To compare our results with theirs we assume that the roughness profile is co-sinusoidal. Then the average of square of inclination angle can be expressed as

$$\theta = \frac{dh}{dx}$$

$$\langle \theta^2 \rangle = \left\langle \left(\frac{dh}{dx}\right)^2 \right\rangle = \langle a^2 k^2 \sin^2 kx \rangle = \frac{1}{\lambda}\int_0^\lambda a^2 k^2 (\sin^2 kx) dx = \frac{1}{\lambda}\int_0^\lambda a^2 k^2 \left(\frac{1-\cos 2kx}{2}\right) dx = \qquad (80)$$

$$\frac{a^2 k^2}{2} = \frac{\delta^2}{2}$$

In order to calculate the maximum reduction in $\left(1 - \frac{v^l}{1-v^l}\langle \theta^2 \rangle\right)$, we select the value of 0.2 for $\delta$ and 0.44 for $v$ for Gold and then obtain

$$1 - \frac{v^l}{1-v^l}\langle\theta^2\rangle = 1 - 0.016 \tag{81}$$

So based on the assumed range of $\delta$ our model suggests that the reduction of the effective surface stress because of the roughness is 1.6 % while Weissmuller and Duan's work shows 10 % reduction with assumption of $\langle\theta^2\rangle = 0.33$. The somewhat larger shifts in the surface stress calculated by Weissmuller and Duan (2008) can only be obtained for extremely large roughness. Since both works (ours and Duan et. al.) assume "small roughness", it is not clear whether our models are applicable for the large range of roughness that lead to the dramatic shifts in surface stress observed by them. .

**Resonance Frequency of Nano-cantilevers**

Nanofabricated cantilever structures have been demonstrated to be extremely versatile sensors and have potential applications in physical, chemical, and biological sciences. Adsorption on surface of such a sensor may induce mass, damping, and stress changes of the cantilever response. One cantilever sensor technique is to monitor changes in the cantilever resonance frequency. The effect of surface stress on the resonance frequency of a cantilever have been modeled analytically by Lu et al. (2005) by incorporating strain-dependant surface stress terms into the equations of motion.

Consider a cantilever used as a sensor. The experimental quantity measured is the surface stress difference, $\Delta\sigma^s = \sigma^s_u - \sigma^s_l$, where $\sigma^s_u$ and $\sigma^s_l$ are the surface stresses on the upper and the lower surfaces, respectively. In the isotropic case, the surface stresses may be written as

$$\sigma^s_u = \tau^o_u + k^s_u(\varepsilon_{ss})_u \quad \text{and} \quad \sigma^s_l = \tau^o_l + k^s_l(\varepsilon_{ss})_l \tag{82}$$

where $\tau^o$ is the strain-independent surface stress, $k^s$ is a constant associated with the surface strain, $\varepsilon_{ss}$ is the surface strain measured from the pre-stressed configuration, and the subscripts $u$ and $l$ always refer to the upper and lower surface, respectively. The surface stress difference can be written as

$$\Delta\sigma^s = \Delta\sigma^o + \Delta\sigma^1 \tag{83}$$

with $\Delta\sigma^o = \tau^o_u - \tau^o_l$ and $\Delta\sigma^1 = k^s_u(\varepsilon_{ss})_u - k^s_l(\varepsilon_{ss})_l$. While the strain-independent part of the surface stress, $\Delta\sigma^o$ can have an impact on the resonance frequency (in a nonlinear setting), it is expected to be small. The strain-dependent part (i.e. surface elasticity) definitely will change the resonance frequency and can be expressed as

$$\frac{(\omega_s)^2 - (\omega_0)^2}{(\omega_0)^2} = 3\frac{k^s_u + k^s_l}{Eh} \tag{84}$$

where $\omega_0$ is the fundamental resonance frequency without considering surface elasticity, $\omega_s$ is the resonance frequency with surface stresses acting, $h$ is the thickness and $E$ is Young's modulus. Liu and Rajapakse (2010) as well came up with the same expression for resonance frequency shift considering surface energy.

To compare the change in resonance frequency of cantilevers with rough surfaces, we consider a beam that has a sinusoidal rough surface on top and flat surface on the bottom. We have

$$k^s_u = \left(k^s\right)^{eff} = k^s - \delta^2 \frac{E}{k(1-v^2)} \left(\frac{9-8v}{8(1-v)}\right). \tag{85}$$

for top surface and $k^s_l = k^s$ for the lower surface. Then the change in resonance frequency can be obtained as

$$\frac{(\omega_{rough})^2 - (\omega_0)^2}{(\omega_0)^2} = \frac{3}{Eh}\left(2k^s - \delta^2 \frac{E}{k(1-v^2)}\left(\frac{9-8v}{8(1-v)}\right)\right). \tag{86}$$

Compared to a cantilever with upper and lower flat surface with resonance frequency

$$\frac{(\omega_s)^2 - (\omega_0)^2}{(\omega_0)^2} = \frac{3}{Eh}(2k^s). \tag{87}$$

Evidently, frequency shift will decrease significantly or even in some cases, may change sign. For instance for copper considering the (001) crystal face (Shenoy, 2005) if we consider a sinusoidal roughness with $ak = 0.2$ and wave length of 10 nm on top surface of cantilever, the change of resonance frequency would be expressed as

$$\frac{(\omega_{rough})^2 - (\omega_0)^2}{(\omega_0)^2} = \frac{3}{Eh}(-16.17) \tag{88}$$

from its value of

$$\frac{(\omega_s)^2 - (\omega_0)^2}{(\omega_0)^2} = \frac{3}{Eh}(-6.32). \tag{89}$$

So quantitatively, the square of resonance frequency is shifted by 2.55 times.

As another example, we consider aluminum with Young's modulus $E$ of 70 GPa, Poisson's ratio $\nu$ of 0.35, $\tau^o \approx 0.91$ N/m and $k^s \approx 4.53$ N/m for the (111) crystal face (Shenoy, 2005). Then the resonance frequency of the cantilever with top rough surface is calculated as

$$\frac{(\omega_{rough})^2 - (\omega_0)^2}{(\omega_0)^2} = \frac{3}{Eh}(3) \quad (90)$$

while in case of considering flat surfaces for cantilever it would be

$$\frac{(\omega_s)^2 - (\omega_0)^2}{(\omega_0)^2} = \frac{3}{Eh}(9.06) \cdot \quad (91)$$

In this case, the square resonance frequency is decreased by three times.

In summary, we have presented simple expressions for homogenized surface stress and surface elasticity for both randomly and periodically rough surfaces. Residual surface stress does not appear to be significantly affected by the presences of roughness----this appears to be in contrast to the conclusions of Weismuller and Duan (2008) although we do notice a dramatic change in the surface elastic modulus. The latter for example, as we demonstrated through simple illustrative quantitative examples, should have significant impact on the way sensing data based on surface effects is interpreted.

**Acknowledgements**

We gratefully acknowledge financial support from University of Houston NSF GK12 program, Grant #0840889 (Program Manager: Drs. Sonia Ortega).

**Appendix A: Useful Relations for Converting Parametric Coordinate to Cartesian Coordinate in Two Dimensions**

The following relations will be useful in converting 2-dimensional parametric coordinate to Cartesian coordinate.

$$h(x) = \delta h_0(x), \quad \delta \ll 1, \quad (92)$$

$$e_s \approx \left(1 - \frac{1}{2}\delta^2 h_{0x}^2\right)e_1 + \delta h_{0x} e_2, \quad e_n \approx -\delta h_{0x} e_1 + \left(1 - \frac{1}{2}\delta^2 h_{0x}^2\right)e_2, \quad \kappa = \delta h_{0xx},$$

$$\frac{\partial x}{\partial s} = \frac{1}{\sqrt{1+h_x^2}} \approx 1 - \frac{1}{2}\delta^2 h_{0x}^2, \quad \frac{\partial y}{\partial s} = \frac{h_x}{\sqrt{1+h_x^2}} \approx \delta h_{0x}, \quad h_{0x} = \frac{\partial h_0(x)}{\partial x},$$

$$u_s \approx \left(1 - \frac{1}{2}\delta^2 h_{0x}^2\right)u_x + \delta u_y h_{0x}, \quad u_n \approx -\delta h_{0x} u_x + \left(1 - \frac{1}{2}\delta^2 h_{0x}^2\right)u_y,$$

$$\varepsilon_{ss} = \varepsilon_{xx} + 2\delta h_{0x}\varepsilon_{xy} + \delta^2 h_{0x}^2 (\varepsilon_{yy} - \varepsilon_{xx}) + O(\delta^3),$$

$$\frac{\partial \varepsilon_{ss}}{\partial s} \approx \frac{\partial \varepsilon_{ss}}{\partial x}\left(1 - \frac{1}{2}\delta^2 h_{0x}^2\right) + \frac{\partial \varepsilon_{ss}}{\partial y}\delta h_{0x} = \frac{\partial \varepsilon_{xx}}{\partial x}\left(1 - \frac{1}{2}\delta^2 h_{0x}^2\right) + 2\delta h_{0xx}\varepsilon_{xy} + 2\delta h_{0x}\frac{\partial \varepsilon_{xy}}{\partial x}$$

$$+ 2\delta^2 h_{0x} h_{0xx}(\varepsilon_{yy} - \varepsilon_{xx}) + \delta^2 h_{0x}^2 \left(\frac{\partial \varepsilon_{yy}}{\partial x} - \frac{\partial \varepsilon_{xx}}{\partial x}\right) + \frac{\partial \varepsilon_{xx}}{\partial y}\delta h_{0x} + 2\delta^2 h_{0x}^2 \frac{\partial \varepsilon_{xy}}{\partial y} + O(\delta^3),$$

$$\frac{\partial u_n}{\partial s} \approx -\delta h_{0xx} u_x + \delta h_{0x}(-\varepsilon_{xx} + \varepsilon_{yy}) - \delta^2 h_{0x} h_{0xx} u_y + \left(1 - \frac{1}{2}\delta^2 h_{0x}^2\right)\frac{\partial u_y}{\partial x} - \delta^2 h_{0x}^2 \frac{\partial u_x}{\partial y},$$

$$\frac{\partial \kappa}{\partial s} = \delta h_{0xxx},$$

$$\frac{\partial^2 u_n}{\partial s^2} \approx -\delta h_{0xxx} u_x - 2\delta h_{0xx}\varepsilon_{xx} + \delta h_{0xx}\varepsilon_{yy} + \delta h_{0x}\left(-\frac{\partial \varepsilon_{xx}}{\partial x} + \frac{\partial \varepsilon_{yy}}{\partial x}\right) - \delta^2 h_{0xx}^2 u_y -$$

$$- \delta^2 h_{0x} h_{0xxx} u_y - 2\delta^2 h_{0x} h_{0xx}\varepsilon_{xy} - 2\delta^2 h_{0x} h_{0xx}\frac{\partial u_x}{\partial y} - 2\delta^2 h_{0x}^2 \frac{\partial \varepsilon_{xy}}{\partial x} + \frac{\partial^2 u_y}{\partial x^2} +$$

$$+ \delta^2 (h_{0x})^2 \left(-\frac{\partial \varepsilon_{xx}}{\partial y} + \frac{\partial \varepsilon_{yy}}{\partial y}\right) + \delta h_{0x}\frac{\partial^2 u_y}{\partial x \partial y}.$$

## Appendix B: General Procedure for Finding Elastic Fields for Cerruti-Boussinesq Problem Using Airy Stress Function

In this appendix, solutions involving general types of loading on half-spaces are considered. Such solutions are developed using Fourier transforms. The media are taken to be elastically isotropic.

Consider a half space defined by $y \leq 0$. The loading is specified by

$$\sigma_{xy} = f(x), \quad \sigma_{yy} = p(x) \quad \text{on} \quad y = 0 \tag{93}$$

which describes a state of general normal and shear loading on the external surface of the half space.

The nature of the loadings, $f(x)$ and $p(x)$, is such that it produces bounded stresses. Infinitely far into the bulk of the medium the stresses must be bounded, so that

$$\sigma_{\alpha\beta} \nrightarrow \infty \quad \text{as} \quad y \to -\infty. \tag{94}$$

As there are no body forces or temperature gradients, the Airy stress function satisfies the simple form of the biharmonic equation

$$\nabla^4 \phi = \frac{\partial^4 \phi}{\partial x^4} + 2\frac{\partial^4 \phi}{\partial x^2 \partial y^2} + \frac{\partial^4 \phi}{\partial y^4} = 0. \tag{95}$$

Introduce the Fourier transform in spatial coordinate $x$,

$$\Phi(\alpha, y) = \int_{-\infty}^{\infty} \phi(x, y) e^{-i\alpha x} dx \tag{96}$$

and it's inverse

$$\phi(x, y) = \frac{1}{2\pi} \int_{-\infty}^{\infty} \Phi(\alpha, y) e^{i\alpha x} d\alpha. \tag{97}$$

Apply the Fourier transform to the biharmonic equation (95), it is found that

$$(i\alpha)^4 \Phi + 2(i\alpha)^2 \frac{\partial^2 \Phi}{\partial y^2} + \frac{\partial^4 \Phi}{\partial y^4} = 0. \tag{98}$$

The acceptable solution for the transform $\Phi(\alpha, y)$, are of the form

$$\Phi(\alpha, y) = (A + By)e^{-|\alpha|y} + (C + Dy)e^{|\alpha|y}. \tag{99}$$

The transform of the stresses are

$$\sigma_{xx} = \frac{\partial^2 \phi}{\partial y^2} \quad \Rightarrow \quad \int_{-\infty}^{\infty} \frac{\partial^2 \phi}{\partial y^2} e^{-i\alpha x} dx = \frac{\partial^2 \Phi}{\partial y^2},$$

$$\sigma_{xy} = -\frac{\partial^2 \phi}{\partial x \partial y} \quad \Rightarrow \quad \int_{-\infty}^{\infty} -\frac{\partial^2 \phi}{\partial x \partial y} e^{-i\alpha x} dx = -i\alpha \frac{\partial \Phi}{\partial y}, \tag{100}$$

$$\sigma_{yy} = \frac{\partial^2 \phi}{\partial x^2} \quad \Rightarrow \quad \int_{-\infty}^{\infty} \sigma_{yy} e^{-i\alpha x} dx = (i\alpha)^2 \Phi(\alpha, y) = -\alpha^2 \Phi(\alpha, y).$$

The inverse transform follow immediately as

$$\sigma_{xx} = \frac{1}{2\pi}\int_{-\infty}^{\infty}\frac{\partial^2\Phi(\alpha,y)}{\partial y^2}e^{i\alpha x}d\alpha,$$

$$\sigma_{xy} = -\frac{1}{2\pi}\int_{-\infty}^{\infty}i\alpha\frac{\partial\Phi(\alpha,y)}{\partial y}e^{i\alpha x}d\alpha, \tag{101}$$

$$\sigma_{yy} = -\frac{1}{2\pi}\int_{-\infty}^{\infty}\alpha^2\Phi(\alpha,y)e^{i\alpha x}d\alpha.$$

Next, invoke the boundary conditions specified above and form the Fourier transform of the normal and tangential loading boundary conditions on $y=0$, we have

$$\int_{-\infty}^{\infty}\sigma_{xy}e^{-i\alpha x}dx = -\int_{-\infty}^{\infty}\frac{\partial^2\phi}{\partial x\partial y}e^{-i\alpha x}dx = -i\alpha\left[\frac{\partial\Phi(\alpha,y)}{\partial y}\right]_{y=0} = \int_{-\infty}^{\infty}f(x)e^{-i\alpha x}dx = f(\alpha) \tag{102}$$

and

$$\int_{-\infty}^{\infty}\sigma_{yy}e^{-i\alpha x}dx = \int_{-\infty}^{\infty}\frac{\partial^2\phi}{\partial x^2}e^{-i\alpha x}dx = -\alpha^2\Phi(\alpha,y=0) = \int_{-\infty}^{\infty}p(x)e^{-i\alpha x}dx = p(\alpha). \tag{103}$$

Since the stresses need to be bounded as $y\to-\infty$, it is clear that

$$A = B = 0 \tag{104}$$

whereas the transformed boundary conditions of (102) and (103) require that

$$-i\alpha(D+|\alpha|C) = f(\alpha) \quad \text{and} \quad -\alpha^2 C = p(\alpha). \tag{105}$$

This leads to

$$C = -\frac{p(\alpha)}{\alpha^2} \quad \text{and} \quad D = -\frac{f(\alpha)}{i\alpha} + \frac{p(\alpha)}{|\alpha|}. \tag{106}$$

The result for $\Phi(\alpha,y)$ is then

$$\Phi(\alpha,y) = \left[-\frac{p(\alpha)}{\alpha^2} + \left(-\frac{f(\alpha)}{i\alpha} + \frac{p(\alpha)}{|\alpha|}\right)y\right]e^{|\alpha|y}. \tag{107}$$

The solution for the stresses is consequently

$$\sigma_{xx}(x,y) = \frac{1}{2\pi}\int_{-\infty}^{\infty}[f(\alpha)(-2\frac{|\alpha|}{i\alpha}+i\alpha y)+p(\alpha)(1+|\alpha|y)]e^{i\alpha x+|\alpha|y}d\alpha,$$

$$\sigma_{xy}(x,y) = \frac{1}{2\pi}\int_{-\infty}^{\infty}[f(\alpha)(1+|\alpha|y)-i\alpha p(\alpha)y]e^{i\alpha x+|\alpha|y}d\alpha, \tag{108}$$

$$\sigma_{yy}(x,y) = \frac{1}{2\pi}\int_{-\infty}^{\infty}[-i\alpha f(\alpha)y+p(\alpha)(1-|\alpha|y)]e^{i\alpha x+|\alpha|y}d\alpha.$$

**Appendix C: Calculation Detail for Solving Boundary Value Problems for Sinusoidal Roughness Case**

After applying far field stress $\sigma = \sigma^\infty e_1 \otimes e_1$, we solve the different order boundary value problems (11) and (12) for half space with flat surface (see Appendix B). By (28), we have the zeroth oreder boundary condition as

$$t_x^{(0)} = k^s \frac{\partial \varepsilon_{xx}^{(0)}}{\partial x} = 0, \qquad t_y^{(0)} = 0. \tag{109}$$

The zeroth order solution would be obtained as

$$\sigma_{xx}^{(0)} = \sigma^\infty, \qquad \sigma_{xy}^{(0)} = \sigma_{yy}^{(0)} = 0,$$
$$\varepsilon_{xx}^{(0)} = \frac{1-\nu^2}{E}\sigma^\infty, \qquad \varepsilon_{xy}^{(0)} = 0, \qquad \varepsilon_{yy}^{(0)} = \frac{-\nu(1+\nu)}{E}\sigma^\infty. \tag{110}$$

By (29), first order B.C. is
$$t_x^{(1)} = h_{0x}\sigma_{xx}^{(0)} = -\sigma^\infty \sin kx,$$
$$t_y^{(1)} = \tau^o h_{0xx} + k^s h_{0xx}\varepsilon_{xx}^{(0)} = -k\left(\tau^o + k^s \frac{(1-\nu^2)}{E}\sigma^\infty\right)\cos kx, \tag{111}$$

and the first order solution is

$$\sigma_{xx}^{(1)} = \left\{-\sigma^\infty(2+ky)e^{ky} - k\left(\tau^o + k^s\frac{(1-\nu^2)}{E}\sigma^\infty\right)(1+ky)e^{ky}\right\}\cos kx,$$

$$\sigma_{xy}^{(1)} = \left\{-\sigma^\infty(ky+1)e^{ky} - k\left(\tau^o + k^s\frac{(1-\nu^2)}{E}\sigma^\infty\right)kye^{ky}\right\}\sin kx,$$

$$\sigma_{yy}^{(1)} = \left\{\sigma^\infty(ky)e^{ky} + k\left(\tau^o + k^s\frac{(1-\nu^2)}{E}\sigma^\infty\right)(ky-1)e^{ky}\right\}\cos kx,$$

$$\varepsilon_{xx}^{(1)} = \frac{1-\nu^2}{E}\left\{-\sigma^\infty(2+ky)e^{ky} - k\left(\tau^o + k^s\frac{(1-\nu^2)}{E}\sigma^\infty\right)(1+ky)e^{ky}\right\}\cos kx \tag{112}$$
$$-\frac{\nu(1+\nu)}{E}\left\{\sigma^\infty(ky)e^{ky} + k\left(\tau^o + k^s\frac{(1-\nu^2)}{E}\sigma^\infty\right)(ky-1)e^{ky}\right\}\cos kx,$$

$$\varepsilon_{yy}^{(1)} = -\frac{\nu(1+\nu)}{E}\left\{-\sigma^\infty(2+ky)e^{ky} - k\left(\tau^o + k^s\frac{(1-\nu^2)}{E}\sigma^\infty\right)(1+ky)e^{ky}\right\}\cos kx$$
$$+\frac{1-\nu^2}{E}\left\{\sigma^\infty(ky)e^{ky} + k\left(\tau^o + k^s\frac{(1-\nu^2)}{E}\sigma^\infty\right)(ky-1)e^{ky}\right\}\cos kx,$$

$$\varepsilon_{xy}^{(1)} = \frac{(1+\nu)}{E}\left\{-\sigma^{\infty}(ky+1)e^{ky} - k\left(\tau^{o} + k^{s}\frac{(1-\nu^{2})}{E}\sigma^{\infty}\right)kye^{ky}\right\}\sin kx.$$

Using zeroth and first order solutions at $y=0$, by (30), the second order B.C. is

$$t_x^{(2)} = (-\sin kx)\left\{-2\sigma^{\infty} - k\left(\tau^{o} + k^{s}\frac{(1-\nu^{2})}{E}\sigma^{\infty}\right)\right\}\cos kx$$

$$-\frac{\cos kx}{k}\left\{-2\sigma^{\infty}k - k^{2}\left(\tau^{o} + k^{s}\frac{(1-\nu^{2})}{E}\sigma^{\infty}\right)\right\}\sin kx$$

$$-\tau^{o}(-\sin kx)(-k\cos kx) - k^{s}(-\sin kx)(-k\cos kx)\left(\frac{1-\nu^{2}}{E}\sigma_{\infty}\right)$$

$$+k^{s}\frac{\cos kx}{k}\left\{3\frac{(1-\nu^{2})}{E}k^{2}\sigma^{\infty} + 2\frac{(1-\nu^{2})}{E}k^{3}\left(\tau^{o} + k^{s}\frac{(1-\nu^{2})}{E}\sigma^{\infty}\right) + \frac{\nu(1+\nu)}{E}k^{2}\sigma^{\infty}\right\}\sin kx \quad (113)$$

$$+2k^{s}(-k\cos kx)\left(-\frac{(1+\nu)}{E}\sigma^{\infty}\sin kx\right) + 2k^{s}(-\sin kx)\left(-\frac{(1+\nu)}{E}k\sigma^{\infty}\cos kx\right)$$

$$+2k^{s}(-\sin kx)(-k\cos kx)\left(\frac{-\nu(1+\nu)}{E}\sigma_{\infty} - \frac{1-\nu^{2}}{E}\sigma_{\infty}\right)$$

$$+k^{s}(-\sin kx)\left(-3\frac{(1-\nu^{2})}{E}k\sigma^{\infty} - 2\frac{(1-\nu^{2})}{E}k^{2}\left(\tau^{o} + k^{s}\frac{(1-\nu^{2})}{E}\sigma^{\infty}\right) - \frac{\nu(1+\nu)}{E}k\sigma^{\infty}\right)\cos kx,$$

$$t_y^{(2)} = (-\sin kx)(-\sigma^{\infty}\sin kx) - \left(\frac{\cos kx}{k}\right)(k\sigma^{\infty}\cos kx)$$

$$+k^{s}(-k\cos kx)\left\{-\frac{(1-2\nu)(1+\nu)}{E}k\left(\tau^{o} + k^{s}\frac{(1-\nu^{2})}{E}\sigma^{\infty}\right) - \sigma^{\infty}\frac{2(1-\nu^{2})}{E}\right\}\cos kx \quad (114)$$

$$+k^{s}(-\sin kx)\left\{\frac{(1-2\nu)(1+\nu)}{E}k^{2}\left(\tau^{o} + k^{s}\frac{(1-\nu^{2})}{E}\sigma^{\infty}\right) + \sigma^{\infty}\frac{2(1-\nu^{2})}{E}k\right\}\sin kx.$$

More simplifying we would have

$$t_x^{(2)} = \beta \sin 2kx, \qquad t_y^{(2)} = \gamma \cos 2kx \tag{115}$$

with

$$\beta = 2\sigma^{\infty} + \frac{1}{2}\tau^{o}k + \frac{5}{2}\frac{1-\nu^{2}}{E}k^{s}k\sigma_{\infty} + 2k^{s}\frac{(1-\nu^{2})}{E}k^{2}\left(\tau^{o} + k^{s}\frac{(1-\nu^{2})}{E}\sigma^{\infty}\right)$$

$$+ \frac{2(1+\nu)}{E}kk^{s}\sigma^{\infty}, \tag{116}$$

$$\gamma = -\sigma^{\infty} + \frac{(1-2\nu)(1+\nu)}{E}k^{2}k^{s}\left(\tau^{o} + k^{s}\frac{(1-\nu^{2})}{E}\sigma^{\infty}\right) + \sigma^{\infty}kk^{s}\frac{2(1-\nu^{2})}{E}.$$

By solving the second order terms we are lead to the following results

$$\sigma_{xx}^{(2)} = 2\beta(1+ky)e^{2ky}\cos 2kx + \gamma(1+2ky)e^{2ky}\cos 2kx,$$

$$\sigma_{yy}^{(2)} = -2\beta ky e^{2ky}\cos 2kx + \gamma(1-2ky)e^{2ky}\cos 2kx,$$

$$\varepsilon_{xx}^{(2)} = \frac{1-\nu^2}{E}\left(2\beta(1+ky)e^{2ky}\cos 2kx + \gamma(1+2ky)e^{2ky}\cos 2kx\right) \quad (117)$$

$$-\frac{\nu(1+\nu)}{E}\left(-2\beta ky e^{2ky}\cos 2kx + \gamma(1-2ky)e^{2ky}\cos 2kx\right).$$

And at $y=0$

$$\varepsilon_{xx}^{(2)}\bigg|_{y=0} = \frac{(1-2\nu)(1+\nu)}{E}\gamma\cos 2kx + \frac{2(1-\nu^2)}{E}\beta\cos 2kx \quad (118)$$

$$= \eta\cos 2kx,$$

with $\eta = \frac{(1-2\nu)(1+\nu)}{E}\gamma + \frac{2(1-\nu^2)}{E}\beta$. In order to find the surface stress on $y=h(x)$, we use the transformation law in Appendix A and Taylor extrapolation that yield $\varepsilon_{ss}$ as

$$[\varepsilon_{ss}]_{y=h(x)} = \left[\varepsilon_{xx} + 2\delta h_{0x}\varepsilon_{xy} + \delta^2(h_{0x})^2(\varepsilon_{yy} - \varepsilon_{xx})\right]_{y=h(x)} =$$

$$= \begin{bmatrix} \varepsilon_{xx}^{(0)} + \delta\left(\varepsilon_{xx}^{(1)} + 2h_{0x}\varepsilon_{xy}^{(0)} + h_0\frac{\partial\varepsilon_{xx}^{(0)}}{\partial y}\right) + \\ \delta^2\left(h_0\frac{\partial\varepsilon_{xx}^{(1)}}{\partial y} + \frac{1}{2}(h_0)^2\frac{\partial^2\varepsilon_{xx}^{(0)}}{\partial y^2} + 2(h_{0x})h_0\frac{\partial\varepsilon_{xy}^{(0)}}{\partial y}\right) \\ -(h_{0x})^2\varepsilon_{xx}^{(0)} + \varepsilon_{xx}^{(2)} + 2h_{0x}\varepsilon_{xy}^{(1)} + (h_{0x})^2\varepsilon_{yy}^{(0)} \end{bmatrix}_{y=0}$$

$$= \frac{1-\nu^2}{E}\sigma^\infty + \delta\left\{-\frac{(1-2\nu)(1+\nu)}{E}k\left(\tau^o + k^s\frac{(1-\nu^2)}{E}\sigma^\infty\right) - \sigma^\infty\frac{2(1-\nu^2)}{E}\right\}\cos kx$$

$$+ \delta^2\left\{\left(-3\frac{(1-\nu^2)}{E}\sigma^\infty - 2\frac{(1-\nu^2)}{E}k\left(\tau^o + k^s\frac{(1-\nu^2)}{E}\sigma^\infty\right) - \frac{\nu(1+\nu)}{E}\sigma^\infty\right)\cos^2 kx \\ + \frac{(1+\nu)}{E}\sigma_\infty\sin^2 kx + \eta\cos 2kx \right\}. \quad (119)$$

**Appendix D: Calculation Detail for Boundary Value Problems of Random Surface Profile**

We apply a far-field stress $\sigma_{xx} = \sigma^\infty e_1 \otimes e_1$ on the half-space, solve the different order boundary value problems (17) and (18) for $\langle u^{(i)}\rangle$ and (19) and (20) for $Qu^{(i)}$ using the approach explained in Appendix B. By (51), we have the zeroth order boundary condition for average terms as

$$\langle t_x^{(0)} \rangle = k^s \frac{\partial \langle \varepsilon_{xx}^{(0)} \rangle}{\partial x} = 0, \qquad \langle t_y^{(0)} \rangle = 0 \tag{120}$$

Zeroth order solution for average terms are

$$\langle \sigma_{xx}^{(0)} \rangle = \sigma^\infty, \qquad \langle \sigma_{xy}^{(0)} \rangle = \langle \sigma_{yy}^{(0)} \rangle = 0 \tag{121}$$

The boundary condition for $Q$-terms (54) would be rewritten as

$$Qt_x^{(0)} = 0, \qquad Qt_y^{(0)} = 0. \tag{122}$$

And the zeroth order solution for $Q$-terms are obtained as

$$Q\sigma^{(0)} = 0 \tag{123}$$

By (52), we have the first order boundary condition for average terms as

$$\langle t_x^{(1)} \rangle = 0, \qquad \langle t_y^{(1)} \rangle = 0 \tag{124}$$

and the first order solution for average terms is

$$\langle \sigma^{(1)} \rangle = 0. \tag{125}$$

Similarly, using first order boundary condition for $Q$-terms (55),

$$Qt_x^{(1)} = h_{0x}\sigma^\infty, \qquad Qt_y^{(1)} = \left[\tau^o + \frac{(1-\nu^2)}{E}(k^s - \tau^o)\sigma^\infty\right]h_{0xx}, \tag{126}$$

first order solution for $Q$-terms are obtained as follows.

$$Q\sigma_{xx}^{(1)}(x,y) = \frac{1}{2\pi}\int_{-\infty}^{\infty}\left[f(\alpha)\left(-2\frac{|\alpha|}{i\alpha}+i\alpha y\right)+p(\alpha)(1+|\alpha|y)\right]e^{i\alpha x+|\alpha|y}d\alpha,$$

$$Q\sigma_{xy}^{(1)}(x,y) = \frac{1}{2\pi}\int_{-\infty}^{\infty}\left[f(\alpha)(1+|\alpha|y)-i\alpha p(\alpha)y\right]e^{i\alpha x+|\alpha|y}d\alpha,$$

$$Q\sigma_{yy}^{(1)}(x,y) = \frac{1}{2\pi}\int_{-\infty}^{\infty}\left[-i\alpha f(\alpha)y+p(\alpha)(1-|\alpha|y)\right]e^{i\alpha x+|\alpha|y}d\alpha,$$

$$Q\varepsilon_{xx}^{(1)}(x,y) = \frac{1-\nu^2}{E}Q\sigma_{xx}^{(1)}(x,y)-\frac{\nu(1+\nu)}{E}Q\sigma_{yy}^{(1)}(x,y) =$$
$$\frac{1-\nu^2}{E}\frac{1}{2\pi}\int_{-\infty}^{\infty}\left[f(\alpha)\left(-2\frac{|\alpha|}{i\alpha}+i\alpha y\right)+p(\alpha)(1+|\alpha|y)\right]e^{i\alpha x+|\alpha|y}d\alpha$$
$$-\frac{\nu(1+\nu)}{E}\frac{1}{2\pi}\int_{-\infty}^{\infty}\left[-i\alpha f(\alpha)y+p(\alpha)(1-|\alpha|y)\right]e^{i\alpha x+|\alpha|y}d\alpha,$$

$$\frac{\partial Q\varepsilon_{xx}^{(1)}(x,y)}{\partial x} = \frac{1-\nu^2}{E}\frac{1}{2\pi}\int_{-\infty}^{\infty}\left[f(\alpha)(-2|\alpha|-\alpha^2 y)+(i\alpha)p(\alpha)(1+|\alpha|y)\right]e^{i\alpha x+|\alpha|y}d\alpha$$
$$-\frac{\nu(1+\nu)}{E}\frac{1}{2\pi}\int_{-\infty}^{\infty}\left[\alpha^2 f(\alpha)y+(i\alpha)p(\alpha)(1-|\alpha|y)\right]e^{i\alpha x+|\alpha|y}d\alpha,$$

(127)

$$Q\varepsilon_{xy}^{(1)}(x,y) = \frac{(1+\nu)}{E}Q\sigma_{xy}^{(1)}(x,y) = \frac{(1+\nu)}{2\pi E}\int_{-\infty}^{\infty}\left[f(\alpha)(1+|\alpha|y)-i\alpha p(\alpha)y\right]e^{i\alpha x+|\alpha|y}d\alpha,$$

$$\frac{\partial Q\varepsilon_{xy}^{(1)}(x,y)}{\partial x} = \frac{(1+\nu)}{2\pi E}\int_{-\infty}^{\infty}\left[f(\alpha)(i\alpha)(1+|\alpha|y)+\alpha^2 p(\alpha)y\right]e^{i\alpha x+|\alpha|y}d\alpha,$$

$$\frac{\partial Q\sigma_{xy}^{(1)}(x,y)}{\partial y} = \frac{1}{2\pi}\int_{-\infty}^{\infty}\left[f(\alpha)(2|\alpha|+\alpha^2 y)-i\alpha p(\alpha)(1+|\alpha|y)\right]e^{i\alpha x+|\alpha|y}d\alpha,$$

$$\frac{\partial Q\sigma_{yy}^{(1)}(x,y)}{\partial y} = \frac{1}{2\pi}\int_{-\infty}^{\infty}\left[-i\alpha f(\alpha)(1+|\alpha|y)+p(\alpha)(-\alpha^2 y)\right]e^{i\alpha x+|\alpha|y}d\alpha,$$

$$\frac{\partial Q\varepsilon_{xx}^{(1)}(x,y)}{\partial y} = \frac{1-\nu^2}{E}\frac{1}{2\pi}\int_{-\infty}^{\infty}\left[i\alpha f(\alpha)(3+|\alpha|y)+p(\alpha)(2|\alpha|+\alpha^2 y)\right]e^{i\alpha x+|\alpha|y}d\alpha$$
$$-\frac{\nu(1+\nu)}{E}\frac{1}{2\pi}\int_{-\infty}^{\infty}\left[-i\alpha f(\alpha)(1+|\alpha|y)-p(\alpha)(\alpha^2 y)\right]e^{i\alpha x+|\alpha|y}d\alpha,$$

$$\frac{\partial^2 Q\varepsilon_{xx}^{(1)}(x,y)}{\partial x \partial y} = \frac{1-\nu^2}{E}\frac{1}{2\pi}\int_{-\infty}^{\infty}\left[-\alpha^2 f(\alpha)(3+|\alpha|y)+(i\alpha)p(\alpha)(2|\alpha|+\alpha^2 y)\right]e^{i\alpha x+|\alpha|y}d\alpha$$
$$-\frac{\nu(1+\nu)}{E}\frac{1}{2\pi}\int_{-\infty}^{\infty}\left[\alpha^2 f(\alpha)(1+|\alpha|y)-p(\alpha)(i\alpha^3 y)\right]e^{i\alpha x+|\alpha|y}d\alpha$$

with $f(\alpha) = \sigma_\infty (i\alpha) h_0(\alpha)$ and $p(\alpha) = \left[\tau^o + \dfrac{(1-\nu^2)}{E} k^s \sigma^\infty\right] (i\alpha)^2 h_0(\alpha)$.

Consequently, by calculating (127) at $y = 0$, the second order B.C. (53) on average fields are rewritten as

$$\begin{aligned}\langle t_x^{(2)} \rangle = & P\left[-2\sigma_\infty \frac{1}{2\pi} h_{0x} \int_{-\infty}^{\infty} |\alpha| h_0(\alpha) e^{i\alpha x} d\alpha + \left(\tau^o + \frac{(1-\nu^2)}{E} k^s \sigma^\infty\right) h_{0x} h_{0xx}(x)\right] \\ & - P\left[2\sigma_\infty \frac{1}{2\pi} h_0 \int_{-\infty}^{\infty} i\alpha |\alpha| h_0(\alpha) e^{i\alpha x} d\alpha - \left(\tau^o + \frac{(1-\nu^2)}{E} k^s \sigma^\infty\right) h_0 h_{0xxx}(x)\right] - \tau^o P[h_{0x} h_{0xx}] \\ & - k^s P[h_{0x} h_{0xx}] \frac{1-\nu^2}{E} \sigma_\infty + 3k^s \frac{1-\nu^2}{E} \sigma_\infty P[h_0(x) h_{0xxx}(x)] - \\ & - 2k^s \frac{1-\nu^2}{E}\left(\tau^o + \frac{(1-\nu^2)}{E} k^s \sigma^\infty\right) P\left[\frac{1}{2\pi} h_0 \int_{-\infty}^{\infty} i\alpha^3 |\alpha| h_0(\alpha) e^{i\alpha x} d\alpha\right] + \\ & + k^s \frac{\nu(1+\nu)}{E} \sigma_\infty P[h_0(x) h_{0xxx}(x)] + 2k^s \frac{(1+\nu)}{E} \sigma_\infty P[h_{0xx}(x) h_{0x}(x)] + \\ & + 2k^s \frac{(1+\nu)}{E} \sigma_\infty P[h_{0x}(x) h_{0xx}(x)] + 2k^s P[h_{0x}(x) h_{0xx}(x)]\left(-\frac{\nu(1+\nu)}{E} \sigma_\infty - \frac{1-\nu^2}{E} \sigma_\infty\right) \\ & + 3k^s \frac{1-\nu^2}{E} \sigma_\infty P[h_{0x}(x) h_{0xx}(x)] \\ & - 2k^s \frac{1-\nu^2}{E}\left(\tau^o + \frac{(1-\nu^2)}{E} k^s \sigma^\infty\right) P\left[\frac{1}{2\pi} h_{0x} \int_{-\infty}^{\infty} \alpha^2 |\alpha| h_0(\alpha) e^{i\alpha x} d\alpha\right] \\ & + k^s \frac{\nu(1+\nu)}{E} \sigma_\infty P[h_{0x}(x) h_{0xx}(x)]\end{aligned}$$

(128)

$$\begin{aligned}\langle t_y^{(2)} \rangle = & \sigma_\infty P[h_0 h_{0xx}(x)] + \sigma_\infty P[h_{0x} h_{0x}(x)] - 2\sigma_\infty \frac{1-\nu^2}{E} k^s P\left[\frac{1}{2\pi} h_{0xx} \int_{-\infty}^{\infty} |\alpha| h_0(\alpha) e^{i\alpha x} d\alpha\right] \\ & + \frac{(1+\nu)(1-2\nu)}{E} k^s \left(\tau^o + \frac{(1-\nu^2)}{E} k^s \sigma^\infty\right) P[h_{0xx}(x) h_{0xx}(x)] \\ & - 2\frac{1-\nu^2}{E} k^s \sigma_\infty P\left[\frac{1}{2\pi} h_{0x} \int_{-\infty}^{\infty} i\alpha |\alpha| h_0(\alpha) e^{i\alpha x} d\alpha\right] + \\ & + \frac{(1+\nu)(1-2\nu)}{E} k^s \left(\tau^o + \frac{(1-\nu^2)}{E} k^s \sigma^\infty\right) P[h_{0x}(x) h_{0xxx}(x)]\end{aligned}$$

(129)

By replacing $h_0(x)$ by its Fourier integral representation $h_0(x) = \dfrac{1}{2\pi}\int_{-\infty}^{\infty} h_0(\xi) e^{i\xi x} d\xi$, (128) and (129) would be rewritten as

$$\begin{aligned}
<t_x^{(2)}> = & -P\left[2\sigma_\infty\left(\frac{1}{2\pi}\right)^2\int_{-\infty}^\infty(i\beta)h_0(\beta)e^{i\beta x}d\beta\int_{-\infty}^\infty|\alpha|h_0(\alpha)e^{i\alpha x}d\alpha\right]+ \\
& +\left(\tau^o+\frac{(1-\nu^2)}{E}k^s\sigma^\infty\right)\langle h_{0x}(x)h_{0xx}(x)\rangle- \\
& -P\left[2\sigma_\infty\left(\frac{1}{2\pi}\right)^2\int_{-\infty}^\infty h_0(\beta)e^{i\beta x}d\beta\int_{-\infty}^\infty(i\alpha)|\alpha|h_0(\alpha)e^{i\alpha x}d\alpha\right] \\
& +\left(\tau^o+\frac{(1-\nu^2)}{E}k^s\sigma^\infty\right)\langle h_0(x)h_{0xxx}(x)\rangle-\tau^o\langle h_{0x}(x)h_{0xx}(x)\rangle \\
& -k^s\frac{1-\nu^2}{E}\sigma_\infty\langle h_{0x}(x)h_{0xx}(x)\rangle+3\frac{1-\nu^2}{E}\sigma_\infty k^s\langle h_0(x)h_{0xxx}(x)\rangle \\
& -2\frac{1-\nu^2}{E}k^s\left(\tau^o+\frac{(1-\nu^2)}{E}k^s\sigma^\infty\right)P\left[\left(\frac{1}{2\pi}\right)^2\int_{-\infty}^\infty h_0(\beta)e^{i\beta x}d\beta\int_{-\infty}^\infty i\alpha^3|\alpha|h_0(\alpha)e^{i\alpha x}d\alpha\right] \\
& +\frac{\nu(1+\nu)}{E}\sigma_\infty k^s\langle h_0(x)h_{0xxx}(x)\rangle+\frac{2(1+\nu)}{E}\sigma_\infty k^s\langle h_{0xx}(x)h_{0x}(x)\rangle+ \\
& +\frac{2(1+\nu)}{E}\sigma_\infty k^s\langle h_{0x}(x)h_{0xx}(x)\rangle+2k^s\langle h_{0xx}(x)h_{0x}(x)\rangle\left(-\frac{\nu(1+\nu)}{E}\sigma_\infty-\frac{1-\nu^2}{E}\sigma_\infty\right) \\
& +3\sigma_\infty\frac{1-\nu^2}{E}k^s\langle h_{0x}(x)h_{0xx}(x)\rangle+\frac{\nu(1+\nu)}{E}\sigma_\infty k^s\langle h_{0x}(x)h_{0xx}(x)\rangle \\
& -2\frac{1-\nu^2}{E}k^s\left(\tau^o+\frac{(1-\nu^2)}{E}k^s\sigma^\infty\right)P\left[\left(\frac{1}{2\pi}\right)^2\int_{-\infty}^\infty(i\beta)h_0(\beta)e^{i\beta x}\int_{-\infty}^\infty\alpha^2|\alpha|h_0(\alpha)e^{i\alpha x}d\alpha\right]
\end{aligned}$$
(130)

$$\begin{aligned}
<t_y^{(2)}> = & \sigma_\infty\langle h_0(x)h_{0xx}(x)\rangle+\sigma_\infty\langle h_{0x}(x)h_{0x}(x)\rangle \\
& -2\sigma_\infty k^s\frac{1-\nu^2}{E}\left(\frac{1}{2\pi}\right)^2P\left[\int_{-\infty}^\infty(i\beta)^2h_0(\beta)e^{i\beta x}d\beta\int_{-\infty}^\infty|\alpha|h_0(\alpha)e^{i\alpha x}d\alpha\right] \\
& +\frac{(1+\nu)(1-2\nu)}{E}k^s\left(\tau^o+\frac{(1-\nu^2)}{E}k^s\sigma^\infty\right)\langle h_{0xx}(x)h_{0xx}(x)\rangle \\
& -P\left[2\sigma_\infty\frac{1-\nu^2}{E}k^s\left(\frac{1}{2\pi}\right)^2\int_{-\infty}^\infty(i\beta)h_0(\beta)e^{i\beta x}d\beta\int_{-\infty}^\infty i\alpha|\alpha|h_0(\alpha)e^{i\alpha x}d\alpha\right]+ \\
& +\frac{(1+\nu)(1-2\nu)}{E}k^s\left(\tau^o+\frac{(1-\nu^2)}{E}k^s\sigma^\infty\right)\langle h_{0x}(x)h_{0xxx}(x)\rangle
\end{aligned}$$
(131)

We should note here that in order to solve second order boundary value problems we must specify the nature of the randomness of the surface. For that purpose we introduce the surface height correlation function $W$

$$\langle h_0(x)h_0(x')\rangle = \eta^2 W(|x-x'|).$$  (132)

where $\eta$ is the root-mean-square departure of the surface from flatness and $W(0)=1$. If $h_0(x)$ with Fourier integral representation of

$$h_0(x) = \frac{1}{2\pi} \int_{-\infty}^{\infty} h_0(\xi) e^{i\xi x} d\xi. \tag{133}$$

be a zero-mean Gaussian random function, then, its Fourier coefficient $h_0(\xi)$ is also a zero-mean Gaussian random variable and possesses the properties (Maradudin, 2007)

$$\langle h_0(\xi) \rangle = 0,$$
$$\langle h_0(\xi) h_0(\xi') \rangle = (2\pi) \eta^2 g(\xi) \delta(\xi + \xi'), \tag{134}$$

where $g(\xi)$ is the one-dimensional Fourier transform of the surface height autocorrelation function $W(|x|)$,

$$g(\xi) = \int_{-\infty}^{\infty} W(|x|) e^{-i\xi x} dx. \tag{135}$$

Here $W(|x|)$ and hence $g(\xi)$ will be assumed to be Gaussian in form

$$W(|x|) = e^{(-x^2/a^2)},$$
$$g(\xi) = \sqrt{\pi} a e^{(-a^2 \xi^2 / 4)}. \tag{136}$$

The characteristic length $a$ is the transverse correlation length of the surface roughness. It is a measure of the average distance between successive 'peaks' or 'valleys' on the surface.

Also, if $h'(x) = L_1 h(x)$ and $h''(x) = L_2 h(x)$ and $L_1$ and $L_2$ be two linear and homogeneous operators, then (Sveshnikov, 1978)

$$\langle h'(x) \rangle = L_1 \langle h(x) \rangle,$$
$$\langle h'_0(x) h'_0(x') \rangle = L_{1x} L_{1x'} W|(x-x')|, \tag{137}$$
$$\langle (L_1 h_0(x))(L_2 h_0(x')) \rangle = L_{1x} L_{2x'} W|(x-x')|.$$

Here the first equation is the expectation value of random variable $h'(x)$, the second equation is the autocorrelation function of a random variable $h'(x)$, and the third equation is the mutual or cross correlation function of two random variables $h'(x)$ and $h''(x)$. As an example, to calculate $\langle h_0(x) h_{0xxx}(x') \rangle$, $L_1 = 1$ and $L_2 = \frac{d^3}{dx^3}$, then

$$\langle h_0(x) h_{0xxx}(x') \rangle = \frac{d^3}{dx'^3} W|x-x'| = \frac{d^3}{dx'^3} e^{\frac{-|x-x'|^2}{a^2}}. \tag{138}$$

Therefore, we obtain

$$\langle h_0(x) h_{0xxx}(x) \rangle = 0, \qquad \langle h_{0x}(x) h_{0xx}(x) \rangle = 0, \tag{139}$$

$$\langle h_{0x}(x)h_{0x}(x)\rangle = \frac{2\eta^2}{a^2}, \qquad \langle h_0(x)h_{0xx}(x)\rangle = -\frac{2\eta^2}{a^2},$$

$$\langle h_{0xx}(x)h_{0xx}(x)\rangle = \frac{12\eta^2}{a^4}, \qquad \langle h_{0x}(x)h_{0xxx}(x)\rangle = -12\frac{\eta^2}{a^4}$$

The second order boundary condition (130) and (131) can be rewritten as

$$\begin{aligned}
<t_x^{(2)}> = &-2\sigma_\infty \left(\frac{1}{2\pi}\right)^2 (2\pi)\eta^2 \int_{-\infty}^{\infty}\int_{-\infty}^{\infty} i\beta |\alpha| g(\alpha)\delta(\alpha+\beta)e^{i(\alpha+\beta)x}d\alpha d\beta \\
&- 2\sigma_\infty \left(\frac{1}{2\pi}\right)^2 (2\pi)\eta^2 \int_{-\infty}^{\infty}\int_{-\infty}^{\infty} i\alpha |\alpha| g(\alpha)\delta(\alpha+\beta)e^{i(\alpha+\beta)x}d\alpha d\beta \\
&- 2\frac{1-v^2}{E}k^s\left(\tau^o + \frac{(1-v^2)}{E}k^s\sigma^\infty\right)\left(\frac{1}{2\pi}\right)^2 (2\pi)\eta^2 \int_{-\infty}^{\infty}\int_{-\infty}^{\infty} i\alpha^3 |\alpha| g(\alpha)\delta(\alpha+\beta)e^{i(\alpha+\beta)x}d\alpha d\beta \\
&- 2\frac{1-v^2}{E}k^s\left(\tau^o + \frac{(1-v^2)}{E}k^s\sigma^\infty\right)\left(\frac{1}{2\pi}\right)^2 (2\pi)\eta^2 \int_{-\infty}^{\infty}\int_{-\infty}^{\infty} (i\beta)\alpha^2 |\alpha| g(\alpha)\delta(\alpha+\beta)e^{i(\alpha+\beta)x}d\alpha d
\end{aligned} \tag{140}$$

$$\begin{aligned}
<t_y^{(2)}> = &-2\sigma_\infty k^s \frac{1-v^2}{E}\left(\frac{1}{2\pi}\right)^2 (2\pi)\eta^2 \int_{-\infty}^{\infty}\int_{-\infty}^{\infty} (i\beta)^2 |\alpha| g(\alpha)\delta(\alpha+\beta)e^{i(\alpha+\beta)x}d\alpha d\beta \\
&- 2\sigma_\infty \frac{1-v^2}{E}k^s\left(\frac{1}{2\pi}\right)^2 (2\pi)\eta^2 \int_{-\infty}^{\infty}\int_{-\infty}^{\infty} (i\beta)(i\alpha) |\alpha| g(\alpha)\delta(\alpha+\beta)e^{i(\alpha+\beta)x}d\alpha d\beta
\end{aligned} \tag{141}$$

By more simplifying we would have

$$<t_x^{(2)}> = 0, \qquad <t_y^{(2)}> = 0 \tag{142}$$

And the second order solution for average stresses would be obtained as

$$<\sigma^{(2)}> = 0 \tag{143}$$

We also need to find the second order solution for $Q$-terms. The second order boundary conditions for $Q$-terms (56) would be rewritten as

$$Qt_x^{(2)} = h_{0x}\left(-2\sigma_\infty \frac{1}{2\pi}\int_{-\infty}^{\infty}|\alpha|h_0(\alpha)e^{i\alpha x}d\alpha + \left(\tau^o + \frac{(1-v^2)}{E}k^s\sigma^\infty\right)h_{0xx}(x)\right)$$

$$-h_0\left(2\sigma_\infty \frac{1}{2\pi}\int_{-\infty}^{\infty}i\alpha|\alpha|h_0(\alpha)e^{i\alpha x}d\alpha - \left(\tau^o + \frac{(1-v^2)}{E}k^s\sigma^\infty\right)h_{0xxx}(x)\right)$$

$$-\tau^o h_{0x}h_{0xx} - k^s h_{0x}h_{0xx}\frac{(1-v^2)}{E}\sigma^\infty$$

$$+k^s h_0\begin{pmatrix}3\dfrac{1-v^2}{E}\sigma_\infty h_{0xxx}(x) - 2\dfrac{1-v^2}{E}\left(\tau^o + \dfrac{(1-v^2)}{E}k^s\sigma^\infty\right)\dfrac{1}{2\pi}\int_{-\infty}^{\infty}i\alpha^3|\alpha|h_0(\alpha)e^{i\alpha x}d\alpha + \\ +\dfrac{v(1+v)}{E}\sigma_\infty h_{0xxx}(x)\end{pmatrix} \quad (144)$$

$$+ 2k^s h_{0xx}\left(\frac{(1+v)}{E}\sigma_\infty h_{0x}(x)\right) + 2k^s h_{0x}\left(\frac{(1+v)}{E}\sigma_\infty h_{0xx}(x)\right) + 2k^s h_{0x}h_{0xx}\left(-\frac{v(1+v)}{E}\sigma_\infty - \frac{(1-v^2)}{E}\sigma^\infty\right)$$

$$+k^s h_{0x}\begin{pmatrix}3\sigma_\infty\dfrac{1-v^2}{E}h_{0xx}(x) - 2\dfrac{1-v^2}{E}\left(\tau^o + \dfrac{(1-v^2)}{E}k^s\sigma^\infty\right)\dfrac{1}{2\pi}\int_{-\infty}^{\infty}\alpha^2|\alpha|h_0(\alpha)e^{i\alpha x}d\alpha + \\ +\sigma_\infty\dfrac{v(1+v)}{E}h_{0xx}(x)\end{pmatrix}$$

$$Qt_y^{(2)} = h_{0x}(x)Q\sigma_{yx}^{(1)} + h_0(x)\sigma_\infty h_{0xx}(x) +$$

$$+ k^s h_{0xx}(x)\left(-2\sigma_\infty \frac{1-v^2}{E}\frac{1}{2\pi}\int_{-\infty}^{\infty}|\alpha|h_0(\alpha)e^{i\alpha x}d\alpha + \frac{(1+v)(1-2v)}{E}\left(\tau^o + \frac{(1-v^2)}{E}k^s\sigma^\infty\right)h_{0xx}(x)\right) \quad (145)$$

$$+ k^s h_{0x}(x)\left(-2\frac{1-v^2}{E}\frac{\sigma_\infty}{2\pi}\int_{-\infty}^{\infty}i\alpha|\alpha|h_0(\alpha)e^{i\alpha x}d\alpha + \frac{(1+v)(1-2v)}{E}\left(\tau^o + \frac{(1-v^2)}{E}k^s\sigma^\infty\right)h_{0xxx}(x)\right)$$

The solutions for the second order $Q$-terms are obtained as

$$Q\sigma_{xx}^{(2)}(x,y) = \frac{1}{2\pi}\int_{-\infty}^{\infty}\left[f(\alpha)\left(-2\frac{|\alpha|}{i\alpha}+i\alpha y\right) + p(\alpha)(1+|\alpha|y)\right]e^{i\alpha x+|\alpha|y}d\alpha$$

$$Q\sigma_{xy}^{(2)}(x,y) = \frac{1}{2\pi}\int_{-\infty}^{\infty}\left[f(\alpha)(1+|\alpha|y) - i\alpha p(\alpha)y\right]e^{i\alpha x+|\alpha|y}d\alpha$$

$$Q\sigma_{yy}^{(2)}(x,y) = \frac{1}{2\pi}\int_{-\infty}^{\infty}\left[-i\alpha f(\alpha)y + p(\alpha)(1-|\alpha|y)\right]e^{i\alpha x+|\alpha|y}d\alpha$$

(146)

$$Q\varepsilon_{xx}^{(2)}(x,y) = \frac{1-v^2}{E}Q\sigma_{xx}^{(1)}(x,y) - \frac{v(1+v)}{E}Q\sigma_{yy}^{(1)}(x,y) =$$

$$\frac{1-v^2}{E}\frac{1}{2\pi}\int_{-\infty}^{\infty}\left[f(\alpha)\left(-2\frac{|\alpha|}{i\alpha}+i\alpha y\right) + p(\alpha)(1+|\alpha|y)\right]e^{i\alpha x+|\alpha|y}d\alpha -$$

$$-\frac{v(1+v)}{E}\frac{1}{2\pi}\int_{-\infty}^{\infty}\left[-i\alpha f(\alpha)y + p(\alpha)(1-|\alpha|y)\right]e^{i\alpha x+|\alpha|y}d\alpha$$

$$Q\varepsilon_{xy}^{(2)}(x,y) = \frac{(1+v)}{E}Q\sigma_{xy}^{(1)}(x,y) = \frac{(1+v)}{2\pi E}\int_{-\infty}^{\infty}\left[f(\alpha)(1+|\alpha|y) - i\alpha p(\alpha)y\right]e^{i\alpha x+|\alpha|y}d\alpha$$

where $f(\alpha)$ and $p(\alpha)$ are the Fourier transform of (144) and (145), respectively.

By using the relations in Appendix A and performing the Taylor series expansion, the surface strain on the rough surface is obtained as

$$[\varepsilon_{ss}]_{y=h(x)} = \left[\varepsilon_{xx} + 2\delta h_{0x}\varepsilon_{xy} + \delta^2(h_{0x})^2(\varepsilon_{yy} - \varepsilon_{xx})\right]_{y=h(x)} =$$

$$= \begin{bmatrix} \varepsilon_{xx}^{(0)} + \delta\left(\varepsilon_{xx}^{(1)} + 2h_{0x}\varepsilon_{xy}^{(0)} + h_0\dfrac{\partial \varepsilon_{xx}^{(0)}}{\partial y}\right) + \\ \delta^2\left(h_0\dfrac{\partial \varepsilon_{xx}^{(1)}}{\partial y} + \dfrac{1}{2}(h_0)^2\dfrac{\partial^2 \varepsilon_{xx}^{(0)}}{\partial y^2} + 2(h_{0x})h_0\dfrac{\partial \varepsilon_{xy}^{(0)}}{\partial y}\right) \\ -(h_{0x})^2\varepsilon_{xx}^{(0)} + \varepsilon_{xx}^{(2)} + 2h_{0x}\varepsilon_{xy}^{(1)} + (h_{0x})^2\varepsilon_{yy}^{(0)} \end{bmatrix}_{y=0}$$

$$= \begin{bmatrix} \langle\varepsilon_{xx}^{(0)}\rangle + \delta\left(Q\varepsilon_{xx}^{(1)}\right) + \\ \delta^2\left(h_0\dfrac{\partial Q\varepsilon_{xx}^{(1)}}{\partial y} - (h_{0x})^2\langle\varepsilon_{xx}^{(0)}\rangle + Q\varepsilon_{xx}^{(2)} + 2h_{0x}Q\varepsilon_{xy}^{(1)} + (h_{0x})^2\langle\varepsilon_{yy}^{(0)}\rangle\right) \end{bmatrix}_{y=0}$$

(147)

and gives us

$$[\varepsilon_{ss}]_{y=h(x)} =$$

$$= \dfrac{1-v^2}{E}\sigma^\infty + \delta\begin{pmatrix} \left\{\dfrac{1}{2\pi}\int_{-\infty}^{\infty}-2|\alpha|\sigma_\infty h_0(\alpha)e^{i\alpha x}d\alpha + \left[\tau^o + \dfrac{(1-v^2)}{E}k^s\sigma^\infty\right]h_{0xx}(x)\right\}\dfrac{(1-v^2)}{E} + \\ +\left\{\tau^o + \dfrac{(1-v^2)}{E}k^s\sigma^\infty h_{0xx}(x)\right\}\dfrac{-v(1+v)}{E} \end{pmatrix}$$

$$+ \delta^2 \begin{pmatrix} h_0(x)\left[3\sigma_\infty\dfrac{1-v^2}{E}h_{0xx}(x) - 2\dfrac{1-v^2}{E}\left(\tau^o + \dfrac{(1-v^2)}{E}k^s\sigma^\infty\right)\dfrac{1}{2\pi}\int_{-\infty}^{\infty}\alpha^2|\alpha|h_0(\alpha)e^{i\alpha x}d\alpha\right] \\ +\sigma_\infty\dfrac{v(1+v)}{E}h_{0xx}(x) \\ -(h_{0x})^2\dfrac{(1-v^2)}{E}\sigma^\infty + Q\varepsilon_{xx}^{(2)} + 2h_{0x}(x)\dfrac{(1+v)}{E}\sigma_\infty h_{0x}(x) + (h_{0x})^2\dfrac{-v(1+v)}{E}\sigma^\infty \end{pmatrix}$$

(148)

In order to find the effective surface stress and effective surface elastic constant we make the ensemble average of total energy of the half space with rough surface equal to energy of a half space with flat surface and effective surface constants.

$$\langle E^{act}(\sigma^\infty)\rangle = E^{eff}(\sigma^\infty) \tag{149}$$

For calculating $(\tau^0)^{eff}$ we need to consider the energy terms up to first order of $\sigma^\infty$ and we call it $E'$. For simplicity of presenting the proceeding calculation, we split $E'$ into two parts.

$$E'_{bulk} = (E'_{bulk})_1 + (E'_{bulk})_2 \tag{150}$$

$$\left(E'_{bulk}\right)_1 = \frac{1}{2}\delta P \int_{-\infty}^{h(x)} \left[ \begin{array}{l} 2\left\{\dfrac{1}{2\pi}\int_{-\infty}^{\infty}\left[\tau^o(i\alpha)^2 h_0(\alpha)\left(1+|\alpha|y\right)\right]e^{i\alpha x+|\alpha|y}d\alpha\right\} \times \dfrac{(1-\nu^2)}{E}\sigma^{\infty} + \\ +2\left\{\dfrac{1}{2\pi}\int_{-\infty}^{\infty}\left[\tau^o(i\alpha)^2 h_0(\alpha)\left(1-|\alpha|y\right)\right]e^{i\alpha x+|\alpha|y}d\alpha\right\} \times \dfrac{-\nu(1+\nu)}{E}\sigma^{\infty} \end{array} \right] dy +$$

$$\frac{\delta^2}{2}\int_{-\infty}^{h(x)} P \left[ \begin{array}{l} \dfrac{(1-\nu^2)}{E}\left(\dfrac{1}{2\pi}\right)^2 \int_{-\infty}^{\infty}\sigma_{\infty}(i\beta)h_0(\beta)\left(-2\dfrac{|\beta|}{i\beta}+i\beta y\right)e^{i\beta x+|\beta|y}d\beta \times \int_{-\infty}^{\infty}\tau^o(i\alpha)^2 h_0(\alpha)\left(1+|\alpha|y\right)e^{i\alpha x+|\alpha|y}d\alpha + \\ -\dfrac{\nu(1+\nu)}{E}\left(\dfrac{1}{2\pi}\right)^2 \int_{-\infty}^{\infty}\sigma_{\infty}(i\beta)h_0(\beta)\left(-2\dfrac{|\beta|}{i\beta}+i\beta y\right)e^{i\beta x+|\beta|y}d\beta \times \int_{-\infty}^{\infty}\tau^o(i\alpha)^2 h_0(\alpha)\left(1-|\alpha|y\right)e^{i\alpha x+|\alpha|y}d\alpha \\ +\dfrac{(1-\nu^2)}{E}\left(\dfrac{1}{2\pi}\right)^2 \dfrac{(1-\nu^2)}{E}k^s\int_{-\infty}^{\infty}\sigma_{\infty}(i\beta)^2 h_0(\beta)\left(1+|\beta|y\right)e^{i\beta x+|\beta|y}d\beta \times \int_{-\infty}^{\infty}\tau^o(i\alpha)^2 h_0(\alpha)\left(1+|\alpha|y\right)e^{i\alpha x+|\alpha|y}d\alpha + \\ -\dfrac{\nu(1+\nu)}{E}\left(\dfrac{1}{2\pi}\right)^2 \dfrac{(1-\nu^2)}{E}k^s\int_{-\infty}^{\infty}\sigma_{\infty}(i\beta)^2 h_0(\beta)\left(1+|\beta|y\right)e^{i\beta x+|\beta|y}d\beta \times \int_{-\infty}^{\infty}\tau^o(i\alpha)^2 h_0(\alpha)\left(1-|\alpha|y\right)e^{i\alpha x+|\alpha|y}d\alpha \\ +\left(\dfrac{1}{2\pi}\right)^2 \dfrac{(1-\nu^2)}{E}\int_{-\infty}^{\infty}\left[\tau^o(i\alpha)^2 h_0(\alpha)\left(1+|\alpha|y\right)\right]e^{i\alpha x+|\alpha|y}d\alpha \times \int_{-\infty}^{\infty}\sigma_{\infty}(i\beta)h_0(\beta)\left(-2\dfrac{|\beta|}{i\beta}+i\beta y\right)e^{i\beta x+|\beta|y}d\beta + \\ +\left(\dfrac{1}{2\pi}\right)^2 \dfrac{(1-\nu^2)}{E}\int_{-\infty}^{\infty}\left[\tau^o(i\alpha)^2 h_0(\alpha)\left(1+|\alpha|y\right)\right]e^{i\alpha x+|\alpha|y}d\alpha \times \dfrac{(1-\nu^2)}{E}k^s\int_{-\infty}^{\infty}\sigma_{\infty}(i\beta)^2 h_0(\beta)\left(1+|\beta|y\right)e^{i\beta x+|\beta|y}d\beta \\ +2\left(\dfrac{1}{2\pi}\right)^2 \dfrac{-\nu(1+\nu)}{E}\int_{-\infty}^{\infty}\left[\tau^o(i\alpha)^2 h_0(\alpha)\left(1+|\alpha|y\right)\right]e^{i\alpha x+|\alpha|y}d\alpha \times \int_{-\infty}^{\infty}-\sigma_{\infty}(i\beta)^2 h_0(\beta)y e^{i\beta x+|\beta|y}d\beta + \\ 2\left(\dfrac{1}{2\pi}\right)^2 \dfrac{-\nu(1+\nu)}{E}\int_{-\infty}^{\infty}\left[\tau^o(i\alpha)^2 h_0(\alpha)\left(1+|\alpha|y\right)\right]e^{i\alpha x+|\alpha|y}d\alpha \times \dfrac{(1-\nu^2)}{E}k^s\int_{-\infty}^{\infty}\sigma_{\infty}(i\beta)^2 h_0(\beta)\left(1-|\beta|y\right)e^{i\beta x+|\beta|y}d\beta \\ +\left(\dfrac{1}{2\pi}\right)^2 \dfrac{(1-\nu^2)}{E}\int_{-\infty}^{\infty}\left[\tau^o(i\alpha)^2 h_0(\alpha)\left(1-|\alpha|y\right)\right]e^{i\alpha x+|\alpha|y}d\alpha \times \int_{-\infty}^{\infty}-\sigma_{\infty}(i\beta)^2 h_0(\beta)y e^{i\beta x+|\beta|y}d\beta + \\ \left(\dfrac{1}{2\pi}\right)^2 \dfrac{(1-\nu^2)}{E}\int_{-\infty}^{\infty}\left[\tau^o(i\alpha)^2 h_0(\alpha)\left(1-|\alpha|y\right)\right]e^{i\alpha x+|\alpha|y}d\alpha \times \dfrac{(1-\nu^2)}{E}k^s\int_{-\infty}^{\infty}\sigma_{\infty}(i\beta)^2 h_0(\beta)\left(1-|\beta|y\right)e^{i\beta x+|\beta|y}d\beta \\ \left(\dfrac{1}{2\pi}\right)^2 \dfrac{-\nu(1+\nu)}{E}\int_{-\infty}^{\infty}\left[\tau^o(i\alpha)^2 h_0(\alpha)\left(1-|\alpha|y\right)\right]e^{i\alpha x+|\alpha|y}d\alpha \times \int_{-\infty}^{\infty}\sigma_{\infty}(i\beta)h_0(\beta)\left(-2\dfrac{|\beta|}{i\beta}+i\beta y\right)e^{i\beta x+|\beta|y}d\beta + \\ \left(\dfrac{1}{2\pi}\right)^2 \dfrac{-\nu(1+\nu)}{E}\int_{-\infty}^{\infty}\left[\tau^o(i\alpha)^2 h_0(\alpha)\left(1-|\alpha|y\right)\right]e^{i\alpha x+|\alpha|y}d\alpha \times \dfrac{(1-\nu^2)}{E}k^s\int_{-\infty}^{\infty}\sigma_{\infty}(i\beta)^2 h_0(\beta)\left(1+|\beta|y\right)e^{i\beta x+|\beta|y}d\beta \\ +\left(\dfrac{1}{2\pi}\right)^2 \dfrac{(1-\nu^2)}{E}\int_{-\infty}^{\infty}\left[\tau^o(i\alpha)^2 h_0(\alpha)\left(1-|\alpha|y\right)\right]e^{i\alpha x+|\alpha|y}d\alpha \times \int_{-\infty}^{\infty}-\sigma_{\infty}(i\beta)^2 h_0(\beta)y e^{i\beta x+|\beta|y}d\beta + \\ \left(\dfrac{1}{2\pi}\right)^2 \dfrac{(1-\nu^2)}{E}\int_{-\infty}^{\infty}\left[\tau^o(i\alpha)^2 h_0(\alpha)\left(1-|\alpha|y\right)\right]e^{i\alpha x+|\alpha|y}d\alpha \times \dfrac{(1-\nu^2)}{E}k^s\int_{-\infty}^{\infty}\sigma_{\infty}(i\beta)^2 h_0(\beta)\left(1-|\beta|y\right)e^{i\beta x+|\beta|y}d\beta \\ +\dfrac{(1+\nu)}{E}\left(\dfrac{1}{2\pi}\right)^2 \left(2\tau^o\dfrac{(1-\nu^2)}{E}k^s\sigma^{\infty}\right)\int_{-\infty}^{\infty}(i\beta)^3 h_0(\beta)y e^{i\beta x+|\beta|y}d\beta \times \int_{-\infty}^{\infty}(i\alpha)^3 h_0(\alpha)y e^{i\alpha x+|\alpha|y}d\alpha + \\ -2\dfrac{(1+\nu)}{E}\left(\dfrac{1}{2\pi}\right)^2 \int_{-\infty}^{\infty}\sigma_{\infty}(i\beta)h_0(\beta)\left(1+|\beta|y\right)e^{i\beta x+|\beta|y}d\beta \times \int_{-\infty}^{\infty}\tau^o(i\alpha)^3 h_0(\alpha)y e^{i\alpha x+|\alpha|y}d\alpha \end{array} \right] \quad (151)$$

More simplifying

$$\left(E'_{bulk}\right)_1 = \frac{1}{2}\delta\int_{-\infty}^{h(x)} P\begin{pmatrix} 2\left\{\frac{1}{2\pi}\int_{-\infty}^{\infty}\left[\tau^o(i\alpha)^2 h_0(\alpha)\left(1+|\alpha|y\right)\right]e^{i\alpha x+|\alpha|y}d\alpha\right\}\times\frac{(1-\nu^2)}{E}\sigma^\infty + \\ +2\left\{\frac{1}{2\pi}\int_{-\infty}^{\infty}\left[\tau^o(i\alpha)^2 h_0(\alpha)\left(1-|\alpha|y\right)\right]e^{i\alpha x+|\alpha|y}d\alpha\right\}\times\frac{-\nu(1+\nu)}{E}\sigma^\infty \end{pmatrix}dy +$$

$$\delta^2\int_{-\infty}^{h(x)} P\begin{pmatrix} \frac{(1-\nu^2)}{E}\left(\frac{1}{2\pi}\right)^2\int_{-\infty}^{\infty}\sigma_\infty(i\beta)h_0(\beta)\left(-2\frac{|\beta|}{i\beta}+i\beta y\right)e^{i\beta x+|\beta|y}d\beta\int_{-\infty}^{\infty}\tau^o(i\alpha)^2 h_0(\alpha)\left(1+|\alpha|y\right)e^{i\alpha x+|\alpha|y}d\alpha + \\ +\frac{-\nu(1+\nu)}{E}\left(\frac{1}{2\pi}\right)^2\int_{-\infty}^{\infty}\sigma_\infty(i\beta)h_0(\beta)\left(-2\frac{|\beta|}{i\beta}+i\beta y\right)e^{i\beta x+|\beta|y}d\beta\int_{-\infty}^{\infty}\tau^o(i\alpha)^2 h_0(\alpha)\left(1-|\alpha|y\right)e^{i\alpha x+|\alpha|y}d\alpha \\ +\frac{(1-\nu^2)}{E}\left(\frac{1}{2\pi}\right)^2\frac{(1-\nu^2)}{E}k^s\int_{-\infty}^{\infty}\sigma_\infty(i\beta)^2 h_0(\beta)\left(1+|\beta|y\right)e^{i\beta x+|\beta|y}d\beta\int_{-\infty}^{\infty}\tau^o(i\alpha)^2 h_0(\alpha)\left(1+|\alpha|y\right)e^{i\alpha x+|\alpha|y}d\alpha + \\ +2\frac{-\nu(1+\nu)}{E}\left(\frac{1}{2\pi}\right)^2\frac{(1-\nu^2)}{E}k^s\int_{-\infty}^{\infty}\sigma_\infty(i\beta)^2 h_0(\beta)\left(1+|\beta|y\right)e^{i\beta x+|\beta|y}d\beta\int_{-\infty}^{\infty}\tau^o(i\alpha)^2 h_0(\alpha)\left(1-|\alpha|y\right)e^{i\alpha x+|\alpha|y}d\alpha \\ +\left(\frac{1}{2\pi}\right)^2\frac{-\nu(1+\nu)}{E}\int_{-\infty}^{\infty}\left[\tau^o(i\alpha)^2 h_0(\alpha)\left(1+|\alpha|y\right)\right]e^{i\alpha x+|\alpha|y}d\alpha\int_{-\infty}^{\infty}-\sigma_\infty(i\beta)^2 h_0(\beta) y e^{i\beta x+|\beta|y}d\beta + \\ +\left(\frac{1}{2\pi}\right)^2\frac{(1-\nu^2)}{E}\int_{-\infty}^{\infty}\left[\tau^o(i\alpha)^2 h_0(\alpha)\left(1-|\alpha|y\right)\right]e^{i\alpha x+|\alpha|y}d\alpha\int_{-\infty}^{\infty}-\sigma_\infty(i\beta)^2 h_0(\beta) y e^{i\beta x+|\beta|y}d\beta + \\ \left(\frac{1}{2\pi}\right)^2\frac{(1-\nu^2)}{E}\frac{(1-\nu^2)}{E}k^s\int_{-\infty}^{\infty}\left[\tau^o(i\alpha)^2 h_0(\alpha)\left(1-|\alpha|y\right)\right]e^{i\alpha x+|\alpha|y}d\alpha\int_{-\infty}^{\infty}\sigma_\infty(i\beta)^2 h_0(\beta)\left(1-|\beta|y\right)e^{i\beta x+|\beta|y}d\beta \\ +\frac{(1+\nu)}{E}\left(\frac{1}{2\pi}\right)^2\left(\tau^o\frac{(1-\nu^2)}{E}k^s\sigma^\infty\right)\int_{-\infty}^{\infty}(i\beta)^3 h_0(\beta) y e^{i\beta x+|\beta|y}d\beta\int_{-\infty}^{\infty}(i\alpha)^3 h_0(\alpha) y e^{i\alpha x+|\alpha|y}d\alpha + \\ -\frac{(1+\nu)}{E}\left(\frac{1}{2\pi}\right)^2\int_{-\infty}^{\infty}\sigma_\infty(i\beta)h_0(\beta)\left(1+|\beta|y\right)e^{i\beta x+|\beta|y}d\beta\int_{-\infty}^{\infty}\tau^o(i\alpha)^3 h_0(\alpha) y e^{i\alpha x+|\alpha|y}d\alpha \end{pmatrix} \quad (152)$$

Calculating the integration in $y$ direction and doing the ensemble average in $x$ direction would yield

$$\left\langle E'_{bulk}\right\rangle_1 = \delta P\begin{pmatrix} \frac{1}{2\pi}\delta h_0(x)\int_{-\infty}^{\infty}\left[\tau^o(i\alpha)^2 h_0(\alpha)\right]e^{i\alpha x}d\alpha\frac{(1-\nu^2)}{E}\sigma^\infty + \\ +\left\{\frac{1}{2\pi}\int_{-\infty}^{\infty}\left[\tau^o(i\alpha)^2 h_0(\alpha)\left(\frac{2}{|\alpha|}+\delta h_0(x)\right)\right]e^{i\alpha x}d\alpha\right\}\frac{-\nu(1+\nu)}{E}\sigma^\infty \end{pmatrix} +$$

$$\delta^2\int_{-\infty}^{h(x)} P\begin{pmatrix} +\frac{(1-\nu^2)}{E}\sigma_\infty\tau^o\left(\frac{1}{2\pi}\right)^2\int_{-\infty}^{\infty}\int_{-\infty}^{\infty}i\beta\left(-2\frac{|\beta|}{i\beta}+i\beta y\right)(i\alpha)^2\left(1+|\alpha|y\right)g(\alpha)\delta(\alpha+\beta)e^{i(\alpha+\beta)x+(|\alpha|+|\beta|)y}d\alpha d\beta + \\ +\frac{-\nu(1+\nu)}{E}\sigma_\infty\tau^o\left(\frac{1}{2\pi}\right)^2\int_{-\infty}^{\infty}\int_{-\infty}^{\infty}i\beta\left(-2\frac{|\beta|}{i\beta}+i\beta y\right)(i\alpha)^2\left(1-|\alpha|y\right)g(\alpha)\delta(\alpha+\beta)e^{i(\alpha+\beta)x+(|\alpha|+|\beta|)y}d\alpha d\beta \\ +\frac{(1-\nu^2)}{E}\frac{(1-\nu^2)}{E}k^s\sigma_\infty\tau^o\left(\frac{1}{2\pi}\right)^2\int_{-\infty}^{\infty}\int_{-\infty}^{\infty}(i\beta)^2\left(1+|\beta|y\right)(i\alpha)^2\left(1+|\alpha|y\right)g(\alpha)\delta(\alpha+\beta)e^{i(\alpha+\beta)x+(|\alpha|+|\beta|)y}d\alpha d\beta + \\ \frac{-2\nu(1+\nu)}{E}\frac{(1-\nu^2)}{E}k^s\sigma_\infty\tau^o\left(\frac{1}{2\pi}\right)^2\int_{-\infty}^{\infty}\int_{-\infty}^{\infty}(i\beta)^2\left(1+|\beta|y\right)(i\alpha)^2\left(1-|\alpha|y\right)g(\alpha)\delta(\alpha+\beta)e^{i(\alpha+\beta)x+(|\alpha|+|\beta|)y}d\alpha d\beta \\ +\frac{\nu(1+\nu)}{E}\sigma_\infty\tau^o\left(\frac{1}{2\pi}\right)^2\int_{-\infty}^{\infty}\int_{-\infty}^{\infty}(i\beta)^2 y(i\alpha)^2\left(1+|\alpha|y\right)g(\alpha)\delta(\alpha+\beta)e^{i(\alpha+\beta)x+(|\alpha|+|\beta|)y}d\alpha d\beta \\ -\frac{(1-\nu^2)}{E}\sigma_\infty\tau^o\left(\frac{1}{2\pi}\right)^2\int_{-\infty}^{\infty}\int_{-\infty}^{\infty}(i\beta)^2 y(i\alpha)^2\left(1-|\alpha|y\right)g(\alpha)\delta(\alpha+\beta)e^{i(\alpha+\beta)x+(|\alpha|+|\beta|)y}d\alpha d\beta \\ +\frac{(1-\nu^2)}{E}\frac{(1-\nu^2)}{E}k^s\sigma_\infty\tau^o\left(\frac{1}{2\pi}\right)^2\int_{-\infty}^{\infty}\int_{-\infty}^{\infty}(i\beta)^2\left(1-|\beta|y\right)(i\alpha)^2\left(1-|\alpha|y\right)g(\alpha)\delta(\alpha+\beta)e^{i(\alpha+\beta)x+(|\alpha|+|\beta|)y}d\alpha d\beta \\ +\frac{(1+\nu)^2}{E^2}\left(\tau^o\frac{(1-\nu^2)}{E}k^s\sigma^\infty\right)\left(\frac{1}{2\pi}\right)^2\int_{-\infty}^{\infty}\int_{-\infty}^{\infty}(i\beta)^3(i\alpha)^3 y^2 g(\alpha)\delta(\alpha+\beta)e^{i(\alpha+\beta)x+(|\alpha|+|\beta|)y}d\alpha d\beta \\ -\frac{(1+\nu)^2}{E^2}\left(\frac{1}{2\pi}\right)^2\sigma_\infty\tau^o\int_{-\infty}^{\infty}\int_{-\infty}^{\infty}(i\beta)\left(1+|\beta|y\right)(i\alpha)^3 y g(\alpha)\delta(\alpha+\beta)e^{i(\alpha+\beta)x+(|\alpha|+|\beta|)y}d\alpha d\beta \end{pmatrix} \quad (153)$$

After doing some algebra we get to

$$\langle E'_{bulk}\rangle_1 = \delta^2 \tau^o \sigma^\infty \left( \frac{-2(1+\nu)(1-2\nu) - 2\nu(1+\nu) + 2(1-\nu^2)}{E} \right) \left( \frac{\eta^2}{a^2} \right) +$$
$$+ \delta^2 \left( 3\frac{(1-\nu^2)}{E}\frac{(1-\nu^2)}{E} - \frac{\nu(1+\nu)}{E}\frac{(1-\nu^2)}{E} \right) k^s \sigma_\infty \tau^o \left( \frac{1}{\sqrt{\pi}} \right) \left( \frac{4\eta^2}{a^3} \right) \quad (154)$$
$$+ \delta^2 \frac{(1+\nu)}{E} \left( \tau^o \frac{(1-\nu^2)}{E} k^s \sigma^\infty \right) \left( \frac{1}{\sqrt{\pi}} \right) \left( \frac{2\eta^2}{a^3} \right).$$

Similarly,

$$\langle E'_{bulk}\rangle_2 = \frac{1}{2}\int_{-\infty}^{h(x)} \langle Q\sigma_{xx}^{(2)}\cdot\langle\varepsilon_{xx}^{(0)}\rangle\rangle + \langle Q\sigma_{yy}^{(2)}\cdot\langle\varepsilon_{yy}^{(0)}\rangle\rangle + \langle Q\varepsilon_{xx}^{(2)}\cdot\langle\sigma_{xx}^{(0)}\rangle\rangle dy \quad (155)$$

where

$$Q\sigma_{xx}^{(2)}(\alpha, y) = \left[ f(\alpha)\left( -2\frac{|\alpha|}{i\alpha} + i\alpha y \right) + p(\alpha)(1+|\alpha|y) \right] e^{|\alpha|y} =$$

$$\left(\frac{1}{2\pi}\right) \begin{Bmatrix} -2\sigma_\infty \int_{-\infty}^{\infty} i(\alpha-\beta)h_0(\alpha-\beta)|\beta|h_0(\beta)d\beta \\ -2\sigma_\infty \int_{-\infty}^{\infty} h_0(\alpha-\beta)(i\beta)|\beta|h_0(\beta)d\beta + \tau^o \int_{-\infty}^{\infty} h_0(\alpha-\beta)(i\beta)^3 h_0(\beta)d\beta \\ + \left( \frac{(4-\nu)(1+\nu)}{E} + \frac{(1-\nu^2)}{E} \right) k^s \sigma^\infty \int_{-\infty}^{\infty} i(\alpha-\beta)h_0(\alpha-\beta)(i\beta)^2 h_0(\beta)d\beta \\ + \left( 4\frac{1-\nu^2}{E} + \frac{\nu(1+\nu)}{E} \right) k^s \sigma_\infty \int_{-\infty}^{\infty} h_0(\alpha-\beta)(i\beta)^3 h_0(\beta)d\beta - \\ -2\frac{1-\nu^2}{E} k^s \left( \tau^o + \frac{(1-\nu^2)}{E} k^s \sigma^\infty \right) \int_{-\infty}^{\infty} h_0(\alpha-\beta)(i\beta^3)|\beta|h_0(\beta)d\beta \\ -2\frac{1-\nu^2}{E} \left( \tau^o + \frac{(1-\nu^2)}{E} k^s \sigma^\infty \right) k^s \int_{-\infty}^{\infty} i(\alpha-\beta)h_0(\alpha-\beta)(\beta^2)|\beta|h_0(\beta)d\beta \end{Bmatrix} \left( -2\frac{|\alpha|}{i\alpha} + i\alpha y \right) e^{|\alpha|y}$$

$$+ \left(\frac{1}{2\pi}\right) \begin{Bmatrix} \sigma_\infty \int_{-\infty}^{\infty} i(\alpha-\beta)h_0(\alpha-\beta)(i\beta)h_0(\beta)d\beta + \sigma_\infty \int_{-\infty}^{\infty} h_0(\alpha-\beta)(i\beta)^2 h_0(\beta)d\beta - \\ -2\sigma_\infty \frac{1-\nu^2}{E} k^s \int_{-\infty}^{\infty} i^2(\alpha-\beta)^2 h_0(\alpha-\beta)|\beta|h_0(\beta)d\beta + \\ \frac{(1+\nu)(1-2\nu)}{E} k^s \left( \tau^o + \frac{(1-\nu^2)}{E} k^s \sigma^\infty \right) \int_{-\infty}^{\infty} i^2(\alpha-\beta)^2 h_0(\alpha-\beta)(i\beta)^2 h_0(\beta)d\beta \\ -2\frac{1-\nu^2}{E} k^s \sigma_\infty \int_{-\infty}^{\infty} i(\alpha-\beta)h_0(\alpha-\beta)(i\beta)|\beta|h_0(\beta)d\beta \\ + \frac{(1+\nu)(1-2\nu)}{E} k^s \left( \tau^o + \frac{(1-\nu^2)}{E} k^s \sigma^\infty \right) \int_{-\infty}^{\infty} i(\alpha-\beta)h_0(\alpha-\beta)(i\beta)^3 h_0(\beta)d\beta \end{Bmatrix} (1+|\alpha|y)e^{|\alpha|y}. \quad (156)$$

$$Q\sigma_{yy}^{(2)}(\alpha, y) = \left[-i\alpha f(\alpha) y + p(\alpha)\left(1-|\alpha|y\right)\right]e^{|\alpha|y} =$$

$$\left(\frac{1}{2\pi}\right)\begin{Bmatrix} -2\sigma_\infty \int_{-\infty}^{\infty} i(\alpha-\beta)h_0(\alpha-\beta)|\beta|h_0(\beta)d\beta - 2\sigma_\infty \int_{-\infty}^{\infty} h_0(\alpha-\beta)(i\beta)|\beta|h_0(\beta)d\beta + \\ +\tau^o \int_{-\infty}^{\infty} h_0(\alpha-\beta)(i\beta)^3 h_0(\beta)d\beta + \\ +\left(\frac{(4-\nu)(1+\nu)}{E} + \frac{(1-\nu^2)}{E}\right)k^s\sigma^\infty \int_{-\infty}^{\infty} i(\alpha-\beta)h_0(\alpha-\beta)(i\beta)^2 h_0(\beta)d\beta \\ +\left(4\frac{1-\nu^2}{E} + \frac{\nu(1+\nu)}{E}\right)k^s\sigma^\infty \int_{-\infty}^{\infty} h_0(\alpha-\beta)(i\beta)^3 h_0(\beta)d\beta - \\ -2\frac{1-\nu^2}{E}k^s\left(\tau^o + \frac{(1-\nu^2)}{E}k^s\sigma^\infty\right)\int_{-\infty}^{\infty} h_0(\alpha-\beta)(i\beta^3)|\beta|h_0(\beta)d\beta \\ -2\frac{1-\nu^2}{E}\left(\tau^o + \frac{(1-\nu^2)}{E}k^s\sigma^\infty\right)k^s \int_{-\infty}^{\infty} i(\alpha-\beta)h_0(\alpha-\beta)(\beta^2)|\beta|h_0(\beta)d\beta \end{Bmatrix}(-i\alpha y)e^{|\alpha|y} +$$

$$\left(\frac{1}{2\pi}\right)\begin{Bmatrix} \sigma_\infty \int_{-\infty}^{\infty} i(\alpha-\beta)h_0(\alpha-\beta)(i\beta)h_0(\beta)d\beta + \sigma_\infty \int_{-\infty}^{\infty} h_0(\alpha-\beta)(i\beta)^2 h_0(\beta)d\beta - \\ -2\sigma_\infty \frac{1-\nu^2}{E}k^s \frac{1}{2\pi}\int_{-\infty}^{\infty} i^2(\alpha-\beta)^2 h_0(\alpha-\beta)|\beta|h_0(\beta)d\beta \\ +\frac{(1+\nu)(1-2\nu)}{E}k^s\left[\tau^o + \frac{(1-\nu^2)}{E}k^s\sigma^\infty\right]\int_{-\infty}^{\infty} i^2(\alpha-\beta)^2 h_0(\alpha-\beta)(i\beta)^2 h_0(\beta)d\beta \\ -2\frac{1-\nu^2}{E}k^s\sigma_\infty \frac{1}{2\pi}\int_{-\infty}^{\infty} i(\alpha-\beta)h_0(\alpha-\beta)(i\beta)|\beta|h_0(\beta)d\beta \\ +\frac{(1+\nu)(1-2\nu)}{E}k^s\left[\tau^o + \frac{(1-\nu^2)}{E}k^s\sigma^\infty\right]\int_{-\infty}^{\infty} i(\alpha-\beta)h_0(\alpha-\beta)(i\beta)^3 h_0(\beta)d\beta \end{Bmatrix}(1-|\alpha|y)e^{|\alpha|y}.$$

(157)

More simplifying we would have

$$\langle E'_{bulk}\rangle_2 = \frac{(1-\nu^2)}{E}\sigma_\infty\left(\frac{1}{2\pi}\right)^2 \times$$

$$\times \int_{-\infty}^{h(x)}\int_{-\infty}^{\infty}\begin{Bmatrix}\begin{Bmatrix}\tau^o(2\pi)\eta^2\delta(\alpha)\sqrt{\pi}a\int_{-\infty}^{\infty}(i\beta)^3 e^{(-a^2\beta^2/4)}d\beta + \\ -2\frac{1-\nu^2}{E}k^s\tau^o(2\pi)\eta^2\delta(\alpha)\sqrt{\pi}a\int_{-\infty}^{\infty}(i\beta^3)|\beta|e^{(-a^2\beta^2/4)}d\beta \\ -2\frac{1-\nu^2}{E}\tau^o k^s(2\pi)\eta^2\delta(\alpha)\sqrt{\pi}a\int_{-\infty}^{\infty}i(\alpha-\beta)(\beta^2)|\beta|e^{(-a^2\beta^2/4)}d\beta\end{Bmatrix}\delta(\alpha)\left(-2\frac{|\alpha|}{i\alpha}+i\alpha y\right) \\ \begin{Bmatrix}+\frac{(1+\nu)(1-2\nu)}{E}k^s\tau^o(2\pi)\eta^2\delta(\alpha)\sqrt{\pi}a\int_{-\infty}^{\infty}i^2(\alpha-\beta)^2(i\beta)^2 e^{(-a^2\beta^2/4)}d\beta \\ +\frac{(1+\nu)(1-2\nu)}{E}k^s\tau^o(2\pi)\eta^2\delta(\alpha)\sqrt{\pi}a\int_{-\infty}^{\infty}i(\alpha-\beta)(i\beta)^3 e^{(-a^2\beta^2/4)}d\beta\end{Bmatrix}\delta(\alpha)(1+|\alpha|y)e^{|\alpha|y}\end{Bmatrix}e^{i\alpha x+|\alpha|y}d$$

$$-\frac{\nu(1+\nu)}{E}\sigma_\infty\left(\frac{1}{2\pi}\right)^2 \times$$

$$\times\int_{-\infty}^{\infty}\begin{Bmatrix}\begin{Bmatrix}\tau^o(2\pi)\eta^2\delta(\alpha)\sqrt{\pi}a\int_{-\infty}^{\infty}h_0(\alpha-\beta)(i\beta)^3 h_0(\beta)d\beta + \\ -2\frac{1-\nu^2}{E}k^s\tau^o(2\pi)\eta^2\delta(\alpha)\sqrt{\pi}a\int_{-\infty}^{\infty}(i\beta^3)|\beta|e^{(-a^2\beta^2/4)}d\beta \\ -2\frac{1-\nu^2}{E}\tau^o k^s(2\pi)\eta^2\delta(\alpha)\sqrt{\pi}a\int_{-\infty}^{\infty}i(\alpha-\beta)(\beta^2)|\beta|e^{(-a^2\beta^2/4)}d\beta\end{Bmatrix}\delta(\alpha)(-i\alpha y)+ \\ \begin{Bmatrix}+\frac{(1+\nu)(1-2\nu)}{E}k^s\tau^o(2\pi)\eta^2\delta(\alpha)\sqrt{\pi}a\int_{-\infty}^{\infty}i^2(\alpha-\beta)^2(i\beta)^2 e^{(-a^2\beta^2/4)}d\beta \\ +\frac{(1+\nu)(1-2\nu)}{E}k^s\tau^o(2\pi)\eta^2\delta(\alpha)\sqrt{\pi}a\int_{-\infty}^{\infty}i(\alpha-\beta)(i\beta)^3 e^{(-a^2\beta^2/4)}d\beta\end{Bmatrix}\delta(\alpha)(1-|\alpha|y)\end{Bmatrix}e^{i\alpha x+|\alpha|y}d\alpha$$

(158)

And finally we get

$$
\begin{aligned}
\langle E'_{bulk}\rangle_2 &= \frac{(1-v^2)}{E}\sigma_\infty\left(\frac{1}{2\pi}\right)\int_{-\infty}^{h(x)}\int_{-\infty}^{\infty}\left\{\begin{array}{l}\eta^2\left\{-2\dfrac{1-v^2}{E}\tau^o k^s\sqrt{\pi}a\left(\dfrac{16i\alpha}{a^4}\right)\right\}\delta(\alpha)\left(-2\dfrac{|\alpha|}{i\alpha}+i\alpha y\right)+\\[2mm]\left[\dfrac{(1+v)(1-2v)}{E}k^s\tau^o\sqrt{\pi}a\left(\dfrac{8\sqrt{\pi}\left(3+\dfrac{a^2\alpha^2}{2}\right)}{a^5}\right)\right]\delta(\alpha)\left(1+|\alpha|y\right)e^{|\alpha|y}\\[2mm]+\dfrac{(1+v)(1-2v)}{E}k^s\tau^o\sqrt{\pi}a\left(\dfrac{-24\sqrt{\pi}}{a^5}\right)\end{array}\right\}e^{i\alpha x+|\alpha|y}d\alpha\\[3mm]
&-\frac{v(1+v)}{E}\sigma_\infty\left(\frac{1}{2\pi}\right)\int_{-\infty}^{h(x)}\int_{-\infty}^{\infty}\eta^2\delta(\alpha)\left\{\begin{array}{l}-2\dfrac{1-v^2}{E}\tau^o k^s\sqrt{\pi}a\left(\dfrac{16i\alpha}{a^4}\right)\delta(\alpha)(-i\alpha y)+\\[2mm]\left[\dfrac{(1+v)(1-2v)}{E}k^s\tau^o\sqrt{\pi}a\left(\dfrac{8\sqrt{\pi}\left(3+\dfrac{a^2\alpha^2}{2}\right)}{a^5}\right)\right]\delta(\alpha)\left(1-|\alpha|y\right)\\[2mm]+\dfrac{(1+v)(1-2v)}{E}k^s\tau^o\sqrt{\pi}a\left(\dfrac{-24\sqrt{\pi}}{a^5}\right)\end{array}\right\}e^{i\alpha x+|\alpha|y}d\alpha=\\
&=0
\end{aligned}
\qquad(159)
$$

Similarly we calculate energy contribution of surface stress and surface strain and show the result as

$$
\begin{aligned}
\langle E'(surface)\rangle &= \frac{1-v^2}{E}\tau^o\sigma^\infty + \delta^2\sigma_\infty\left[\frac{2(1-v^2)}{E}\right]\tau^o\left(-\frac{2\eta^2}{a^2}\right)+\\
&+\delta^2\tau^o\left[\left(\frac{1}{\sqrt{\pi}}\right)\sigma^\infty k^s\frac{(1-2v)(1+v)^2(1-v)-2(1-v^2)^2}{E^2}\left(\frac{16\eta^2}{a^3}\right)\right]\\
&+\delta^2\tau^o(k^s)^2\sigma^\infty\frac{(1-v^2)(1+v)^2\left((1-v)(1-3v)+v^2\right)}{E^3}\left(\frac{12\eta^2}{a^4}\right)
\end{aligned}
\qquad(160)
$$

At the end, the effective surface tension is obtained as

$$
\begin{aligned}
(\tau^o)^{eff} &= \left(\frac{E}{1-v^2}\right)\frac{\partial\langle E'\rangle(\sigma^\infty)}{\partial\sigma^\infty}\bigg|_{\sigma^\infty=0}=\\
&=\tau^o-\delta^2\tau^o\left(\frac{4\eta^2}{a^2}+\frac{1}{\sqrt{\pi}}(k^s)\frac{(1+8v)(1+v)}{E}\left(\frac{2\eta^2}{a^3}\right)-(k^s)^2\sigma^\infty\frac{(1+v)^2(1-2v)^2}{E^2}\left(\frac{12\eta^2}{a^4}\right)\right)
\end{aligned}
\qquad(161)
$$

In order to calculate the effective surface elasticity constant, we precede similarly as for finding effective surface stress except that this time we consider only the energy terms that are second order in $\sigma^\infty$. The detail calculation is not shown here. The final result would be obtained as

$$\left(k^{s}\right)^{eff} = \left(\frac{E}{1-\nu^{2}}\right)^{2} \frac{\partial^{2} \langle E''\rangle (\sigma^{\infty})}{\left(\partial \sigma^{\infty}\right)^{2}}\bigg|_{\sigma^{\infty}=0}$$

$$\left(k^{s}\right) + \delta^{2} \left[ \begin{array}{l} -\dfrac{(9-8\nu)E}{4(1-\nu)(1-\nu^{2})}\left(\dfrac{1}{\sqrt{\pi}}\right)\left(\dfrac{\eta^{2}}{a}\right) + \left(k^{s}\right)\left(\dfrac{8\eta^{2}}{a^{2}}\right) \\ + \dfrac{(-57\nu+11)(1+\nu)}{4E}\left(k^{s}\right)^{2}\left(\dfrac{1}{\sqrt{\pi}}\right)\left(\dfrac{\eta^{2}}{a^{3}}\right) + \dfrac{(1-2\nu)^{2}(1+\nu)^{2}}{E}\left(k^{s}\right)^{3}\left(\dfrac{6\eta^{2}}{a^{4}}\right) \end{array} \right]. \tag{162}$$

## Appendix E: Finding Effective Surface Stress and Effective Surface Elastic Constant Considering Asymmetry Term, $\tau^{0}\nabla_{s}u$

In this appendix, the impact of asymmetry term of surface stress $\tau^{0}\nabla_{s}u$ in the effective surface stress and effective surface elastic constant for the case of sinusoidal rough profile would be presented. The surface constitutive law for the considered model problem can be explained as

$$S = \mathbb{C}_{s}\left(\varepsilon^{s} - \varepsilon_{s}^{0}\right) + \tau^{0}\nabla_{s}u \quad \text{on} \quad \partial B, \tag{163}$$

with boundary condition

$$\left(\mathbb{C}\nabla u\right)e_{n} = \text{div}_{s}\left[\mathbb{C}_{s}\left(\varepsilon^{s} - \varepsilon_{s}^{0}\right) + \tau^{0}\nabla_{s}u\right] \quad \text{on} \quad \partial B. \tag{164}$$

$\nabla_{s}u$ that is an out of plane component of surface stress, can be expressed as

$$\nabla_{s}u = (\frac{\partial u_{s}}{\partial s} - \kappa u_{n})e_{s} \otimes e_{s} + (\kappa u_{s} + \frac{\partial u_{n}}{\partial s})e_{s} \otimes e_{n} \quad \text{on} \quad \partial B \tag{165}$$

We remark that

$$(\frac{\partial u_{s}}{\partial s} - \kappa u_{n}) = \varepsilon_{ss}. \tag{166}$$

The surface stress can be expressed as summation of the in-plane and out-of-plane components denoted by $\sigma^{s}$ and $w^{s}$, respectively.

$$S = \sigma^{s}(e_{s} \otimes e_{s}) + w^{s}(e_{s} \otimes e_{n})$$
$$\sigma^{s} = \tau^{o} + (k^{s} + \tau^{0})\varepsilon_{ss}, \quad w^{s} = \tau^{0}(\kappa u_{s} + \frac{\partial u_{n}}{\partial s}) \quad \text{on} \quad \partial B \tag{167}$$

Thus, the boundary condition on $\partial B$ would be explained as

$$(\mathbb{C}\nabla u)e_n = \left(\tau^o \kappa + \tau^o \frac{\partial \kappa}{\partial s}u_s + \tau^o \kappa \frac{\partial u_s}{\partial s} + \left(k^s + \tau^0\right)\kappa\varepsilon_{ss} + \tau^0 \frac{\partial^2 u_n}{\partial s^2}\right)e_n +$$
$$+\left(\left(k^s + \tau^0\right)\frac{\partial \varepsilon_{ss}}{\partial s} - \tau^0 \kappa^2 u_s - \tau^0 \kappa \frac{\partial u_n}{\partial s}\right)e_s. \tag{168}$$

Below we convert the above boundary condition to Cartesian coordinates. Inserting (10) and (27) into (168) and keeping terms up to $O(\delta^2)$, we find the boundary conditions on the nominal flat surface for $u^{(i)}$ $(i = 0, 1, 2)$, i.e., the right hand side of (12) as follows.

$$t_x^{(0)} = \left(k^s + \tau^0\right)\frac{\partial \varepsilon_{xx}^{(0)}}{\partial x}, \qquad t_y^{(0)} = \tau^0 \frac{\partial^2 u_y^{(0)}}{\partial x^2} \tag{169}$$

$$t_x^{(1)} = h_{0x}\sigma_{xx}^{(0)} - h_0 \frac{\partial \sigma_{xy}^{(0)}}{\partial y} - \tau^0 h_{0x} \frac{\partial^2 u_y^{(0)}}{\partial x^2} + \left(k^s + \tau^0\right)\frac{\partial \varepsilon_{xx}^{(1)}}{\partial x} + \left(k^s + \tau^0\right)h_0 \frac{\partial^2 \varepsilon_{xx}^{(0)}}{\partial x \partial y}$$
$$+ 2\left(k^s + \tau^0\right)h_{0xx}\varepsilon_{xy}^{(0)} + 2\left(k^s + \tau^0\right)h_{0x}\frac{\partial \varepsilon_{xy}^{(0)}}{\partial x} + \left(k^s + \tau^0\right)h_{0x}\frac{\partial \varepsilon_{xx}^{(0)}}{\partial y} - \tau^0 h_{0xx}\frac{\partial u_y^{(0)}}{\partial x}$$

$$t_y^{(1)} = h_{0x}\sigma_{yx}^{(0)} - h_0 \frac{\partial \sigma_{yy}^{(0)}}{\partial y} + \tau^o h_{0xx} + (k^s + 2\tau^0)h_{0xx}\varepsilon_{xx}^{(0)} - 2\tau^0 h_{0xx}\varepsilon_{xx}^{(0)} + \tau^0 h_{0xx}\varepsilon_{yy}^{(0)} \tag{170}$$
$$+ \tau^0 h_{0x}\left(-\frac{\partial \varepsilon_{xx}^{(0)}}{\partial x} + \frac{\partial \varepsilon_{yy}^{(0)}}{\partial x}\right) + \tau^0 \frac{\partial^2 u_y^{(1)}}{\partial x^2} + \tau^0 h_0 \frac{\partial^3 u_y^{(0)}}{\partial x^2 \partial y} + \tau^0 h_{0x} \frac{\partial^2 u_y^{(0)}}{\partial x \partial y} +$$
$$+ \left(k^s + \tau^0\right)h_{0x}\frac{\partial \varepsilon_{xx}^{(0)}}{\partial x}$$

$$t_x^{(2)} = h_{0x}\sigma_{xx}^{(1)} + h_0 h_{0x}\frac{\partial \sigma_{xx}^{(0)}}{\partial y} + \frac{1}{2}h_{0x}^2\sigma_{xy}^{(0)} - h_0\frac{\partial \sigma_{xy}^{(1)}}{\partial y} - \frac{1}{2}h_0^2\frac{\partial^2 \sigma_{xy}^{(0)}}{\partial y^2} - \tau^o h_{0x}h_{0xx} -$$

$$- \left(k^s + 2\tau^0\right)h_{0x}h_{0xx}\varepsilon_{xx}^{(0)} + 2\tau^0 h_{0x}h_{0xx}\varepsilon_{xx}^{(0)} - \tau^0 h_{0x}h_{0xx}\varepsilon_{yy}^{(0)} - \tau^0 h_{0x}h_{0x}\left(-\frac{\partial \varepsilon_{xx}^{(0)}}{\partial x} + \frac{\partial \varepsilon_{yy}^{(0)}}{\partial x}\right)$$

$$-\tau^0 h_{0x}\frac{\partial^2 u_y^{(1)}}{\partial x^2} - \tau^0 h_{0x}h_0\frac{\partial^2 u_y^{(0)}}{\partial x \partial y} + \left(k^s + \tau^0\right)\frac{\partial \varepsilon_{xx}^{(2)}}{\partial x} + \left(k^s + \tau^0\right)h_0\frac{\partial^2 \varepsilon_{xx}^{(1)}}{\partial x \partial y} +$$

$$+\frac{1}{2}\left(k^s + \tau^0\right)(h_0)^2\frac{\partial^3 \varepsilon_{xx}^{(0)}}{\partial x \partial y^2} - \left(k^s + \tau^0\right)h_{0x}^2\frac{\partial \varepsilon_{xx}^{(0)}}{\partial x} + 2\left(k^s + \tau^0\right)h_{0xx}\varepsilon_{xy}^{(1)} +$$

$$+2\left(k^s + \tau^0\right)h_{0xx}h_0\frac{\partial \varepsilon_{xy}^{(0)}}{\partial y} + 2\left(k^s + \tau^0\right)h_{0x}\frac{\partial \varepsilon_{xy}^{(1)}}{\partial x} + 2\left(k^s + \tau^0\right)h_{0x}h_0\frac{\partial^2 \varepsilon_{xy}^{(0)}}{\partial x \partial y} +$$

$$+2\left(k^s + \tau^0\right)h_{0x}h_{0xx}\left(\varepsilon_{yy}^{(0)} - \varepsilon_{xx}^{(0)}\right) + \left(k^s + \tau^0\right)h_{0x}^2\left(\frac{\partial \varepsilon_{yy}^{(0)}}{\partial x} - \frac{1}{2}\frac{\partial \varepsilon_{xx}^{(0)}}{\partial x}\right) +$$

$$+\left(k^s + \tau^0\right)h_{0x}\frac{\partial \varepsilon_{xx}^{(1)}}{\partial y} + \left(k^s + \tau^0\right)h_{0x}h_0\frac{\partial^2 \varepsilon_{xx}^{(0)}}{\partial y^2} + 2\left(k^s + \tau^0\right)(h_{0x})^2\frac{\partial \varepsilon_{xy}^{(0)}}{\partial y}$$

$$-\tau^0 h_{0x}h_{0xx}\left(-\varepsilon_{xx}^{(0)} + \varepsilon_{yy}^{(0)}\right) - \tau^0 h_{0xx}\frac{\partial u_y^{(1)}}{\partial x} - \tau^0 h_{0xx}h_0\frac{\partial^2 u_y^{(0)}}{\partial x \partial y}$$

$$t_y^{(2)} = h_{0x}\sigma_{yx}^{(1)} + h_{0x}h_0\frac{\partial \sigma_{yx}^{(0)}}{\partial y} + \frac{1}{2}h_{0x}^2\sigma_{yy}^{(0)} - h_0\frac{\partial \sigma_{yy}^{(1)}}{\partial y} - \frac{1}{2}h_0^2\frac{\partial^2 \sigma_{yy}^{(0)}}{\partial y^2} \qquad (171)$$

$$+\left(k^s + 2\tau^0\right)\left(h_{0xx}\varepsilon_{xx}^{(1)} + h_{0xx}h_0\frac{\partial \varepsilon_{xx}^{(0)}}{\partial y} + 2h_{0x}h_{0xx}\varepsilon_{xy}^{(0)}\right) - 2\tau^0 h_{0xx}\varepsilon_{xx}^{(1)}$$

$$-2\tau^0 h_{0xx}h_0\frac{\partial \varepsilon_{xx}^{(0)}}{\partial y} + \tau^0 h_{0xx}\varepsilon_{yy}^{(1)} + \tau^0 h_{0xx}h_0\frac{\partial \varepsilon_{yy}^{(0)}}{\partial y} + \tau^0 h_{0x}\left(-\frac{\partial \varepsilon_{xx}^{(1)}}{\partial x} + \frac{\partial \varepsilon_{yy}^{(1)}}{\partial x}\right)$$

$$+\tau^0 h_{0x}h_0\left(-\frac{\partial^2 \varepsilon_{xx}^{(0)}}{\partial x \partial y} + \frac{\partial^2 \varepsilon_{yy}^{(0)}}{\partial x \partial y}\right) - \frac{1}{2}\tau^0 h_{0x}h_{0xx}\frac{\partial u_y^{(0)}}{\partial x} - \tau^0 h_{0xx}^2 u_y^{(0)} - \frac{1}{2}\tau^0\left(h_{0x}^2\right)\frac{\partial^2 u_y^{(0)}}{\partial x^2}$$

$$+\tau^0\frac{\partial^2 u_y^{(2)}}{\partial x^2} + \tau^0 h_0\frac{\partial^3 u_y^{(1)}}{\partial x^2 \partial y} + \frac{1}{2}\tau^0(h_0)^2\frac{\partial^4 u_y^{(0)}}{\partial x^2 \partial y^2} - 2\tau^0 h_{0xx}h_{0x}\varepsilon_{xy}^{(0)} - 2\tau^0 h_0 h_{0x}\frac{\partial \varepsilon_{xy}^{(0)}}{\partial x}$$

$$-2\tau^0 h_{0x}h_{0xx}\frac{\partial u_x^{(0)}}{\partial y} + \tau^0(h_{0x})^2\left(-\frac{\partial \varepsilon_{xx}^{(0)}}{\partial y} + \frac{\partial \varepsilon_{yy}^{(0)}}{\partial y}\right) + \tau^0 h_{0x}\frac{\partial^2 u_y^{(1)}}{\partial x \partial y} + \tau^0 h_{0x}h_0\frac{\partial^3 u_y^{(0)}}{\partial x \partial y^2}$$

$$+h_{0x}\left(k^s + \tau^0\right)\frac{\partial \varepsilon_{xx}^{(1)}}{\partial x} + h_{0x}h_0\left(k^s + \tau^0\right)\frac{\partial^2 \varepsilon_{xx}^{(0)}}{\partial x \partial y} + 2\left(k^s + \tau^0\right)h_{0x}h_{0xx}\varepsilon_{xy}^{(0)} +$$

$$+2\left(k^s + \tau^0\right)h_{0x}h_{0x}\frac{\partial \varepsilon_{xy}^{(0)}}{\partial x} + \left(k^s + \tau^0\right)h_{0x}h_{0x}\frac{\partial \varepsilon_{xx}^{(0)}}{\partial y} - \tau^0 h_{0x}h_{0xx}\frac{\partial u_y^{(0)}}{\partial x}$$

Similar to the previous cases, we proceed to solve the different order boundary value problems.

Zero order B.C.
$$t_x^{(0)} = 0, \quad t_y^{(0)} = 0. \tag{172}$$

Zero order solution
$$\sigma_{xx}^{(0)} = \sigma^\infty, \quad \sigma_{xy}^{(0)} = \sigma_{yy}^{(0)} = 0$$
$$\varepsilon_{xx}^{(0)} = \frac{1-\nu^2}{E}\sigma^\infty, \quad \varepsilon_{xy}^{(0)} = 0, \quad \varepsilon_{yy}^{(0)} = \frac{-\nu(1+\nu)}{E}\sigma^\infty, \tag{173}$$

First order B.C.
$$t_x^{(1)} = h_{0x}\sigma_{xx}^{(0)} = -\sigma^\infty \sin kx,$$

$$t_y^{(1)} = \tau^o h_{0xx} + k^s h_{0xx}\varepsilon_{xx}^{(0)} + \tau^o h_{0xx}\varepsilon_{yy}^{(0)} = \tag{174}$$
$$= -k\left(\tau^o + k^s \frac{(1-\nu^2)}{E}\sigma^\infty - \tau^o \frac{\nu(1-\nu)}{E}\sigma^\infty\right)\cos kx,$$

First order solution
$$\sigma_{xx}^{(1)} = \left\{-\sigma^\infty(2+ky)e^{ky} - k\left(\tau^o + k^s \frac{(1-\nu^2)}{E}\sigma^\infty - \tau^o \frac{\nu(1-\nu)}{E}\sigma^\infty\right)(1+ky)e^{ky}\right\}\cos kx$$

$$\sigma_{xy}^{(1)} = \left\{-\sigma^\infty(ky+1)e^{ky} - k\left(\tau^o + k^s \frac{(1-\nu^2)}{E}\sigma^\infty - \tau^o \frac{\nu(1-\nu)}{E}\sigma^\infty\right)kye^{ky}\right\}\sin kx$$

$$\sigma_{yy}^{(1)} = \left\{\sigma^\infty(ky)e^{ky} + k\left(\tau^o + k^s \frac{(1-\nu^2)}{E}\sigma^\infty - \tau^o \frac{\nu(1-\nu)}{E}\sigma^\infty\right)(ky-1)e^{ky}\right\}\cos kx$$

$$\varepsilon_{xx}^{(1)} = \frac{1-\nu^2}{E}\left\{-\sigma^\infty(2+ky)e^{ky} - k\left(\tau^o + k^s \frac{(1-\nu^2)}{E}\sigma^\infty - \tau^o \frac{\nu(1-\nu)}{E}\sigma^\infty\right)(1+ky)e^{ky}\right\}\cos kx - \tag{175}$$
$$- \frac{\nu(1+\nu)}{E}\left\{\sigma^\infty(ky)e^{ky} + k\left(\tau^o + k^s \frac{(1-\nu^2)}{E}\sigma^\infty - \tau^o \frac{\nu(1-\nu)}{E}\sigma^\infty\right)(ky-1)e^{ky}\right\}\cos kx$$

$$\varepsilon_{yy}^{(1)} = -\frac{\nu(1+\nu)}{E}\left\{-\sigma^\infty(2+ky)e^{ky} - k\left(\tau^o + k^s \frac{(1-\nu^2)}{E}\sigma^\infty - \tau^o \frac{\nu(1-\nu)}{E}\sigma^\infty\right)(1+ky)e^{ky}\right\}\cos kx +$$
$$+ \frac{1-\nu^2}{E}\left\{\sigma^\infty(ky)e^{ky} + k\left(\tau^o + k^s \frac{(1-\nu^2)}{E}\sigma^\infty - \tau^o \frac{\nu(1-\nu)}{E}\sigma^\infty\right)(ky-1)e^{ky}\right\}\cos kx$$

$$\varepsilon_{xy}^{(1)} = \frac{(1+\nu)}{E}\left\{-\sigma^\infty(ky+1)e^{ky} - k\left(\tau^o + k^s\frac{(1-\nu^2)}{E}\sigma^\infty - \tau^o\frac{\nu(1-\nu)}{E}\sigma^\infty\right)kye^{ky}\right\}\sin kx$$

$$\frac{\partial u_y^{(1)}}{\partial x} = \frac{\nu(1+\nu)}{E}k\left\{-\sigma^\infty\left(\frac{1}{k}+y\right)e^{ky} - k\left(\tau^o + k^s\frac{(1-\nu^2)}{E}\sigma^\infty - \tau^o\frac{\nu(1-\nu)}{E}\sigma^\infty\right)(y)e^{ky}\right\}\sin kx +$$

$$-\frac{1-\nu^2}{E}k\left\{\sigma^\infty\left(-\frac{1}{k}+y\right)e^{ky} + k\left(\tau^o + k^s\frac{(1-\nu^2)}{E}\sigma^\infty - \tau^o\frac{\nu(1-\nu)}{E}\sigma^\infty\right)\left(-\frac{1}{k}+y\right)e^{ky}\right\}\sin kx$$

$$u_x^{(1)} = \frac{1-\nu^2}{E}\left\{-\frac{1}{k}\sigma^\infty(2+ky)e^{ky} - \left(\tau^o + k^s\frac{(1-\nu^2)}{E}\sigma^\infty - \tau^o\frac{\nu(1-\nu)}{E}\sigma^\infty\right)(1+ky)e^{ky}\right\}\sin kx -$$

$$-\frac{\nu(1+\nu)}{E}\left\{\sigma^\infty(y)e^{ky} + \left(\tau^o + k^s\frac{(1-\nu^2)}{E}\sigma^\infty - \tau^o\frac{\nu(1-\nu)}{E}\sigma^\infty\right)(ky-1)e^{ky}\right\}\sin kx$$

The first order solutions on the boundary $y = 0$ are needed in order to find the second order boundary conditions.

$$\varepsilon_{xx}^{(1)} = \left\{-\frac{(1-2\nu)(1+\nu)}{E}k\left(\tau^o + k^s\frac{(1-\nu^2)}{E}\sigma^\infty - \tau^o\frac{\nu(1-\nu)}{E}\sigma^\infty\right) - \sigma^\infty\frac{2(1-\nu^2)}{E}\right\}\cos kx$$

$$\frac{\partial \varepsilon_{xx}^{(1)}}{\partial x} = \left\{\frac{(1-2\nu)(1+\nu)}{E}k^2\left(\tau^o + k^s\frac{(1-\nu^2)}{E}\sigma^\infty - \tau^o\frac{\nu(1-\nu)}{E}\sigma^\infty\right) + \sigma^\infty\frac{2(1-\nu^2)}{E}k\right\}\sin kx$$

$$\frac{\partial \sigma_{xy}^{(1)}}{\partial y} = \left\{-2\sigma^\infty k - k^2\left(\tau^o + k^s\frac{(1-\nu^2)}{E}\sigma^\infty - \tau^o\frac{\nu(1-\nu)}{E}\sigma^\infty\right)\right\}\sin kx$$

$$\sigma_{xx}^{(1)} = \left\{-2\sigma^\infty - k\left(\tau^o + k^s\frac{(1-\nu^2)}{E}\sigma^\infty - \tau^o\frac{\nu(1-\nu)}{E}\sigma^\infty\right)\right\}\cos kx \tag{176}$$

$$\frac{\partial \varepsilon_{xx}^{(1)}}{\partial y} = -\left\{3\frac{(1-\nu^2)}{E}k\sigma^\infty + 2\frac{(1-\nu^2)}{E}k^2\left(\tau^o + k^s\frac{(1-\nu^2)}{E}\sigma^\infty - \tau^o\frac{\nu(1-\nu)}{E}\sigma^\infty\right) + \frac{\nu(1+\nu)}{E}k\sigma^\infty\right\}\cos kx$$

$$\frac{\partial^2 \varepsilon_{xx}^{(1)}}{\partial x \partial y} = \left\{3\frac{(1-\nu^2)}{E}k^2\sigma^\infty + 2\frac{(1-\nu^2)}{E}k^3\left(\tau^o + \frac{(1-\nu^2)}{E}\sigma^\infty - \tau^o\frac{\nu(1-\nu)}{E}\sigma^\infty\right) + \frac{\nu(1+\nu)}{E}k^2\sigma^\infty\right\}\sin kx$$

$$\sigma_{xy}^{(1)} = -\sigma^\infty \sin kx$$

$$\varepsilon_{xy}^{(1)} = -\frac{(1+\nu)}{E}\sigma^\infty \sin kx$$

$$\frac{\partial \varepsilon_{xy}^{(1)}}{\partial x} = -\frac{(1+\nu)}{E} k \sigma^\infty \cos kx$$

$$\frac{\partial \sigma_{yy}^{(1)}}{\partial y} = k \sigma^\infty \cos kx$$

$$\frac{\partial u_y^{(1)}}{\partial x} = -\frac{\nu(1+\nu)}{E} \sigma^\infty \sin kx + \frac{1-\nu^2}{E}\left\{\sigma^\infty + k\left(\tau^o + \frac{(1-\nu^2)}{E}\sigma^\infty - \tau^o \frac{\nu(1-\nu)}{E}\sigma^\infty\right)\right\} \sin kx$$

$$\varepsilon_{yy}^{(1)} = -\frac{\nu(1+\nu)}{E}\left\{-2\sigma^\infty - k\left(\tau^o + k^s \frac{(1-\nu^2)}{E}\sigma^\infty - \tau^o \frac{\nu(1-\nu)}{E}\sigma^\infty\right)\right\} \cos kx +$$

$$-\frac{1-\nu^2}{E}\left\{k\left(\tau^o + k^s \frac{(1-\nu^2)}{E}\sigma^\infty - \tau^o \frac{\nu(1-\nu)}{E}\sigma^\infty\right)\right\} \cos kx$$

Second order B.C.:

$$t_x^{(2)} = (-\sin kx)\left\{-2\sigma^\infty - k\left(\tau^o + k^s \frac{(1-\nu^2)}{E}\sigma^\infty - \tau^o \frac{\nu(1-\nu)}{E}\sigma^\infty\right)\right\} \cos kx$$

$$-\frac{\cos kx}{k}\left\{-2\sigma^\infty k - k^2\left(\tau^o + k^s \frac{(1-\nu^2)}{E}\sigma^\infty - \tau^o \frac{\nu(1-\nu)}{E}\sigma^\infty\right)\right\} \sin kx - \tau^o(-\sin kx)(-k\cos kx)$$

$$+(-3k^s - \tau^0)(-\sin kx)(-k\cos kx)\left(\frac{1-\nu^2}{E}\sigma_\infty\right) + 2k^s(-\sin kx)(-k\cos kx)\left(-\frac{\nu(1-\nu)}{E}\sigma_\infty\right)$$

$$+k^s \frac{\cos kx}{k}\left\{3\frac{(1-\nu^2)}{E}k^2\sigma^\infty + 2\frac{(1-\nu^2)}{E}k^3\left(\tau^o + k^s \frac{(1-\nu^2)}{E}\sigma^\infty - \tau^o \frac{\nu(1-\nu)}{E}\sigma^\infty\right) + \frac{\nu(1+\nu)}{E}k^2\sigma^\infty\right\}\sin kx \quad (177)$$

$$+2k^s(-k\cos kx)\left(-\frac{(1+\nu)}{E}\sigma^\infty \sin kx\right) + 2k^s(-\sin kx)\left(-\frac{(1+\nu)}{E}k\sigma^\infty \cos kx\right)$$

$$+k^s(-\sin kx)\left(-3\frac{(1-\nu^2)}{E}k\sigma^\infty - 2\frac{(1-\nu^2)}{E}k^2\left(\tau^o + k^s \frac{(1-\nu^2)}{E}\sigma^\infty - \tau^o \frac{\nu(1-\nu)}{E}\sigma^\infty\right) - \frac{\nu(1+\nu)}{E}k\sigma^\infty\right)\cos$$

$$-\tau^o\left(k\frac{\nu(1+\nu)}{E}\sigma^\infty \sin 2kx - k\frac{1-\nu^2}{E}\left\{\sigma^\infty + k\left(\tau^o + k^s \frac{(1-\nu^2)}{E}\sigma^\infty - \tau^o \frac{\nu(1-\nu)}{E}\sigma^\infty\right)\right\}\sin 2kx\right)$$

$$t_y^{(2)} = (-\sin kx)(-\sigma^\infty \sin kx) - \left(\frac{\cos kx}{k}\right)(k\sigma^\infty \cos kx)$$

$$+ (k^s - \tau^0)(-k\cos kx)\left\{-\frac{(1-2\nu)(1+\nu)}{E}k\left(\tau^o + k^s\frac{(1-\nu^2)}{E}\sigma^\infty - \tau^o\frac{\nu(1-\nu)}{E}\sigma^\infty\right) - \sigma^\infty\frac{2(1-\nu^2)}{E}\right\}\cos kx$$

$$+ \tau^o(-k\cos kx)\left(\begin{array}{l}-\frac{\nu(1+\nu)}{E}\left\{-2\sigma^\infty - k\left(\tau^o + k^s\frac{(1-\nu^2)}{E}\sigma^\infty - \tau^o\frac{\nu(1-\nu)}{E}\sigma^\infty\right)\right\} + \\ -\frac{1-\nu^2}{E}k\left(\tau^o + k^s\frac{(1-\nu^2)}{E}\sigma^\infty - \tau^o\frac{\nu(1-\nu)}{E}\sigma^\infty\right)\end{array}\right)\cos kx +$$

$$+ \tau^o(-\sin kx)\left(\begin{array}{l}-\frac{\nu(1+\nu)}{E}\left\{-2\sigma^\infty - k\left(\tau^o + k^s\frac{(1-\nu^2)}{E}\sigma^\infty - \tau^o\frac{\nu(1-\nu)}{E}\sigma^\infty\right)\right\} + \\ -\frac{1-\nu^2}{E}k\left(\tau^o + k^s\frac{(1-\nu^2)}{E}\sigma^\infty - \tau^o\frac{\nu(1-\nu)}{E}\sigma^\infty\right)\end{array}\right)(-k\sin kx) \quad (178)$$

$$+ \tau^o\left(\frac{\cos kx}{k}\right)\left(\begin{array}{l}-\frac{\nu(1+\nu)}{E}\left\{-2\sigma^\infty - k\left(\tau^o + k^s\frac{(1-\nu^2)}{E}\sigma^\infty - \tau^o\frac{\nu(1-\nu)}{E}\sigma^\infty\right)\right\} + \\ -\frac{1-\nu^2}{E}\left\{k\left(\tau^o + k^s\frac{(1-\nu^2)}{E}\sigma^\infty - \tau^o\frac{\nu(1-\nu)}{E}\sigma^\infty\right)\right\}\end{array}\right)(-k^2\cos kx) +$$

$$+ \tau^o(-\sin kx)\left(\begin{array}{l}-\frac{\nu(1+\nu)}{E}\left\{-2\sigma^\infty - k\left(\tau^o + k^s\frac{(1-\nu^2)}{E}\sigma^\infty - \tau^o\frac{\nu(1-\nu)}{E}\sigma^\infty\right)\right\} + \\ -\frac{1-\nu^2}{E}\left\{k\left(\tau^o + k^s\frac{(1-\nu^2)}{E}\sigma^\infty - \tau^o\frac{\nu(1-\nu)}{E}\sigma^\infty\right)\right\}\end{array}\right)(-k\sin kx)$$

$$+ (k^s - \tau^o)(-\sin kx)\left\{\frac{(1-2\nu)(1+\nu)}{E}k^2\left(\tau^o + k^s\frac{(1-\nu^2)}{E}\sigma^\infty - \tau^o\frac{\nu(1-\nu)}{E}\sigma^\infty\right) + \sigma^\infty\frac{2(1-\nu^2)}{E}k\right\}\sin kx$$

By more simplifying we would have

$$t_x^{(2)} = \beta \sin 2kx,$$

$$\beta = \left(\begin{array}{l}2\sigma^\infty + \frac{1}{2}k\tau^o + \left(\frac{5}{2}k^s + \frac{7}{2}\tau^0\right)k\frac{(1-\nu^2)}{E}\sigma^\infty + (k^s + \tau^0)k\frac{-\nu(1-\nu) + 2(1+\nu)}{E}\sigma^\infty + \\ \frac{(1-\nu^2)}{E}\left(\begin{array}{l}k^2(2k^s + 3\tau^o)\tau^o + k^s(2k^s + 3\tau^o)k^2\frac{(1-\nu^2)}{E}\sigma^\infty - \\ -\tau^o(2k^s + 3\tau^o)k^2\frac{\nu(1-\nu)}{E}\sigma^\infty\end{array}\right) - \frac{\nu(1+\nu)}{E}k^s k\sigma^\infty\end{array}\right) \quad (179)$$

$$t_y^{(2)} = \gamma \cos 2kx,$$

$$\gamma = \left(\begin{array}{l}-\sigma^\infty + k^s\left\{\frac{(1-2\nu)(1+\nu)}{E}k^2\left(\tau^o + k^s\frac{(1-\nu^2)}{E}\sigma^\infty - \tau^o\frac{\nu(1-\nu)}{E}\sigma^\infty\right) - \sigma^\infty k\frac{2(1-\nu^2)}{E}\right\} \\ -2\tau^o\left(\begin{array}{l}-\frac{\nu(1+\nu)}{E}k\left\{-2\sigma^\infty - k\left(\tau^o + k^s\frac{(1-\nu^2)}{E}\sigma^\infty - \tau^o\frac{\nu(1-\nu)}{E}\sigma^\infty\right)\right\} + \\ -\frac{1-\nu^2}{E}k^2\left(\tau^o + k^s\frac{(1-\nu^2)}{E}\sigma^\infty - \tau^o\frac{\nu(1-\nu)}{E}\sigma^\infty\right)\end{array}\right)\end{array}\right) \quad (180)$$

By solving the second order terms we are lead to the following results

$$\sigma_{xx}^{(2)} = 2\beta(1+ky)e^{2ky}\cos 2kx + \gamma(1+2ky)e^{2ky}\cos 2kx$$

$$\sigma_{yy}^{(2)} = -2\beta ky e^{2ky}\cos 2kx + \gamma(1-2ky)e^{2ky}\cos 2kx$$

$$\varepsilon_{xx}^{(2)} = \frac{1-v^2}{E}\left(2\beta(1+ky)e^{2ky}\cos 2kx + \gamma(1+2ky)e^{2ky}\cos 2kx\right) - \frac{v(1+v)}{E}\left(-2\beta ky e^{2ky}\cos 2kx + \gamma(1-2ky)e^{2ky}\cos 2kx\right)$$

(181)

And at $y=0$

$$\varepsilon_{xx}^{(2)}\Big|_{y=0} = \frac{(1-2v)(1+v)}{E}\gamma\cos 2kx + \frac{2(1-v^2)}{E}\beta\cos 2kx = \eta\cos 2kx$$

$$\eta = \frac{(1-2v)(1+v)}{E}\gamma + \frac{2(1-v^2)}{E}\beta$$

(182)

In order to find the surface stress on $y = h(x)$, we use the transformation law in Appendix A and Taylor extrapolation that yield $\varepsilon_{ss}$ as

$$[\varepsilon_{ss}]_{y=h(x)} = \left[\varepsilon_{xx} + 2\delta h_{0x}\varepsilon_{xy} + \delta^2(h_{0x})^2(\varepsilon_{yy} - \varepsilon_{xx})\right]_{y=h(x)} =$$

$$\left[\begin{array}{l}\varepsilon_{xx}^{(0)} + \delta\left(\varepsilon_{xx}^{(1)} + 2h_{0x}\varepsilon_{xy}^{(0)} + h_0\frac{\partial\varepsilon_{xx}^{(0)}}{\partial y}\right) + \\ \delta^2\left(h_0\frac{\partial\varepsilon_{xx}^{(1)}}{\partial y} + \frac{1}{2}(h_0)^2\frac{\partial^2\varepsilon_{xx}^{(0)}}{\partial y^2} + 2(h_{0x})h_0\frac{\partial\varepsilon_{xy}^{(0)}}{\partial y} - (h_{0x})^2\varepsilon_{xx}^{(0)} + \varepsilon_{xx}^{(2)} + \right. \\ \left. 2h_{0x}\varepsilon_{xy}^{(1)} + (h_{0x})^2\varepsilon_{yy}^{(0)}\right)\end{array}\right]_{y=0}$$

$$= \frac{1-v^2}{E}\sigma^\infty + \delta\left\{-\frac{(1-2v)(1+v)}{E}k\left(\tau^o + k^s\frac{(1-v^2)}{E}\sigma^\infty - \tau^o\frac{v(1-v)}{E}\sigma^\infty\right) - \sigma^\infty\frac{2(1-v^2)}{E}\right\}\cos kx +$$

$$+\delta^2\left\{\begin{array}{l}-3\frac{(1-v^2)}{E}\sigma^\infty - 2\frac{(1-v^2)}{E}k\left(\tau^o + k^s\frac{(1-v^2)}{E}\sigma^\infty - \tau^o\frac{v(1-v)}{E}\sigma^\infty\right) - \frac{v(1+v)}{E}\sigma^\infty\cos^2 kx \\ +\frac{(1+v)}{E}\sigma_\infty\sin^2 kx + \eta\cos 2kx\end{array}\right\}$$

(183)

Proceeding similarly as in for previous parts we calculate the total energy of the rough half space and then find the effective surface stress and effective surface elastic constant as below. The detail calculations are not presented here.

$$\left(\tau^0\right)^{eff} = \tau^o - \delta^2 \tau^o \left( \begin{array}{c} \dfrac{3}{4} + kk^s \dfrac{(1+8\nu)(1+\nu)}{8E} - k\tau^o \left(\dfrac{\nu(3-\nu)}{8E}\right) - \\ -\dfrac{(1-2\nu)^2(1+\nu)^2}{2E^2} k^2 \left(k^s\right)^2 + -\dfrac{(1-2\nu)^2(1+\nu)^2}{E^2} k^2 \tau^o \dfrac{\nu}{2(1+\nu)} \end{array} \right) \quad (184)$$

If $\dfrac{kk^s}{E} \ll 1$, equations (148) can be further simplified as

$$\left(\tau^0\right)^{eff} = \tau^o \left(1 - \dfrac{3}{4}\delta^2\right) \quad (185)$$

And

$$\left(k^s\right)^{eff} = k^s + $$

$$+\delta^2 \left( \begin{array}{c} -\dfrac{E}{k(1-\nu^2)}\left[\dfrac{9-8\nu}{8(1-\nu)}\right] + \dfrac{1}{4}k^s + k(k^s)^2 \dfrac{(1+\nu)}{E}\left[\dfrac{-24\nu+7}{8}\right] + k(\tau^o)^2 \left[\dfrac{\nu^2(7-8\nu)}{8E(1+\nu)}\right] \\ +k^s\tau^o k\left[-\dfrac{\nu(7-12\nu)}{4E}\right] + \dfrac{(1-2\nu)^2(1+\nu)^2}{2E^2} k^2 k^s \left((k^s)^2 + \left(\tau^o\right)^2 \nu^2 - 2k^s \dfrac{\nu}{(1+\nu)}\tau^o\right) \end{array} \right) \quad (186)$$

Again, if $\dfrac{kk^s}{E} \ll 1$, equations (186) can be further simplified as

$$\left(k^s\right)^{eff} = k^s + \delta^2 \left(-\dfrac{E}{k(1-\nu^2)}\left[\dfrac{9-8\nu}{8(1-\nu)}\right]\right) \quad (187)$$

We conclude that the effect of asymmetry term in effective surface stress and effective surface elastic constant is negligible.